\documentclass[11pt]{article}
\PassOptionsToPackage{numbers,square,sort&compress}{natbib}
\usepackage[numbers,square,sort&compress]{natbib}

\makeatletter
\def\NAT@sort{\@ne}
\def\NAT@cmprs{\@ne}
\makeatother

\usepackage{hyperref}

\hypersetup{
   colorlinks=true,
   linkcolor={blue!100!black},
   citecolor={blue!100!black},
   urlcolor={blue!100!black},
   plainpages=false
}

\usepackage{float}

\usepackage{booktabs} %
\usepackage{multirow} %
\usepackage{threeparttable} %
\usepackage{siunitx} %
\usepackage{makecell} %
\usepackage{tabularx} %

\usepackage{tikz}
\usetikzlibrary{shapes,arrows} 

\usepackage[label font=it]{subfig}
\usepackage{caption}
\usepackage{multirow,tabularx}

\newsavebox{\twofigures}
\usepackage[export]{adjustbox}
\usepackage{calc}
\usepackage[version=4]{mhchem}
\usepackage{siunitx}
\usepackage{longtable,tabularx}
\usepackage{tabularray}
\usepackage{lmodern,pdftexcmds,amsmath}

\usepackage[title]{appendix}
\usepackage{titlesec}
\titleformat{\paragraph}
{\normalfont\normalsize}{\theparagraph}{1em}{}
\titlespacing*{\paragraph}
{0pt}{3.25ex plus 1ex minus .2ex}{1.5ex plus .2ex}
\counterwithin*{equation}{section}
\DeclareMathAlphabet{\mathsfbi}{OT1}{\sfdefault}{bx}{sl}
\DeclareMathVersion{sfletters}
\SetSymbolFont{letters}{sfletters}{OML}{ntxsfmi}{b}{it}

\makeatletter
\newcommand{\mathbfsbilow}[1]{%
  \text{\mathversion{sfletters}$\m@th#1$}%
}
\DeclareRobustCommand{\tensor}[1]{%
  \begingroup
  \ifcat\noexpand #1\relax
    \edef\greek@test{\detokenize{#1}}%
    \edef\greek@test{\expandafter\@cdr\greek@test\@nil}%
    \edef\greek@test{\expandafter\@car\greek@test\@nil}%
    \edef\x{\the\lccode\expandafter`\greek@test}%
    \edef\y{\number\expandafter`\greek@test}%
    \ifnum\x=\y\relax
      \mathbfsbilow{#1}%
    \else
      \mathsfbi{#1}%
    \fi
  \else
    \mathsfbi{#1}%
  \fi
  \endgroup
}
\makeatother

\usepackage{orcidlink}

\usepackage{tabularx, caption}
 \makeatletter
 \newcommand*{\compress}{\@minipagetrue}
 \makeatother

\newsavebox{\bigimage}

\usepackage{graphicx}
\usepackage{newtxtext}
\usepackage{newtxmath}

\usepackage{algorithm}
\usepackage[noend]{algpseudocode}

\hypersetup{
    colorlinks = true,
     citecolor  = blue,
    allcolors=blue,
}

\newcommand{\RomanNumeralCaps}[1]

\usepackage{multirow,tabularx}

\usepackage[export]{adjustbox}
\usepackage{calc}
\usepackage[version=4]{mhchem}
\usepackage{siunitx}
\usepackage{longtable,tabularx}
\usepackage{tabularray}
 \UseTblrLibrary{amsmath}
 \usepackage{lmodern,pdftexcmds,amsmath}
\usepackage{booktabs}
\usepackage{tabularx}
\usepackage{array} 
\usepackage[noblocks]{authblk}
\usepackage[margin=1in]{geometry}
\usepackage[title]{appendix}

\newcommand\affiliation[1]{\gdef\@affiliation{\let\aff\aff@inst#1}}
\gdef\@affiliation{}
\def\email#1{Email address for correspondence: #1}
\def\aff#1{\ignorespaces\textsuperscript{#1}}
\def\corresp#1{\unskip\thanks{#1}}
\numberwithin{equation}{section}

\renewenvironment{abstract}
{\begin{quote}
\noindent \rule{\linewidth}{.5pt}\par{\bfseries \abstractname.}}
{\medskip\noindent \rule{\linewidth}{.5pt}
\end{quote}
}

\hypersetup{
   colorlinks=true,
   linkcolor={blue!100!black},
   citecolor={blue!100!black},
   plainpages=false,
}

\title{\bf A Physics-Informed B-Spline Framework for Continuous Approximation of Flow Data}

\author[1]{\bf Junoh Jung\corresp{\email{jjung@anl.gov}}}
\author[1]{\bf David Lenz}
\author[1]{\bf Emil Constantinescu}
\author[1]{\bf  Tom Peterka}

\affil[1]{\normalsize Mathematics and Computer Science Division, Argonne National Laboratory, Lemont, IL 60439, USA \vspace{-1cm}}

\date{}
\begin{document}
\maketitle

\begin{abstract} Continuous approximations of flow data are useful for downstream analysis, differentiation, and visualization, but purely data-driven reconstructions do not, in general, preserve the governing physics. This limitation becomes particularly important when the input data are physically inconsistent, whether due to low-fidelity discretizations or unmodeled discrepancies. In such cases, the reconstructed fields may exhibit inaccurate PDE residuals, violated balance laws, or unreliable derived quantities. To address this limitation, we propose a physics-informed B-spline framework for the continuous approximation of flow data that embeds physical constraints directly into the reconstruction process. The method constructs compact, continuously differentiable representations of discrete fields using tensor-product B-splines and determines the spline control points by solving an optimization problem that balances data fidelity with residuals of the governing partial differential equations (PDEs), together with initial and boundary conditions. By leveraging exact analytical derivatives of the B-spline basis, the framework enables efficient and accurate evaluation of physical residuals without storing full-resolution fields. We refer to this approach as physics-informed multivariate functional approximation (PI-MFA), where multivariate functional approximation (MFA) is our spline-based framework for continuous representation and visualization. Numerical studies on the one-dimensional convection-diffusion equation, two-dimensional coupled Burgers equations, and two-dimensional incompressible Navier–Stokes equations show that PI-MFA reduces PDE residuals and improves global balance-law consistency. Compared with standard and regularized MFA, the proposed method produces more physically faithful reconstructions and, in cases involving physically inconsistent data, lower approximation errors, while also offering computational advantages over the tested physics-informed neural network architectures. Overall, PI-MFA preserves the compactness, local support, and exact differentiability of classical spline spaces while producing reliable continuous flow fields for scientific analysis and visualization. \\  
\end{abstract}

\section{Introduction}\label{sec:intro} 

B-splines and non-uniform rational B-splines (NURBS) provide a classical framework for smooth approximation and geometric modeling because they combine local support, partition of unity, and controllable continuity within a piecewise-polynomial basis \citep{piegl1997nurbs, deboor2001spline, schumaker2007spline}. Through tensor product construction, these properties extend naturally to multivariate curves, surfaces, and volumes, making spline spaces attractive for representing spatiotemporal scientific fields. Beyond computer-aided design (CAD), spline bases have long been used for interpolation, least-squares fitting, smoothing, scattered-data reconstruction, where analytic derivatives and sparse structured operators are especially valuable \citep{Eilers1996, johnson2009scattered,merchel2022fast}.

In computational mechanics, these same advantages motivated the development of isogeometric analysis (IGA), in which spline or NURBS spaces are used to represent both geometry and solution fields \citep{hughes2005isogeometric,bazilevs2006isogeometric,cottrell2007studies}. Their higher inter-element continuity, exact CAD geometry, and flexible refinement have made spline spaces a major framework for partial differential equation (PDE) discretization, while isogeometric collocation has further shown that spline smoothness can be exploited efficiently in strong-form formulations \citep{auricchio2010isogeometric,auricchio2012isogeometric,schillinger2013isogeometric,montardini2017optimal}. 
The present work has a different objective. Rather than computing a PDE solution directly, as in IGA, we seek a post hoc continuous reconstruction of already-generated discrete data that remains compact and differentiable for downstream analysis.

This objective is increasingly important because large-scale simulations enabled by modern high-performance computing now routinely generate spatiotemporal fields with billions of degrees of freedom. Although fine-resolution finite element, finite volume, and finite difference discretizations improve predictive capability, they also produce massive snapshot data that are expensive to store, transfer, and interrogate in postprocessing. This data bottleneck is particularly severe in large-scale transient simulations, for example, in computational fluid dynamics  or fluid–structure interaction problems, where reliable analysis often requires dense temporal sampling and repeated evaluation of derived quantities such as gradients, fluxes, and residual-based indicators.

Continuous data reconstruction addresses this challenge by replacing discrete simulation outputs with a compact functional representation. One such framework based on B-splines, multivariate functional approximation (MFA) \citep{peterka2018mfa}, enables evaluation at arbitrary resolution, analytic differentiation, and downstream analysis directly on the reconstructed model without accessing the raw simulation data. Recent work has significantly expanded the MFA framework. These advances include adaptive regularization to suppress spurious oscillations \citep{lenz_adaptive_2023}, scalable rendering pipelines for large B-spline models \citep{sun_scalable_2023,sun_mfa-dvr_2024}, adaptive refinement on varying-resolution tensor-product meshes \citep{peterka2023adaptive}, continuity recovery across parallel block interfaces \citep{grindeanu_cluster19}, domain-decomposition-based solvers \citep{mahadevan_jcs24}, and topological feature extraction on the continuous representation \citep{ma_topoinvis24,sun_ldav24}.

However, existing MFA models are not required to satisfy the governing physics that generated the data. From the perspective of computational mechanics, this limitation is important because visually plausible reconstructions can still be physically inadmissible. In fluid problems, for example, violations of conservation laws, boundary conditions, or initial conditions can lead to misleading gradients, incorrect fluxes, or spurious structures when users interrogate high-order derivatives or integrate quantities over space-time. Moreover, physically meaningful diagnostics such as mass balance, residual norms, and constraint satisfaction are often not explicitly stored in the raw output and therefore cannot be recovered reliably from a purely data-driven MFA.

This issue is further amplified in emerging AI/ML-coupled simulation and surrogate modeling workflows. In such settings, inexpensive low-fidelity models, such as coarse-grid simulations, reduced-order approximations, or cheaper numerical schemes, are often used to accelerate simulation speed, while data-driven models are trained to correct their discrepancy relative to high-fidelity data \citep{guo_multifidelity_2022,howard_multifidelity_2023,kiener_datadriven_2023,Kang2023Learning,sousa_enhancing_2024, kang_differentiable_2025, Jung2026Learning}. However, the approximated solution from low-fidelity simulation frequently remains the field that must be stored, transferred, reconstructed, and visualized during training, deployment, and downstream analysis. Consequently, under-resolution, discretization-induced imbalance, and numerical non-conservation present in the low-cost simulation can persist in the represented field. A related challenge arises when the available data is produced by approximating governing equations with missing or inaccurate source terms, closure relations, or other model-form deficiencies, so that the sampled field is not fully consistent with the target PDE \citep{garg_physics-integrated_2022}. Such cases further motivate the need for a physics-informed reconstruction framework that does more than just fit models to pointwise values.

Physics-informed modeling offers a natural way to address this limitation by embedding governing PDEs, boundary conditions (BCs), and initial conditions (ICs) directly into the approximation. Physics-informed neural and operator-learning methods have shown that enforcing governing laws can improve robustness and reconstruction fidelity, especially under sparse or noisy data \citep{raissi2019physics,yu2022gradient,shu_physics-informed_2023}. Related efforts in scientific visualization have likewise demonstrated the value of physics-aware reconstruction for producing more interpretable fields \citep{lutjens_flood_2020,chakravarty_hydraulic_2021,banerjee_picv_2024,ohashi_multifield_2025}. However, these previous approaches have not been designed as a compact spline-based functional representation for large scientific datasets, so a gap remains between data-driven MFA and physics-constrained reconstruction.

\begin{figure}
	\centering
	\includegraphics[width=1\textwidth]{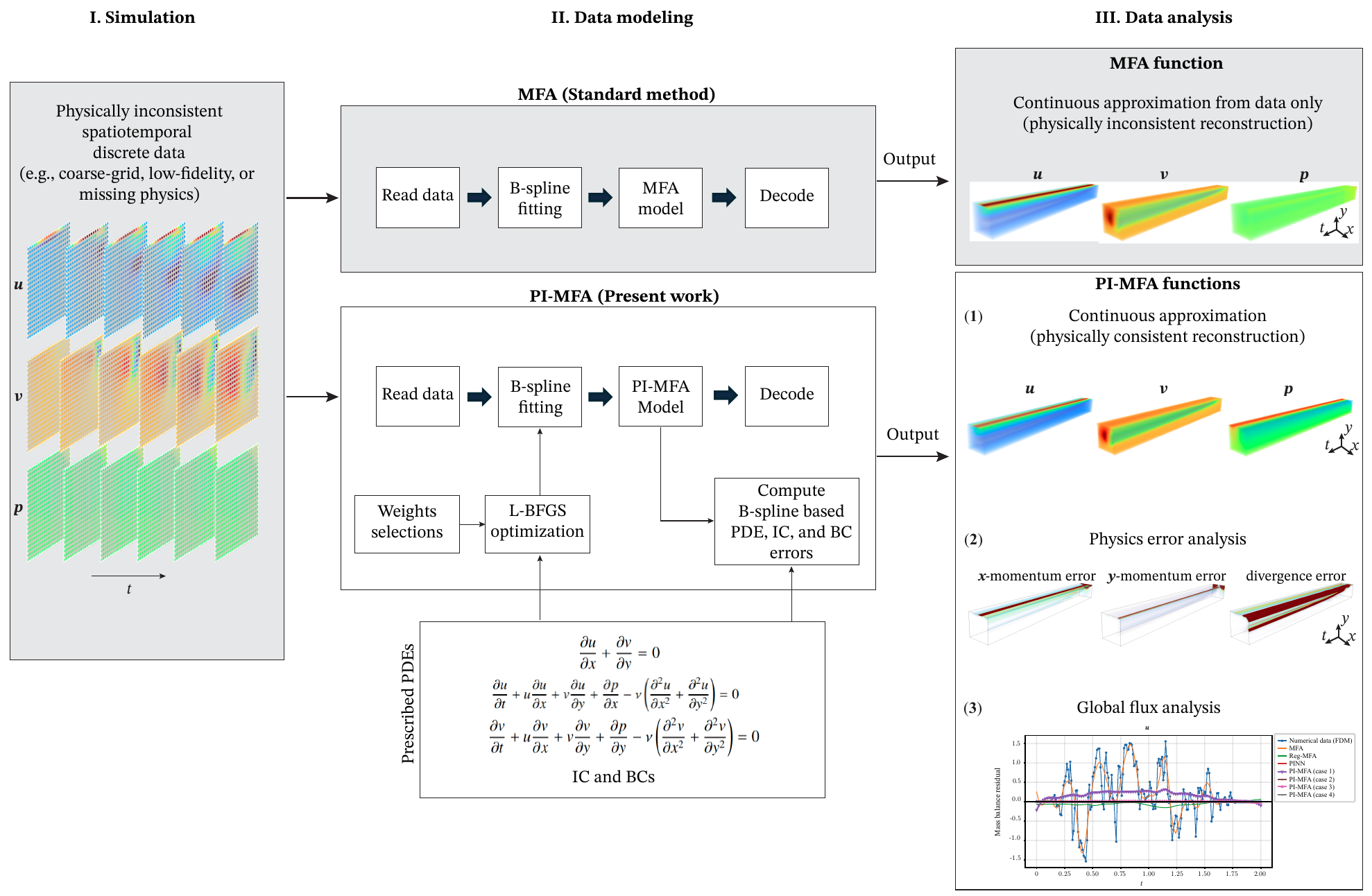}
  \caption{Schematic of the proposed PI-MFA. Boxes and components shown in the gray-shaded background represent existing work, whereas those in the white background highlight the new contributions of this study.}
  \label{fig:Scematics_PIMFA}
\end{figure}

To fill this gap, we propose PI-MFA, a physics-informed continuous reconstruction framework built upon the MFA architecture for the post hoc reconstruction of already-generated simulation data. As illustrated schematically in Fig.~\ref{fig:Scematics_PIMFA}, PI-MFA is used to construct a compact tensor product B-spline representation by solving a weighted optimization problem that explicitly couples data fidelity with strong-form PDE, IC, and BC residuals. By leveraging analytic B-spline derivatives, these governing equations and constraints are evaluated and assembled directly in coefficient space to optimize the B-spline control points.

Importantly, PI-MFA is not another direct spline-based PDE solver. Whereas isogeometric analysis, isogeometric collocation, and related spline-based discretizations \citep{hughes2005isogeometric,auricchio2010isogeometric,schillinger2013isogeometric,montardini2017optimal} use spline or NURBS spaces to compute PDE solutions from the governing equations, PI-MFA addresses a post hoc reconstruction problem. Given already-generated spatiotemporal data, it constructs a compact, dataset-specific tensor-product B-spline representation. The novelty lies in jointly optimizing the B-spline control points against data fidelity and PDE, IC, and BC residuals, so that the resulting representation is steered toward improved physical consistency. This distinguishes PI-MFA from classical spline smoothing and regularized MFA, where derivative-based terms primarily serve as roughness penalties or stabilizers \citep{Eilers1996,johnson2009scattered,lenz_adaptive_2023}, and from physics-informed neural surrogates such as PINNs \citep{raissi2019physics,yu2022gradient}, which use globally parameterized neural networks and are often designed to learn mappings or solutions through training. For a given dataset, PI-MFA instead yields a compact, locally supported, continuously differentiable spline model that supports storage reduction, efficient derivative and residual evaluation, balance diagnostics, and downstream analysis of multifield space-time data. Table~\ref{tab:method_summary} compares PI-MFA with representative spline-based PDE discretization (IGA), MFA, regularized MFA, and physics-informed neural-network methodologies, and highlights the distinction between PI-MFA and the reconstruction baselines used in this work.

\begin{table*}[t]
\centering
\scriptsize
\setlength{\tabcolsep}{2.5pt}
\renewcommand{\arraystretch}{1.18}
\caption{Comparison of PI-MFA with related spline-based PDE discretization, data-reconstruction, regularized reconstruction, and physics-informed neural-network approaches.}
\label{tab:method_summary}
\begin{threeparttable}
\begin{tabularx}{\textwidth}{@{}
>{\raggedright\arraybackslash}p{0.105\textwidth}
>{\raggedright\arraybackslash}p{0.135\textwidth}
>{\raggedright\arraybackslash}p{0.115\textwidth}
>{\centering\arraybackslash}p{0.055\textwidth}
>{\centering\arraybackslash}p{0.070\textwidth}
>{\raggedright\arraybackslash}p{0.145\textwidth}
>{\centering\arraybackslash}p{0.050\textwidth}
>{\raggedright\arraybackslash}p{0.070\textwidth}
>{\raggedright\arraybackslash}X
@{}}
\toprule
\textbf{Method} &
\textbf{Primary goal} &
\textbf{Parametrization / DoFs} &
\textbf{Uses existing data} &
\textbf{Forward PDE solver} &
\textbf{Physics term} &
\textbf{Local support} &
\textbf{Gradient} &
\textbf{Output / storage} \\
\midrule

IGA  &
Forward PDE discretization using CAD-compatible smooth bases &
B-spline/NURBS solution coefficients &
No &
Yes &
PDE discretization with boundary conditions &
Yes &
Analytic &
Numerical PDE solution in a spline/NURBS space. \\

\midrule

MFA &
Post hoc continuous reconstruction of sampled scientific data &
Tensor-product B-spline control points &
Yes &
No &
None &
Yes &
Analytic &
Compact local spline representation. \\

Reg-MFA &
Smooth post hoc reconstruction of sampled scientific data &
Tensor-product B-spline control points &
Yes &
No &
Roughness penalty &
Yes &
Analytic &
Compact local spline representation. \\

PINN &
Physics-informed solution, inverse problem, or reconstruction &
Neural-network weights &
Optional &
Optional &
PDE, IC, and BC residuals evaluated by automatic differentiation &
No &
AD &
Compact global neural parametrization. \\

\midrule

\textbf{PI-MFA} &
\textbf{Post hoc physics-aware reconstruction of already generated flow data} &
\textbf{Tensor-product B-spline control points} &
\textbf{Yes} &
\textbf{No} &
\textbf{Data misfit plus PDE, IC/BC, and auxiliary residual penalties} &
\textbf{Yes} &
\textbf{Analytic} &
\textbf{Compact local spline model with derivative, residual, and balance diagnostics.} \\

\bottomrule
\end{tabularx}

\end{threeparttable}
\end{table*}

We demonstrate the framework on the one-dimensional convection-diffusion equation, the two-dimensional Burgers equations, and the two-dimensional incompressible Navier-Stokes equations. Using data from low-fidelity simulations and models with intentional model-form discrepancies, we demonstrate that PI-MFA substantially reduces strong-form residuals and improves global balance consistency. Furthermore, depending on the source of the inconsistency, the method can enhance overall reconstruction accuracy and successfully recover unobserved variables, such as pressure, from velocity data alone.

The main contributions of this paper are as follows.
\begin{itemize}
\item We present a physics-informed extension of MFA that incorporates PDE, BC, and IC information into spline-based functional approximation, producing physically consistent continuous reconstructions while preserving the compactness and differentiability of standard MFA \citep{peterka2018mfa}. 

\item We formulate a unified objective that couples data fidelity and physical constraints. Analytic derivatives of the B-spline basis enable efficient assembly of the exact PDE residuals, block Jacobians, and objective gradients with respect to the control points, thereby supporting robust L-BFGS minimization.

\item Through canonical fluid benchmarks (1D convection-diffusion, 2D Burgers, and 2D Navier-Stokes), we demonstrate that PI-MFA systematically reduces strong-form residuals compared with data-only MFA, regularized MFA, and standard PINN. Furthermore, it restores global balance laws under model-form discrepancies, filters numerical noise from low-fidelity data, and demonstrates pressure inference in the cavity example from velocity-constrained momentum relations, up to a pressure reference condition.
\end{itemize}

The remainder of the paper is organized as follows. Section~\ref{sec:bspline_mfa} reviews B-splines and the standard MFA formulation. Section~~\ref{section:PI-MFA} presents the PI-MFA formulation, including the residual, Jacobian, and gradient constructions together with the L-BFGS solution strategy. Section~\ref{sec:results} reports numerical results for the convection-diffusion, Burgers, and Navier-Stokes test problems. Section~\ref{conclusion} summarizes the paper and discusses directions for future work.

\section{B-splines and MFA}\label{sec:bspline_mfa} 
In this section we review B-splines and the standard MFA formulation~\citep{peterka2018mfa} that forms the basis of our physics-informed extension. We also briefly describe regularized MFA~\citep{lenz_adaptive_2023}, which will serve as a comparative baseline in our numerical studies.
\subsection{B-splines and spline curves}\label{subsec:bspline_curve} 
B-splines are highly-differentiable functions with a compact and locally supported basis, and closed-form derivatives that support derivative-based diagnostics and PDE residual evaluation. Let $\{(v_j,Q_j)\}_{j=0}^{M-1}$ denote samples of a $D$-component quantity $Q_j\in\mathbb{R}^D$ at parameter values $v_j$ (e.g., time, arc length, or a mapped spatial coordinate). A degree-$p$ B-spline representation of the underlying curve/field is 
\begin{equation} 
C(u)=\sum_{i=0}^{n-1} N_{i,p}(u)\,P_i, \qquad P_i\in\mathbb{R}^D,\quad u\in[u_0,u_{n+p}], \label{eq:bspline-curve} 
\end{equation} 
where $\{P_i\}_{i=0}^{n-1}$ are the (unknown) control points and $\{N_{i,p}\}_{i=0}^{n-1}$ are B-spline basis functions defined over a non-decreasing knot vector  $\mathcal U=(u_0,u_1,\dots,u_{n+p})$. The basis functions are piecewise polynomials. Each $N_{i,p}$ is a polynomial of degree $p$ on each knot span $[u_k,u_{k+1})$, has compact support on $[u_i,u_{i+p+1})$, and is nonnegative; and the basis functions form a partition of unity. Knot multiplicity controls smoothness: if a knot has multiplicity $r$, the basis continuity across that knot is $C^{p-r}$. The basis function $N_{i,p}$ is defined via the standard Cox-de Boor recursion~\citep{cox1972bspline,deboor1972bspline}. For $p=0$, 
\begin{equation} 
N_{i,0}(u)= \begin{cases} 1, & u_i \le u < u_{i+1},\\ 0, & \text{otherwise}, \end{cases} 
\end{equation} 
and for $p\ge 1$, 
\begin{equation} N_{i,p}(u) = \frac{u-u_i}{u_{i+p}-u_i}\,N_{i,p-1}(u) + \frac{u_{i+p+1}-u}{u_{i+p+1}-u_{i+1}}\,N_{i+1,p-1}(u), \label{eq:cox-de-boor} 
\end{equation} 
with the convention that terms with zero denominators are set to zero. A key advantage for mechanics applications is that exact derivatives are easily computable. For $p\ge 1$, 
\begin{equation} \frac{d}{du}N_{i,p}(u) = \frac{p}{u_{i+p}-u_i}\,N_{i,p-1}(u) - \frac{p}{u_{i+p+1}-u_{i+1}}\,N_{i+1,p-1}(u), \label{eq:bspline-deriv} 
\end{equation} 
and therefore 
\begin{equation} C'(u)=\sum_{i=0}^{n-1} N'_{i,p}(u)\,P_i. \label{eq:curve-deriv} 
\end{equation} 
Higher-order derivatives follow similarly and are computed stably via recursion, allowing direct evaluation of spatial/temporal derivatives required by PDE operators. 
\subsection{Tensor product B-splines}\label{subsec:tensor-product-bsplines} 
MFA~\citep{peterka2018mfa} represents multidimensional, multivariate fields (e.g., space–time solution snapshots) using tensor-product B-splines. Here, we have $d\ge 2$ that denotes the number of parametric coordinates (e.g., $d=2$ for $(x,t)$, $d=3$ for $(x,y,t)$). For each coordinate $\ell=1,\dots,d$, we define a knot vector  $\mathcal U^{(\ell)}=(u^{(\ell)}_0,\dots,u^{(\ell)}_{n_\ell+p_\ell})$,  with degree $p_\ell$ and $n_\ell$ basis functions, and $N^{(\ell)}_{i_\ell,p_\ell}(u_\ell)$ is the corresponding univariate basis. With $\mathbf{u}=(u_1,\dots,u_d)$ and a multi-index $\mathbf{i}=(i_1,\dots,i_d)$, the tensor-product spline is 
\begin{equation} 
C(\mathbf{u}) = \sum_{i_1=0}^{n_1-1}\cdots\sum_{i_d=0}^{n_d-1} \Bigg(\prod_{\ell=1}^{d} N^{(\ell)}_{i_\ell,p_\ell}(u_\ell)\Bigg)\,P_{\mathbf{i}}, \qquad P_{\mathbf{i}}\in\mathbb{R}^{D}. \label{eq:tensor-bspline} 
\end{equation} 
This separable construction yields surfaces ($d=2$), volumes ($d=3$), and higher-dimensional space-time representations. Importantly, partial derivatives are again analytic and inherit tensor separability. Differentiating with respect to coordinate $u_\ell$ yields 
\begin{equation} 
\frac{\partial C}{\partial u_\ell}(\mathbf{u}) = \sum_{\mathbf{i}} \left(\frac{d}{du_\ell}N^{(\ell)}_{i_\ell,p_\ell}(u_\ell) \prod_{k\ne \ell} N^{(k)}_{i_k,p_k}(u_k)\right) P_{\mathbf{i}}, \qquad \ell=1,\dots,d. \label{eq:tensor-deriv} 
\end{equation} 
This property is essential for PI–MFA because PDE residual terms can be assembled directly from spline derivatives, without resorting to any additional differentiation of the original data. 

\subsection{The MFA data-fitting problem}\label{subsec:The_MFA_data-fitting_problem} 

Next, we formulate the MFA data-fitting problem based on the tensor-product B-splines defined in Eq.~\eqref{eq:tensor-bspline}. Let the sample locations in the spline parameter domain be
\(\mathbf v_j=(v_{j,1},\dots,v_{j,d})\), \(j=1,\dots,M\), and let
\(Q\in\mathbb{R}^{M\times D}\) collect the corresponding sampled field values, with the \(j\)-th row denoted by \(Q_j^\top\in\mathbb{R}^D\). We flatten the multi-index \(\mathbf i=(i_1,\dots,i_d)\) into a single linear index \(\bar{i}=1,\dots,n_c\), where \(n_c=\prod_{\ell=1}^d n_\ell\) is the total number of tensor-product control points. Evaluating the tensor-product spline at the sample points gives
\begin{equation} \label{eq:tensor-product}
C(\mathbf v_j)=\sum_{\bar{i}=1}^{n_c} A_{j\bar{i}} P_{\bar{i}}, 
\qquad 
A_{j\bar{i}}=\prod_{\ell=1}^d 
N^{(\ell)}_{i_\ell,p_\ell}(v_{j,\ell}), 
\end{equation} 
where \(P_{\bar{i}}\in\mathbb{R}^{D}\) is the control point associated with the tensor-product basis function indexed by \(\bar{i}\). Collecting the evaluations at all sample locations yields the matrix form \(AP\), where
\(A\in\mathbb{R}^{M\times n_c}\) and \(P\in\mathbb{R}^{n_c \times D}\). The standard MFA fitting problem is therefore
\begin{equation}\label{eq:least-square_MFA}  
\min_P \|Q-AP\|_F^2, 
\end{equation} 
where \(\|\cdot\|_F\) denotes the Frobenius norm. By the local support of the B-spline basis, at most \(\prod_{\ell=1}^d (p_\ell+1)\) tensor-product basis functions are nonzero at any sample location, so each row of \(A\) contains only a small number of nonzero entries. Thus, once the knot vectors, basis functions, and sample locations are fixed, \(A\) is fixed, and standard MFA model fitting is a linear least-squares minimization problem in the control points. The physics-informed extension introduced in Section~\ref{section:PI-MFA} retains the same spline representation but augments the objective with PDE, BC, and IC residual terms. For linear governing equations with linear constraints, the resulting objective remains quadratic in the control points, while for nonlinear governing equations, the PDE residuals make the PI-MFA optimization nonlinear.


\subsection{Regularized MFA}\label{subsec:reg_mfa}

Regularized spline approximation predates MFA by several decades. Classical penalized B-spline and spline-smoothing formulations combine least-squares fitting with roughness penalties on spline derivatives or on finite differences of neighboring coefficients, thereby suppressing overfitting and spurious oscillations while preserving the locality and sparsity of the B-spline basis \citep{Eilers1996,johnson2009scattered}. From this perspective, regularized MFA may be viewed as the tensor product scientific-data analogue of classical regularized B-spline approximation. The adaptive regularization framework developed for MFA \citep{lenz_adaptive_2023} specializes these ideas to compact spline models of scientific fields and provides an important baseline for the present work.

Although B-spline bases are smooth and locally supported, a purely data-driven MFA fit may still develop spurious oscillations when the control lattice is overly flexible relative to the sampling density or when the target field contains noise or underresolved features. A standard remedy is to augment the least-squares data-fit objective in Eq.~\eqref{eq:least-square_MFA} with penalty terms that control variation of the spline coefficients across the parametric directions.

Based on the notations we defined in Section~\ref{subsec:The_MFA_data-fitting_problem}, the regularized MFA problem is then written as
\begin{equation}
\min_{P} \mathcal{L}_{\mathrm{reg}}(P)
=
\|Q-AP\|_{F}^{2}
+
\mathcal{R}(P),
\label{eq:reg_mfa_obj}
\end{equation}
where $\mathcal{R}(P)$ is a roughness penalty. In the present work, we compare our proposed method with the first- and second-order regularization in each parametric direction,
\begin{equation}
\mathcal{R}(P)
=
\sum_{\ell=1}^{d} \|D_{\ell}^{(1)}P\|_{F}^{2}
+
\sum_{\ell=1}^{d} \|D_{\ell}^{(2)}P\|_{F}^{2},
\label{eq:reg_mfa_penalty}
\end{equation}
where $D_{\ell}^{(1)}$ and $D_{\ell}^{(2)}$ denote linear operators respectively measuring first- and second-order variation along the parametric coordinate $u_{\ell}$. These operators may be assembled from analytic B-spline derivative matrices or equivalently from local finite-difference-type stencils on the control lattice. The first-order term limits excessive local variation, whereas the second-order term penalizes curvature and is particularly effective in damping high-frequency oscillations. Since the spline representation remains linear in the control points, the regularized problem preserves the quadratic least-squares structure of standard MFA. Hence, regularization improves stability and suppresses oscillatory artifacts without changing the basic algebraic structure of MFA.

However, regularized MFA remains a data-driven approximation method. The additional penalty terms promote smoothness, but they do not explicitly enforce governing equations, BCs, or ICs. Regularized MFA controls roughness geometrically, whereas the PI-MFA formulation introduced next incorporates physics-based constraints to steer the reconstruction toward physical consistency.

\section{Physics-informed MFA}\label{section:PI-MFA} 
We now extend the standard MFA framework of Section~\ref{sec:bspline_mfa} to a physics-informed setting by embedding PDE, BC, and IC information into the fitting algorithm. The key idea is to retain MFA's compact, tensor-product B-spline representation while augmenting the fitting objective with physics-based residuals that are evaluated by using analytic spline derivatives.

\subsection{Physics-informed MFA formulation}\label{subsec:pi-mfa-formulation}

We consider a physical system governed by a set of partial differential equations (PDEs), expressed in the general operator form
\begin{equation}
\mathcal{N}[f](z)=\mathbf{0},
\qquad z\in\Omega\times(0,T],
\label{eq:generic_pde}
\end{equation}
where $\mathcal{N}[f](z)\in\mathbb{R}^{n_{\mathrm{eq}}}$ represents the PDE residual vector of the governing equations for a physical field $f$. Here, $z=(\mathbf{x},t)$ denotes a coordinate point in the spatiotemporal domain $\Omega\times[0,T]$, with $\Omega\subset\mathbb{R}^{d-1}$ representing the spatial domain.

To approximate the true solution $f$, we employ the continuous B-spline reconstructed field $C$ defined in Eq.~\eqref{eq:tensor-product}. This field is parameterized by a tensor of control points, $P$. To facilitate standard unconstrained optimization, we flatten these control points into a single state vector
\[
\mathbf{p}:=\operatorname{vec}(P)\in\mathbb{R}^{n_cD}.
\]

The PI-MFA framework relies on four distinct sets of collocation points to enforce data fidelity and physical constraints across the domain
\begin{subequations}
\begin{align}
\mathcal{D} &= \{(z_k^d,\mathbf{f}_k^d)\}_{k=1}^{M_{\mathrm{data}}}, & z_k^d&=(\mathbf{x}_k^d,t_k^d), \\
\mathcal{R} &= \{z_k^r\}_{k=1}^{M_{\mathrm{pde}}}, & z_k^r&=(\mathbf{x}_k^r,t_k^r), \\
\mathcal{B} &= \{(z_k^b,\mathbf{f}_k^b)\}_{k=1}^{M_{\mathrm{BC}}}, & z_k^b&=(\mathbf{x}_k^b,t_k^b), \\
\mathcal{I} &= \{(\mathbf{x}_k^0,\mathbf{f}_k^0)\}_{k=1}^{M_{\mathrm{IC}}}.
\end{align}
\end{subequations}
where $M$ is the number of the data points. Here, the vectors $\mathbf{f}_k^d$, $\mathbf{f}_k^b$, and $\mathbf{f}_k^0$ denote the known target field values at the observation data, boundary, and initial-condition points, respectively. Evaluating the parameterized field $C(z; \mathbf{p})$ at these discrete points yields the following pointwise residual functions
\begin{subequations}
\begin{align}
R_{\mathrm{data},k}(\mathbf{p}) &= C(z_k^d;\mathbf{p})-\mathbf{f}_k^d, \\
R_{\mathrm{pde},k}(\mathbf{p})  &= \mathcal{N}[C](z_k^r;\mathbf{p}), \\
R_{\mathrm{BC},k}(\mathbf{p})   &= C(z_k^b;\mathbf{p})-\mathbf{f}_k^b, \\
R_{\mathrm{IC},k}(\mathbf{p})   &= C(\mathbf{x}_k^0,0;\mathbf{p})-\mathbf{f}_k^0.
\end{align}
\end{subequations}

These pointwise residuals are subsequently aggregated into corresponding mean squared error (MSE) loss components
\begin{equation}
\mathcal{L}_{\star}(\mathbf{p}) = \frac{1}{M_{\star}}\sum_{k=1}^{M_{\star}}\|R_{\star,k}(\mathbf{p})\|_2^2, \qquad \star\in\{\mathrm{data},\mathrm{pde},\mathrm{BC},\mathrm{IC}\}.
\end{equation}
The total loss is then formulated as a weighted sum of these individual components
\begin{equation}
\mathcal{L}_{\mathrm{total}}(\mathbf{p}) = \lambda_{\mathrm{data}}\mathcal{L}_{\mathrm{data}}(\mathbf{p}) + \lambda_{\mathrm{pde}}\mathcal{L}_{\mathrm{pde}}(\mathbf{p}) + \lambda_{\mathrm{BC}}\mathcal{L}_{\mathrm{BC}}(\mathbf{p}) + \lambda_{\mathrm{IC}}\mathcal{L}_{\mathrm{IC}}(\mathbf{p}),
\label{eq:total_loss}
\end{equation}
where $\lambda_{\star}$ are user-defined penalty weights that balance the influence of the governing physics against the data and boundary and initial constraints.

To efficiently minimize this composite objective using gradient-based optimizers (e.g., L-BFGS), we consolidate the pointwise residuals into stacked global vectors
\[
\mathbf{r}_\star(\mathbf{p}) = \begin{bmatrix} R_{\star,1}(\mathbf{p})\\ \vdots\\ R_{\star,n_\star}(\mathbf{p}) \end{bmatrix}, \qquad \star\in\{\mathrm{data},\mathrm{pde},\mathrm{BC},\mathrm{IC}\},
\]
alongside their corresponding global Jacobian matrices with respect to the control variables,
\[
J_\star(\mathbf{p}) = \frac{\partial \mathbf{r}_\star}{\partial \mathbf{p}}.
\]

By applying the chain rule, the exact analytical gradient of each individual loss component takes the form
\begin{equation}
\nabla_{\mathbf{p}}\mathcal{L}_\star(\mathbf{p}) = \frac{2}{M_\star}J_\star(\mathbf{p})^\top \mathbf{r}_\star(\mathbf{p}), \qquad \star\in\{\mathrm{data},\mathrm{pde},\mathrm{BC},\mathrm{IC}\}.
\label{eq:generic_loss_gradient}
\end{equation}
Consequently, the full expression for the gradient of the total objective function is efficiently assembled as
\begin{equation}
\nabla_{\mathbf{p}}\mathcal{L}_{\mathrm{total}}(\mathbf{p}) = \frac{2\lambda_{\mathrm{data}}}{M_{\mathrm{data}}}J_{\mathrm{data}}^\top \mathbf{r}_{\mathrm{data}} + \frac{2\lambda_{\mathrm{pde}}}{M_{\mathrm{pde}}}J_{\mathrm{pde}}^\top \mathbf{r}_{\mathrm{pde}} + \frac{2\lambda_{\mathrm{BC}}}{M_{\mathrm{BC}}}J_{\mathrm{BC}}^\top \mathbf{r}_{\mathrm{BC}} + \frac{2\lambda_{\mathrm{IC}}}{M_{\mathrm{IC}}}J_{\mathrm{IC}}^\top \mathbf{r}_{\mathrm{IC}}.
\label{eq:generic_total_gradient}
\end{equation}

For the linear constraints, specifically those arising from observation data, Dirichlet boundary conditions, and initial conditions, the residuals can be expressed directly in matrix-vector form
\begin{equation}
\mathbf{r}_{\mathrm{data}}=N_{\mathrm{data}}\mathbf{p}-\mathbf{q}_{\mathrm{data}}, \qquad \mathbf{r}_{\mathrm{BC}}=N_{\mathrm{BC}}\mathbf{p}-\mathbf{q}_{\mathrm{BC}}, \qquad \mathbf{r}_{\mathrm{IC}}=N_{\mathrm{IC}}\mathbf{p}-\mathbf{q}_{\mathrm{IC}},
\label{eq:generic_linear_terms}
\end{equation}
where $N_{\star}$ represents the B-spline basis matrix constructed at the corresponding sample coordinates, and $\mathbf{q}_{\star}$ stores the concatenated target values. This linear relationship conveniently yields constant Jacobians
\[
J_{\mathrm{data}}=N_{\mathrm{data}}, \qquad J_{\mathrm{BC}}=N_{\mathrm{BC}}, \qquad J_{\mathrm{IC}}=N_{\mathrm{IC}},
\]
which only need to be precomputed once prior to optimization. In contrast, the structure of the PDE Jacobian depends entirely on the governing physics. For a linear PDE, $J_{\mathrm{pde}}$ remains a constant matrix. However, for nonlinear governing equations, the Jacobian $J_{\mathrm{pde}}(\mathbf{p})$ dynamically depends on the current state of the control points and is updated at each optimization iteration.

\subsection{Optimization with L-BFGS}\label{subsec:Optimization-with-L-BFGS}

We minimize the PI-MFA objective in Eq.~\eqref{eq:total_loss}
with respect to the vectorized spline control points
$\mathbf p=\mathrm{vec}(P)$ using the limited-memory BFGS method
(L-BFGS) with a strong Wolfe line search \citep{Liu_1989}. The analytical
gradients derived in each example are supplied directly to the optimizer. These
gradients are exact for the finite-dimensional B-spline objective because the
field values and all required spatial and temporal derivatives are evaluated
analytically from the B-spline basis. Thus, no finite-difference approximation
of the objective gradient is required.

In the implementation, the full global residual vector, weight matrix, and
global Jacobian are not assembled as dense objects. Instead, the data, PDE,
BC, IC, and auxiliary constraint contributions
are evaluated and accumulated blockwise. For the linear data, Dirichlet
BC, and IC terms used in this work,
the corresponding Jacobians are fixed sparse basis-evaluation matrices. For
linear PDEs, the PDE Jacobian is also fixed. For nonlinear systems, the PDE
Jacobian depends on the current spline state and is updated at each optimization
iteration.

At iteration $k$, let $\mathbf p_k$ denote the current control point
vector and let $\mathbf g_k = \nabla_{\mathbf p} \mathcal{L}(\mathbf p_k)$ be the analytical gradient of the PI-MFA objective. Given a search direction
$\mathbf d_k$ and a step length $\alpha_k$ accepted by the strong Wolfe line
search, the coefficients are updated as $\mathbf p_{k+1} = \mathbf p_k + \alpha_k \mathbf d_k$. The L-BFGS curvature pair is then defined by
\begin{equation}
    \mathbf s_k = \mathbf p_{k+1}-\mathbf p_k,
    \qquad
    \mathbf y_k = \mathbf g_{k+1}-\mathbf g_k,
    \qquad
    \rho_k = \frac{1}{\mathbf y_k^\top \mathbf s_k}.
    \label{eq:lbfgs_curvature_pair}
\end{equation}
These curvature pairs approximate the action of the inverse Hessian on the
gradient. In the limited-memory variant, the inverse-Hessian matrix is not
formed explicitly. Instead, only the most recent $m$ pairs
$\{(\mathbf s_i,\mathbf y_i)\}$ are stored, and the search direction is computed
by the standard two-loop recursion,
\[
    \mathbf d_k = -H_k^{(m)}\mathbf g_k ,
\]
where $H_k^{(m)}$ denotes the implicit limited-memory inverse-Hessian
approximation. The strong Wolfe line search promotes positive curvature pairs,
$\mathbf y_k^\top \mathbf s_k>0$, which helps preserve the positive definiteness
of this implicit approximation.

In all examples, the data-only MFA solution is used as the initial control points. At each iteration, the current spline fields and their
analytical derivatives are evaluated at the data and collocation points, the
residual-block contributions and analytical gradient are assembled, and L-BFGS
updates its implicit curvature approximation using the stored history pairs. The
implementation uses the L-BFGS++ library \citep{qiu_lbfgspp}.

Although the PI-MFA objective has a weighted nonlinear least-squares structure,
we use L-BFGS rather than Gauss--Newton or Levenberg--Marquardt in the present
work. The main practical reason is scalability with respect to the number of
tensor-product space-time control points. In the Burgers and Navier-Stokes examples, the unknown vector contains the stacked control points of multiple coupled fields, so Gauss--Newton-type methods would require repeated solution of large linearized least-squares systems, either with explicitly assembled Jacobians or through matrix-free iterations. L-BFGS requires only objective and gradient evaluations, has a limited-memory footprint, and provides a single optimization strategy that applies uniformly to the linear and nonlinear examples considered here. A
systematic comparison with Gauss--Newton, Levenberg--Marquardt, and other structure-exploiting solvers is interesting, but outside the scope of the present study.

\section{Results}\label{sec:results}
In this section we evaluate the performance of the proposed physics-informed MFA  framework on three canonical partial differential equations: the one-dimensional convection–diffusion equation, the two-dimensional Burgers' equations, and the two-dimensional incompressible Navier-Stokes equations. The MFA framework inherently constructs continuous representations in space and time. Thus, 1D spatial data defined over $x$ is naturally reconstructed in space–time as $f(x,t)$, and 2D spatial data defined over $(x,y)$ is reconstructed as $f(x,y,t)$.

The performance of the PI-MFA framework is assessed with three metrics: the reconstruction error relative to the target data, the strong-form PDE residuals, and the global mass-balance discrepancy (evaluated only for the 1D convection-diffusion and 2D Burgers cases).

To quantify the reconstruction accuracy, we define a scalar error metric based on the MSE. The MSE between the approximated field ($f_{\mathrm{approx}}$) and a reference solution ($f_{\mathrm{ref}}$, representing either the analytical solution or the high-fidelity simulation data) is computed over the full space-time grid as
\begin{equation}\label{eq:MSE}
\mathrm{E}=\frac{1}{M_{total}}\sum_{i,n}(f_{\mathrm{approx}}(x_{i},t_{n})-f_{\mathrm{ref}}(x_{i},t_{n}))^{2}.
\end{equation}

We benchmark PI-MFA against purely data-driven MFA, regularized MFA (Reg-MFA), and PINNs. To ensure a fair comparison with our spline-based methods, the PINNs are implemented as reconstruction baselines trained separately on each fixed dataset, rather than as standard forward solvers. For all test cases, the PINN objective function combines a data misfit term with penalties for the strong-form PDE, ICs, and BCs, evaluated on the identical observational data. Regarding network architectures, the 1D convection–diffusion baseline employs a multilayer perceptron (MLP) with tanh activations and five hidden layers of 64 neurons, trained via Adam followed by L-BFGS. The 2D Burgers and 2D Navier–Stokes baselines utilize similar tanh MLPs but with six hidden layers of 128 neurons. More details on the hyperparameters are provided in Appendix~\ref{app:pinn_baseline_settings}.

\subsection{One-dimensional convection-diffusion equation}
\label{subsec:1d-convdiff}
\begin{figure}
	\centering
	\includegraphics[width=.9\textwidth]{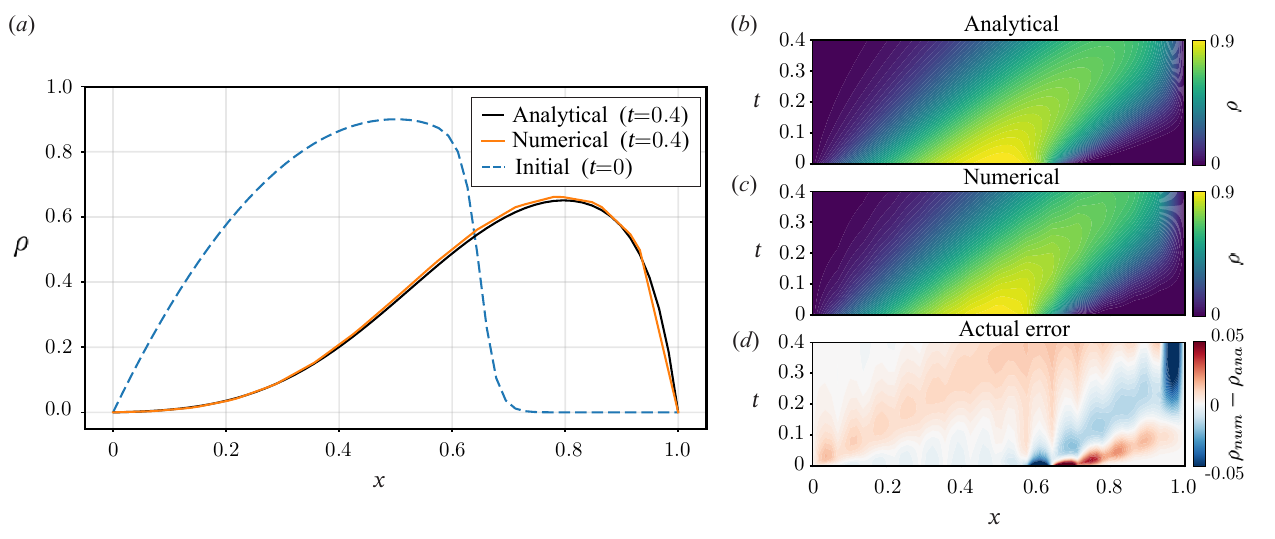}
  \caption{One-dimensional convection-diffusion problem on $x\in[0,1]$ with homogeneous Dirichlet boundary conditions.
  Left: initial condition at $t=0$ (blue dashed) and terminal-time profiles at $T=0.4$ obtained from a non-conservative finite difference simulation (orange) and the analytical reference solution (black).
  Right: spatiotemporal fields of the analytical solution $\rho_{\mathrm{ana}}$, the numerical solution $\rho_{\mathrm{num}}$, and the pointwise error $e(x,t)=\rho_{\mathrm{num}}(x,t)-\rho_{\mathrm{ana}}(x,t)$.}
  \label{fig:1dconvdiff_Case1}
\end{figure}

We consider the one-dimensional convection-diffusion equation
\begin{equation}
\frac{\partial \rho}{\partial t} + a\,\frac{\partial \rho}{\partial x}
= \nu\,\frac{\partial^2 \rho}{\partial x^2},
\qquad x\in[0,1],\;\; t\in[0,T],
\label{eq:1Dconvdiff}
\end{equation}
where $\rho(x,t)$ denotes a transported scalar, $a$ is the (constant) convection speed, and $\nu$ is the diffusion coefficient.
In the following experiments, we set $a=1$ and $\nu=0.05$. Homogeneous Dirichlet boundary conditions are imposed,
$\rho(0,t)=\rho(1,t)=0$ for all $t\in[0,T]$. In order to ensure compatibility with these boundary conditions, the initial profile is multiplied by the vanishing factor
$\phi(x)=4x(1-x)$, which satisfies $\phi(0)=\phi(1)=0$. In this work, we use the front-type initial condition
\begin{equation}\label{eq:initial-condition}
\rho(x,0)
= \frac{A_{\mathrm{f}}}{2}\left(1+\tanh\!\left(\frac{x_f-x}{w}\right)\right)\phi(x),
\qquad
A_{\mathrm{f}}=0.9,\;\; x_f=0.65,\;\; w=0.03,
\end{equation}
as shown in Fig.~\ref{fig:1dconvdiff_Case1}(\textit{a}).
Additional initial conditions used for supplementary tests are provided in Appendix~\ref{app:1Dconv_differentICs}.

The solution for Eq.~\eqref{eq:1Dconvdiff} with constant coefficients and homogeneous Dirichlet boundary conditions is obtained by removing the advective drift via an exponential substitution and then applying a sine-series
eigenfunction expansion~\citep{Crank1975Diffusion,Haberman2013AppliedPDE}. Introducing
\begin{equation}
\alpha=\frac{a}{2\nu},\qquad u(x,t)=e^{-\alpha x}\rho(x,t),
\end{equation}
one obtains the reaction-diffusion equation $u_t=\nu u_{xx}-\frac{a^2}{4\nu}u$ with $u(0,t)=u(1,t)=0$. The solution
can be written as the modal series
\begin{equation}\label{eq:analytical-solution}
\rho_{\mathrm{ana}}(x,t)=e^{\alpha x}\sum_{n=1}^{\infty} b_n \sin(n\pi x)\,
\exp\!\left(-\left[\nu(n\pi)^2+\frac{a^2}{4\nu}\right]t\right),
\qquad
b_n = 2\int_0^1 e^{-\alpha x}\rho(x,0)\sin(n\pi x)\,dx.
\end{equation}

In this work, Eq.~\eqref{eq:analytical-solution} was evaluated by truncating the modal expansion after \(N_m=900\) sine modes. The coefficients
\(b_n\) were computed using a composite trapezoidal rule on a uniform grid with
\(N_q=4097\) quadrature nodes on \([0,1]\). As a convergence check, we repeated
the evaluation with \(N_m=1800\) and \(N_q=8193\). On the full space--time grid
used for the reported error metrics, the resulting change in
\(\rho_{\mathrm{ana}}\) was at most \(5.59\times 10^{-5}\) in the \(L^\infty\)
norm and \(1.27\times 10^{-6}\) in the relative discrete \(L^2\) norm.

As an additional diagnostic of physics consistency, we monitor the integrated mass balance implied by the
conservative (flux) form of Eq.~\eqref{eq:1Dconvdiff}. Assuming $\rho$ is sufficiently smooth to interchange differentiation
and integration, Eq.~\eqref{eq:1Dconvdiff} can be written as
$\rho_t + \partial_x J = 0$, with total flux $J(x,t)=a\,\rho(x,t)-\nu\,\rho_x(x,t)$.
Integrating over a space-time control volume $R=[x_0,x_1]\times[t_0,t_1]$ and applying the fundamental theorem of
calculus in both $x$ and $t$, we get  the integral balance
\begin{equation}
\int_{x_0}^{x_1} \rho(x,t_1)\,\mathrm{d}x - \int_{x_0}^{x_1} \rho(x,t_0)\,\mathrm{d}x
+ \int_{t_0}^{t_1}\big[J(x_1,t)-J(x_0,t)\big]\,\mathrm{d}t = 0,
\end{equation}
or, equivalently, after separating advective and diffusive contributions,
\begin{equation}
\int_{x_0}^{x_1}\big[\rho(x,t_1)-\rho(x,t_0)\big]\,\mathrm{d}x
+ a\int_{t_0}^{t_1}\big[\rho(x_1,t)-\rho(x_0,t)\big]\,\mathrm{d}t
= \nu\int_{t_0}^{t_1}\big[\rho_x(x_1,t)-\rho_x(x_0,t)\big]\,\mathrm{d}t.
\end{equation}
This identity states that the change in total ``mass'' $\int_{x_0}^{x_1}\rho\,\mathrm{d}x$ between $t_0$ and $t_1$ is
exactly balanced by the net advective and diffusive fluxes through the spatial boundaries, accumulated over
$[t_0,t_1]$. For the full domain $[0,1]$ with homogeneous Dirichlet boundaries, the advective boundary contribution
vanishes (since $\rho(0,t)=\rho(1,t)=0$), and the balance reduces to
\begin{equation}
\int_{0}^{1}\big[\rho(x,t_1)-\rho(x,t_0)\big]\,\mathrm{d}x
- \nu\int_{t_0}^{t_1}\big[\rho_x(1,t)-\rho_x(0,t)\big]\,\mathrm{d}t=0,\label{eq:1dmassbalance}
\end{equation}
which provides an integral measure of how well a candidate field satisfies flux consistency.

\subsubsection{Dataset generation and standard MFA baseline}\label{subsubsec:1D_datageneration_and_MFA}
To emulate discretization-induced conservation errors, we generate a synthetic observation dataset ($\rho_{\mathrm{num}}$) using an intentionally degraded finite difference approximation of the strong form of Eq.~\eqref{eq:1Dconvdiff}. Specifically, the solution is advanced in time via the explicit forward Euler method on a coarsened spatial grid, employing standard centered-difference approximations for both the convective ($\rho_x$) and diffusive ($\rho_{xx}$) terms. The simulation is performed with $N_x = 60$ spatial nodes and integrated over the interval $t \in [0, 0.4]$ using a time step of $\Delta t = 10^{-3}$, yielding $N_t = 400$ temporal snapshots. This low-fidelity dataset is generated using an intentionally degraded finite-difference discretization that exhibits a measurable global flux-balance error under the diagnostic in Eq.~\eqref{eq:1dmassbalance}. The resulting pointwise error, defined as $e(x,t) = \rho_{\mathrm{num}}(x,t) - \rho_{\mathrm{ana}}(x,t)$, is illustrated in Fig.~\ref{fig:1dconvdiff_Case1}. Ultimately, the primary objective of the proposed PI-MFA framework is to mitigate this discretization-induced discrepancy while recovering a continuous field that is more consistent with the governing physics.

Next, we apply the standard MFA to identify suitable control-point resolutions for the subsequent PI-MFA study. In general, increasing the number of control points improves data-fitting accuracy, but beyond a certain resolution overfitting and spurious oscillations may appear. Following previous MFA studies~\citep{peterka2018mfa, lenz_fourier-informed_2023, lenz_adaptive_2023}, we use the spline degree to $p=3$ in both space and time, which provides a reasonable balance between smoothness and flexibility at the resolutions considered here. Now, we approximate $\rho(x,t)$ by a tensor product B-spline expansion 
\begin{equation} \label{eqn:standardMFA}
\rho_{\mathrm{MFA}}(x,t;P_\rho) = \sum_{i=0}^{n_x-1} \sum_{j=0}^{n_t-1} N_i^{(x)}(x)\,N_j^{(t)}(t)\,(P_\rho)_{ij}, 
\end{equation}
where $N_i^{(x)}$ and $N_j^{(t)}$ are B-spline basis functions of degree $p=3$ in space and time, respectively, and $P_\rho$ is the control-point vector. The total number of control points is $n_x \times n_t$, and we vary this resolution to study the trade-off between approximation quality and model complexity. 

The individual loss terms $\mathcal{L}_{\mathrm{data}}$, $\mathcal{L}_{\mathrm{pde}}$, $\mathcal{L}_{\mathrm{IC}}$, and $\mathcal{L}_{\mathrm{BC}}$ are each computed by using the MSE definition in Eq.~\eqref{eq:MSE}. In order to establish a baseline for model comparison, the total loss $\mathcal{L}_{\mathrm{total}}$ is formulated as the unweighted sum of these components (i.e., $\lambda_{*}=1$), preserving the MSE-based structure. This unweighted setup isolates the raw magnitude of each loss term, independent of any scaling weights. Furthermore, while the continuous nature of MFA and PI-MFA allows for error evaluation on any arbitrary grid, we evaluate the error at the exact spatial and temporal locations of the input simulation data.

\begin{figure}
	\centering
	\includegraphics[width=.9\textwidth]{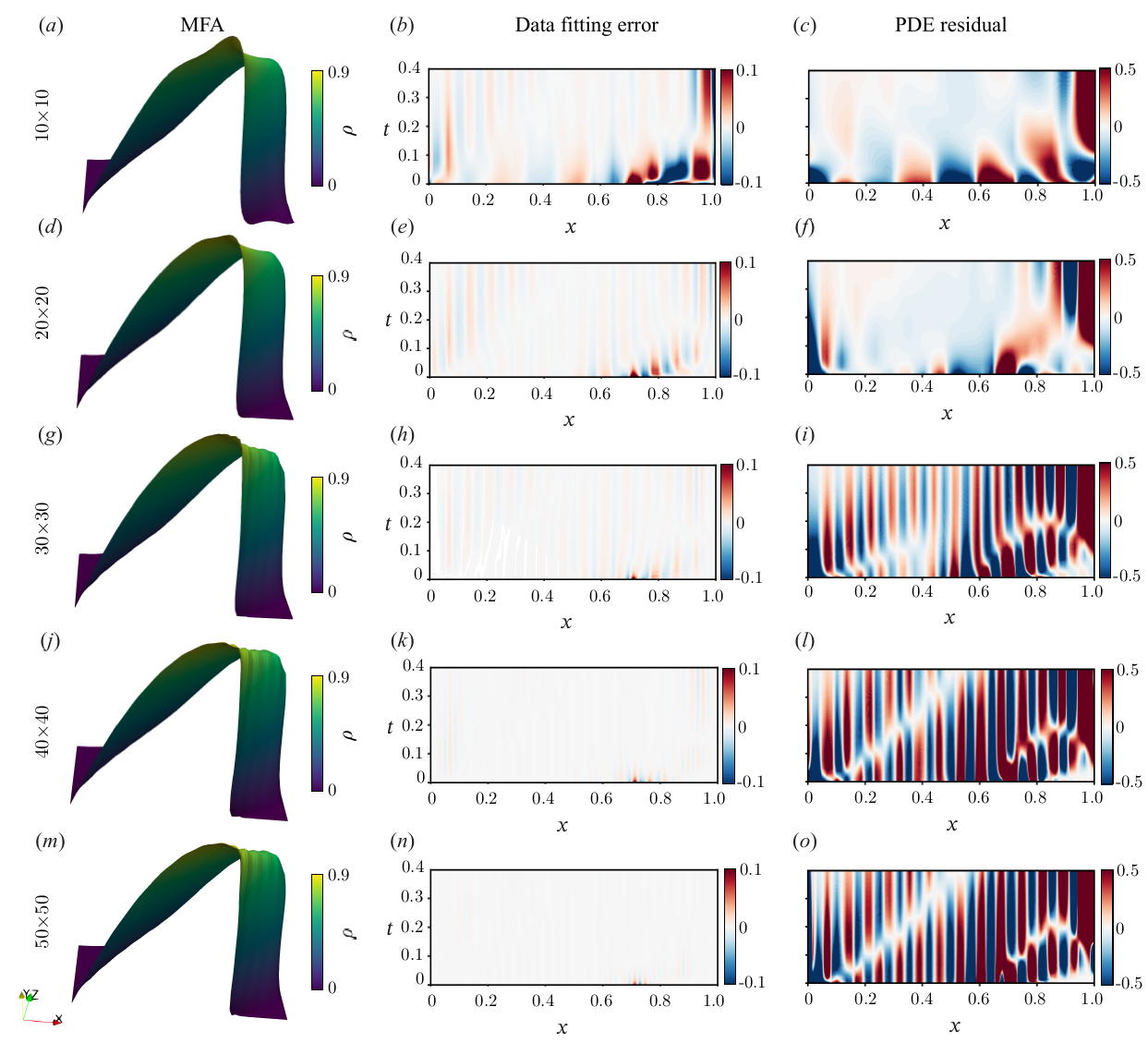}
  \caption{Results of standard MFA data fitting, shown as a function of the number of control points. The number of control points in space and time \((N_x, N_t)\) is indicated at the top of each panel.}\label{Fig_1DConvDiff_MFA}
\end{figure}

We consider control-point grids of $10\times 10$, $20\times 20$, $30\times 30$, $40\times 40$, and $50\times 50$. The corresponding MFA reconstructions, data-fitting errors, and PDE residuals are shown in Fig.~\ref{Fig_1DConvDiff_MFA}. For resolutions finer than $30\times 30$, noticeable oscillations re-emerge in the solution. These oscillations are due to the original input data generated from a coarse-grid simulation. Table~\ref{tab:1DConvDiff_loss_terms} reports the loss components for the purely data-driven MFA fits (i.e., $\lambda_{\mathrm{pde}} = \lambda_{\mathrm{BC}} = \lambda_{\mathrm{IC}} = 0$) across the varying control-point resolutions. While the data-fitting error naturally decreases as the number of control points increases, the PDE loss remains relatively large. This indicates that a purely data-driven fit does not inherently enforce physical consistency, and it is further supported by the PDE residuals exhibited in Fig.~\ref{Fig_1DConvDiff_MFA}(\textit{c}, \textit{f}, \textit{i}, \textit{l}, \textit{o}). Balancing this trade-off between minimizing data-fitting error and PDE residual's oscillatory behavior, we select a $40\times 40$ control-point grid as the baseline resolution for the subsequent PI-MFA experiments.

\begin{table}[ht]
\centering
\small
\setlength{\tabcolsep}{6pt}
\renewcommand{\arraystretch}{1.15}
\caption{Individual and total loss components as a function of control-point resolution ($N_x \times N_t$) for the 1D convection-diffusion equation. The losses (MSE) are reported for the purely data-driven MFA baseline.}
\label{tab:1DConvDiff_loss_terms}
\begin{tabular}{
  l 
  S[table-format=1.2e-1] 
  S[table-format=1.2e-1] 
  S[table-format=1.2e-2] 
  S[table-format=1.2e-2] 
  S[table-format=1.2e-2]
}
\toprule
& \multicolumn{5}{c}{\textbf{Control-point resolution} $N_x\times N_t$} \\
\cmidrule(lr){2-6}
\textbf{Loss term}
& \multicolumn{1}{c}{$10\times10$}
& \multicolumn{1}{c}{$20\times20$}
& \multicolumn{1}{c}{$30\times30$}
& \multicolumn{1}{c}{$40\times40$}
& \multicolumn{1}{c}{$50\times50$} \\
\midrule
$\mathcal{L}_{\text{data}}$  & 4.53e-5 & 7.54e-6 & 3.55e-6  & 1.54e-6  & 3.49e-7  \\
$\mathcal{L}_{\text{pde}}$   & 3.38e-1 & 1.87e0  & 1.41e0   & 7.68e-1  & 1.20   \\
$\mathcal{L}_{\text{IC}}$    & 3.34e-4 & 5.32e-5 & 8.22e-6  & 1.67e-5  & 3.93e-5  \\
$\mathcal{L}_{\text{BC}}$    & 6.36e-5 & 6.59e-7 & 8.19e-11 & 4.34e-10 & 3.99e-20 \\
\midrule
\textbf{$\mathcal{L}_{\text{total}}$} & 3.38e-1 & 1.87e0  & 1.41e0   & 7.68e-1  & 1.20   \\
\bottomrule
\end{tabular}

\end{table}

\subsubsection{PI-MFA formulation for the 1D convection-diffusion system}
Following Section~\ref{subsec:pi-mfa-formulation}, we construct PI-MFA models using observation data sampled from the numerical simulation at a subset of space-time locations, together with collocation points in the interior for enforcing the PDE, BCs, and ICs. The loss function for the control points $P_\rho$ is
\begin{equation}
\begin{aligned}  \mathcal{L}(P_{\rho}) =\;&  \frac{\lambda_{\mathrm{data}}}{M_{\mathrm{data}}} \bigl\| N_{\mathrm{data}} P_{\rho} - Q_{\mathrm{data}} \bigr\|_2^2  + \frac{\lambda_{\mathrm{pde}}}{M_{\mathrm{pde}}} \bigl\| r_{\mathrm{PDE}}(P_{\rho}) \bigr\|_2^2 \\  &+ \frac{\lambda_{\mathrm{BC}}}{M_{\mathrm{BC}}} \bigl\| N_{\mathrm{BC}} P_{\rho} - Q_{\mathrm{BC}} \bigr\|_2^2  + \frac{\lambda_{\mathrm{IC}}}{M_{\mathrm{IC}}} \bigl\| N_{\mathrm{IC}} P_{\rho} - Q_{\mathrm{IC}} \bigr\|_2^2,
\end{aligned}\label{eq:loss-1d}
\end{equation}
where $N_{\mathrm{data}}$ evaluates the spline basis at data locations and $Q_{\mathrm{data}}$ stores the corresponding observation data, $N_{\mathrm{BC}}$ and $Q_{\mathrm{BC}}$ encode Dirichlet boundary values, $N_{\mathrm{IC}}$ and $Q_{\mathrm{IC}}$ encode the IC in Eq.~\eqref{eq:initial-condition}, and $r_{\mathrm{PDE}}(P_\rho)$ is the PDE residual evaluated at interior collocation points. The weights $\lambda_{\mathrm{data}},\lambda_{\mathrm{pde}},\lambda_{\mathrm{BC}},\lambda_{\mathrm{IC}} \ge 0$ balance data fidelity and physics enforcement.

For the 1D convection–diffusion equation in Eq.~\eqref{eq:1Dconvdiff}, the PDE residual at a set of interior collocation points can be written in matrix form as
\begin{equation}  r_{\mathrm{PDE}}(P_{\rho})  = N_t P_{\rho} + a\,N_x P_{\rho} - \nu\,N_{xx} P_{\rho},  \label{eq:1D_PDE}
\end{equation}
where $N_t$, $N_x$, and $N_{xx}$ are basis matrices collecting, respectively, the first derivative with respect to time, the first derivative with respect to $x$, and the second derivative with respect to $x$ of the tensor product B-spline basis, evaluated at all interior collocation points. By linearity of the MFA ansatz, the Jacobian of the PDE residual with respect to the control points is independent of $P_\rho$ and given by
\begin{equation}  J_{\mathrm{PDE}}  = \frac{\partial r_{\mathrm{PDE}}}{\partial P_{\rho}}  = N_t + a\,N_x - \nu\,N_{xx}.  \label{eq:1D_Jacobian}
\end{equation}
Consistent with the general formulation in Section~\ref{subsec:pi-mfa-formulation}, the gradient of the loss in Eq.~\eqref{eq:loss-1d} becomes
\begin{equation}
\begin{aligned}  \nabla_{P_\rho} \mathcal{L}(P_\rho) =\;&  2 \frac{\lambda_{\mathrm{data}}}{M_{\mathrm{data}}}\, N_{\mathrm{data}}^{\top}    (N_{\mathrm{data}} P_\rho - Q_{\mathrm{data}})  + 2 \frac{\lambda_{\mathrm{pde}}}{M_{\mathrm{pde}}}\, J_{\mathrm{PDE}}^{\top} r_{\mathrm{PDE}}(P_\rho) \\  &+ 2 \frac{\lambda_{\mathrm{BC}}}{M_{\mathrm{BC}}}\, N_{\mathrm{BC}}^{\top}    (N_{\mathrm{BC}} P_\rho - Q_{\mathrm{BC}})  + 2 \frac{\lambda_{\mathrm{IC}}}{M_{\mathrm{IC}}}\, N_{\mathrm{IC}}^{\top}    (N_{\mathrm{IC}} P_\rho - Q_{\mathrm{IC}}).\end{aligned}\label{eq:1D_grad}
\end{equation}
All basis matrices ($N_\ast$) and the Jacobian $J_{\mathrm{PDE}}$ are precomputed from the analytic B-spline derivatives and reused throughout the L-BFGS iterations. Although this example involves a linear PDE, our ultimate goal is to apply the approach to nonlinear PDEs. Therefore, to validate the method in a representative setting, we use L-BFGS here rather than solving the problem directly as a linear least-squares system, as is done in the standard MFA method.

\subsubsection{Mitigation of discretization-induced conservation errors}

To use the standard MFA model, we set $\lambda_{\mathrm{pde}} = \lambda_{\mathrm{BC}} = \lambda_{\mathrm{IC}} = 0$ in Eq.~\eqref{eq:loss-1d} which is the same as in 
Eq.~\eqref{eq:least-square_MFA}, so that the control points are determined only by the data misfit term. This baseline quantifies the best possible fit to the simulation data for a given spline resolution, independent of physics enforcement. In addition, we consider the regularized MFA approach~\citep{lenz_adaptive_2023} as another baseline described in Section~\ref{subsec:reg_mfa}. This method incorporates first- and second-derivative regularization terms, which promote smoother solutions. We also compare our method with a standard PINN~\citep{raissi2019physics}. The PINN architecture consists of layers $[2,64,64,64,64,64,1]$, and it is trained by using the Adam optimizer. For consistency, the same dataset is used in the PINN data term.

\begin{figure}
	\centering
	\includegraphics[width=.95\textwidth]{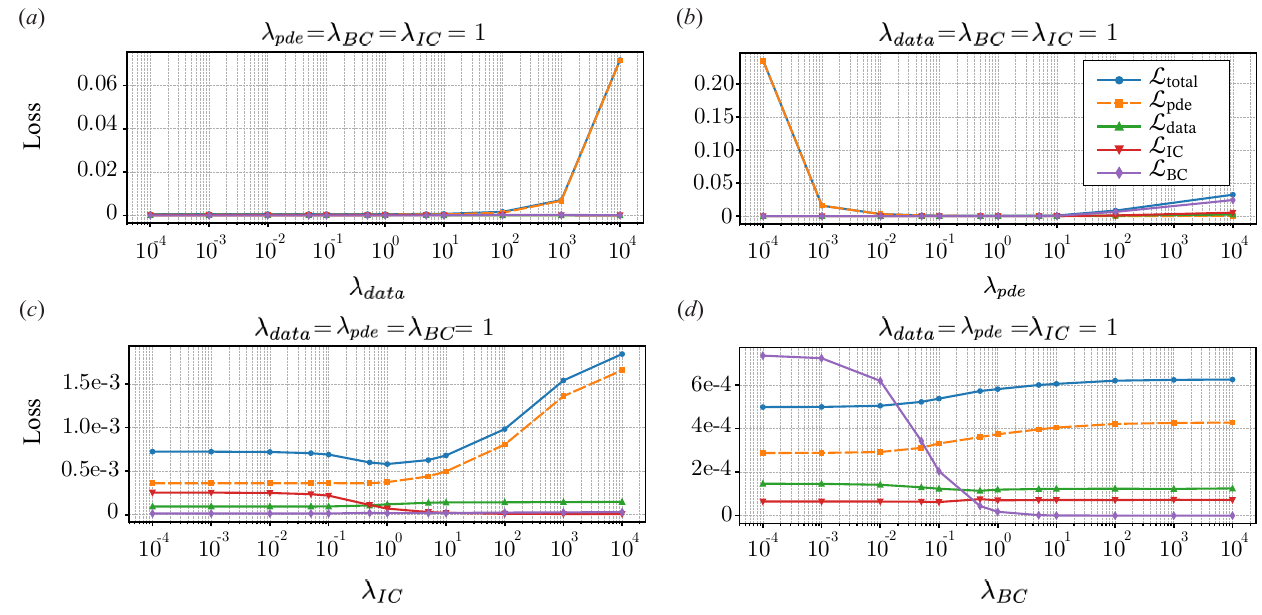}
  \caption{One-dimensional sweeps to choose the loss weights 
$\lambda$ for the 1D convection–diffusion equation: (\textit{a}) vary 
$\lambda_{data}$ while holding the other $\lambda$ values fixed; (\textit{b}) vary $\lambda_{pde}$; (\textit{c}) vary $\lambda_{IC}$; and (\textit{d}) vary $\lambda_{BC}$, each with all other weights fixed.}\label{Fig_1DConvDiff_1Dsweep_T3}
\end{figure}

\begin{figure}
	\centering
	\includegraphics[width=.45\textwidth]{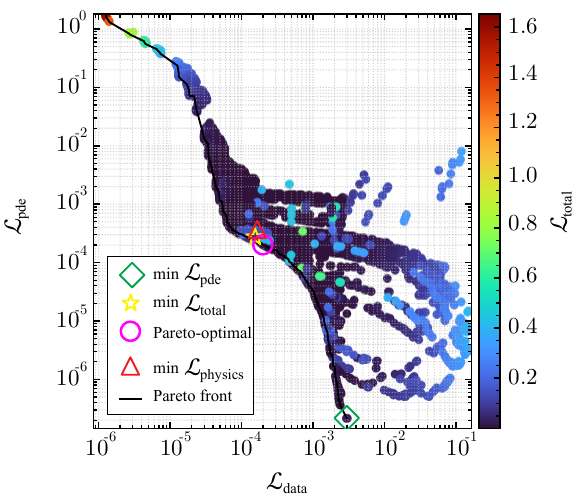}
  \caption{Results of the 4D parameter sweep over $(\lambda_{\mathrm{data}}, \lambda_{\mathrm{pde}}, \lambda_{\mathrm{IC}}, \lambda_{\mathrm{BC}})$, shown in the $(\mathcal{L}_{\mathrm{data}}, \mathcal{L}_{\mathrm{pde}})$ plane. The black curve denotes the Pareto front. Specific configurations are highlighted by distinct markers: the green diamond ($\min \mathcal{L}_{\mathrm{pde}}$), the yellow star ($\min \mathcal{L}_{\mathrm{total}}$), and the red triangle ($\min \mathcal{L}_{\mathrm{physics}}$). The magenta circle identifies the balanced Pareto-optimal configuration that minimizes the distance to the theoretical origin $(0, 0)$.}\label{Fig_lambda_summary}
\end{figure}


Our next task is to choose the loss weights $\lambda$ that balance the relative contributions of the different components in the PI-MFA objective. We begin with a sensitivity study,  that is, a 1D sweep in which we vary one weight at a time while holding the other three fixed. Figure~\ref{Fig_1DConvDiff_1Dsweep_T3} summarizes the 1D sweeps. Panel (\textit{a}) varies $\lambda_{\mathrm{data}}$, panel (\textit{b}) varies $\lambda_{\mathrm{pde}}$, panel (\textit{c}) varies $\lambda_{\mathrm{IC}}$, and panel (\textit{d}) varies $\lambda_{\mathrm{BC}}$, with the remaining three weights held fixed in each case. As shown in Table~\ref{tab:losses_1d}, the PDE residual ($\mathcal{L}_{\mathrm{pde}}$) is generally orders of magnitude larger than the data-fitting error, so $\mathcal{L}_{\mathrm{pde}}$ dominates $\mathcal{L}_{\mathrm{total}}$ when no physics is enforced. As expected, increasing a weight $\lambda_i$ decreases its corresponding loss term $\mathcal{L}_i$ but typically increases the competing terms. In particular, we observe strong sensitivity to $\lambda_{\mathrm{pde}}$. For example, decreasing $\lambda_{\mathrm{pde}}$ below $10^{-2}$ leads to a sharp increase in both $\mathcal{L}_{\mathrm{pde}}$ and the total loss $\mathcal{L}_{\mathrm{total}}$. In contrast, $\mathcal{L}_{\mathrm{total}}$ is comparatively insensitive to $\lambda_{\mathrm{data}}$ for $\lambda_{\mathrm{data}} \lesssim 1$, whereas very large $\lambda_{\mathrm{data}}$ improves the data fit at the expense of a higher PDE residual. $\lambda_{\mathrm{IC}}$ and $\lambda_{\mathrm{BC}}$ are not dominant for changing $\mathcal{L}_{\mathrm{total}}$, as shown in Fig.~\ref{Fig_1DConvDiff_1Dsweep_T3}(\textit{c}) and (\textit{d}). Overall, these trends indicate that the dominant trade-off is between the data-fitting and PDE-residual terms, motivating us to treat $(\mathcal{L}_{\mathrm{data}},\mathcal{L}_{\mathrm{pde}})$ as the primary objective in the subsequent analysis.

To account for couplings among all four weights, we perform a full  4D sweep over ($\lambda_{\mathrm{data}}$, $\lambda_{\mathrm{pde}}$, $\lambda_{\mathrm{IC}}$, $ and \lambda_{\mathrm{BC}}$). To effectively balance between data fidelity ($\mathcal{L}_{\mathrm{data}}$) and physical consistency ($\mathcal{L}_{\mathrm{pde}}$), we consider Pareto optimality \citep{Pareto1971} rather than simply minimizing the unweighted sum of the loss components, $\mathcal{L}_{\mathrm{total}}$. This perspective is widely used in physics-informed training and loss balancing for multiobjective problems \citep{Marler2004Survey,Marler2010WeightedSum,Rohrhofer2021ParetoPINN,Heldmann2023BiObjectivePINN,Bischof2025MultiObjectivePINN}. Figure~\ref{Fig_lambda_summary} presents the results of this 4D sweep, projected onto the $(\mathcal{L}_{\mathrm{pde}},\mathcal{L}_{\mathrm{data}})$ plane, where the color mapping indicates the total loss $\mathcal{L}_{\mathrm{total}}$. This visualization highlights the fundamental trade-off between data fidelity and physics enforcement, which largely governs the overall behavior of $\mathcal{L}_{\mathrm{total}}$. From this parameter space, we select a Pareto-optimal configuration (identified in linear scale as the magenta circle in Fig.~\ref{Fig_lambda_summary}). For comparative analysis, we also evaluate the configurations that yield the minimum $\mathcal{L}_{\mathrm{pde}}$ and the minimum physics loss, $\min \mathcal{L}_{\mathrm{physics}}$, where $\mathcal{L}_{\mathrm{physics}} = \mathcal{L}_{\mathrm{pde}} + \mathcal{L}_{\mathrm{IC}} + \mathcal{L}_{\mathrm{BC}}$ encapsulates all non-data constraints.

\begin{figure}
	\centering
	\includegraphics[width=.6\textwidth]{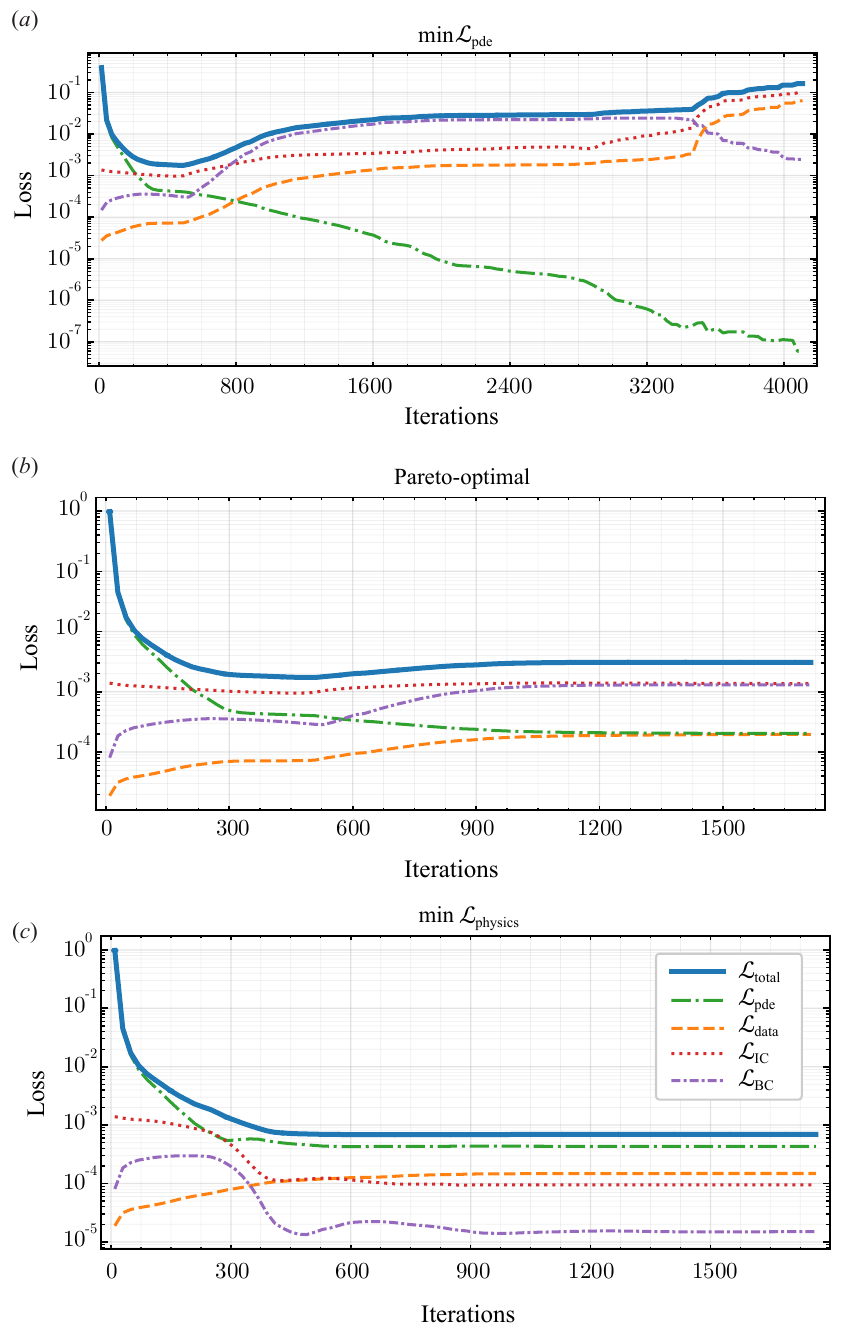}
  \caption{Convergence histories of the optimization losses for three representative weight configurations: $\min \mathcal{L}_{\mathrm{physics}}$, $\min \mathcal{L}_{\mathrm{pde}}$, and the Pareto-optimal case. These configurations correspond to the specific $\lambda$ selections identified in the parameter sweep (Fig.~\ref{Fig_lambda_summary}). The $\min \mathcal{L}_{\mathrm{data}}$ configuration is omitted here because of its near-instantaneous convergence.}\label{Fig_iterplots}
\end{figure}

In order to accelerate the optimization process, all PI-MFA runs are warm-started using the control points already computed from the purely data-driven MFA least-squares solution. Figure~\ref{Fig_iterplots} illustrates the resulting loss convergence histories for three representative configurations: $\min \mathcal{L}_{\mathrm{pde}}$, the Pareto-optimal case, and $\min \mathcal{L}_{\mathrm{physics}}$. The $\min \mathcal{L}_{\mathrm{data}}$ configuration is omitted from this visual comparison because the warm-started initialization causes it to converge almost instantaneously. Among these cases, the configuration targeting $\min \mathcal{L}_{\mathrm{pde}}$ requires the longest optimization trajectory. This extended iteration count reflects the numerical stiffness introduced by enforcing PDE consistency without adequately balancing the boundary constraints. Nevertheless, the optimization terminates successfully once $\mathcal{L}_{\mathrm{pde}}$ reaches a sufficiently small threshold.

The final losses and the corresponding weight choices are reported in Table~\ref{tab:losses_1d}. The data-only MFA baseline achieves an excellent data fit ($\mathcal{L}_{\mathrm{data}}=1.54\times 10^{-6}$) but incurs a massive PDE residual ($\mathcal{L}_{\mathrm{pde}}=7.68\times 10^{-1}$), confirming that physics consistency is not recovered without explicit enforcement. To establish a comprehensive machine-learning baseline, we also evaluate a standard PINN. While the PINN incorporates the governing equations and significantly reduces the PDE residual compared with standard MFA ($\mathcal{L}_{\mathrm{pde}}=3.43\times 10^{-3}$), it requires 855 iterations to train and relatively struggles to enforce the initial and boundary conditions because of the lack of local support on the coarse grid. This results in a comparatively high total loss ($\mathcal{L}_{\mathrm{total}}=3.10\times 10^{-2}$).

In contrast, the proposed PI-MFA configurations systematically improve physical consistency with fewer iterations than PINNs. The PI-MFA case minimizing $\mathcal{L}_{\mathrm{pde}}$ explicitly drives the PDE residual down to $2.17\times 10^{-7}$ but at the cost of increased data mismatch and larger initial and boundary constraint residuals. The Pareto-optimal selection provides a well-balanced compromise. It attains a near-minimal PDE residual ($2.02\times 10^{-4}$) while maintaining excellent data fidelity ($\mathcal{L}_{\mathrm{data}}=1.99\times 10^{-4}$), demonstrating the framework's capability to successfully navigate the inherent trade-off between fitting the data and enforcing the continuous physics. Furthermore, the case minimizing $\mathcal{L}_{\mathrm{physics}}$ suppresses all physical constraints concurrently, yielding the smallest total loss ($\mathcal{L}_{\mathrm{total}}=6.08\times 10^{-4}$) among the compared configurations, outperforming the PINN by nearly two orders of magnitude.

\begin{table}[t]
  \centering

  \footnotesize
  \setlength{\tabcolsep}{4.5pt} 
  \renewcommand{\arraystretch}{1.15}

  \sisetup{
    input-exponent-markers = eE,
    output-exponent-marker = \mathrm{e},
    group-digits = false,
    detect-weight = true,
    detect-family = true
  }
    \caption{Final loss components and selected hyperparameter values ($\lambda$) for the data-only MFA baseline, regularized MFA (Reg-MFA), a physics-informed neural network (PINN), and three representative PI-MFA runs ($\min \mathcal{L}_{\mathrm{pde}}$, Pareto-optimal, and $\min \mathcal{L}_{\mathrm{physics}}$) chosen from the 4D $\lambda$-sweep in Fig.~\ref{Fig_lambda_summary}. The PINN model was trained utilizing the $\min \mathcal{L}_{\mathrm{physics}}$ weights. Reported losses are evaluated at the final optimizer iterate.}
            \label{tab:losses_1d}
  \begin{threeparttable}
  \begin{tabular}{lcccccc}
    \toprule
      & \multicolumn{1}{c}{\textbf{MFA}} 
      & \multicolumn{1}{c}{\textbf{Reg-MFA}}
      & \multicolumn{1}{c}{\textbf{PINN}}
      & \multicolumn{3}{c}{\textbf{PI-MFA}} \\
     \cmidrule(lr){2-2}\cmidrule(lr){3-3}\cmidrule(lr){4-4}\cmidrule(lr){5-7}
    & \multicolumn{1}{c}{(data only)}
    & \multicolumn{1}{c}{(smoother)}
    & \multicolumn{1}{c}{(min $\mathcal{L}_{\mathrm{physics}}$ in PI-MFA)}
    & \multicolumn{1}{c}{min $\mathcal{L}_{\mathrm{pde}}$}
    & \multicolumn{1}{c}{Pareto-optimal}
    & \multicolumn{1}{c}{min $\mathcal{L}_{\mathrm{physics}}$} \\
  \midrule

    Iterations 
      & -- & -- & 855 & 4120 & 1725 & 1835 \\

    $\lambda_{\mathrm{data}}$ 
      & 1 & 1 & $10^{-1}$ & $10^{-4}$ & 5 & $10^{-1}$ \\

    $\lambda_{\mathrm{pde}}$
      & 0 & 0 & $10^{2}$ & $10^{4}$ & 10 & $10^{2}$ \\

    $\lambda_{\mathrm{IC}}$
      & 0 & 0 & $10^{2}$ & $5\times10^{-2}$ & $5\times10^{-2}$ & $10^{2}$ \\

    $\lambda_{\mathrm{BC}}$
      & 0 & 0 & $10^{2}$ & 1 & $10^{-1}$ & $10^{2}$ \\
    \midrule

    $\mathcal{L}_{\mathrm{data}}$ 
      & {1.54e-6} & {1.28e-6} & {3.44e-3} & {2.96e-3} & {1.99e-4} & {1.66e-4} \\

    $\mathcal{L}_{\mathrm{pde}}$
      & {7.68e-1} & {1.66e+0}  & {3.43e-3} & {2.17e-7} & {2.02e-4} & {3.71e-4} \\

    $\mathcal{L}_{\mathrm{IC}}$
      & {1.67e-5} & {1.42e-3} & {8.57e-3} & {8.70e-3} & {3.86e-4} & {5.65e-5} \\

    $\mathcal{L}_{\mathrm{BC}}$
      & {4.34e-10}& {8.20e-11}& {1.56e-2} & {2.33e-2} & {1.31e-3} & {1.47e-5} \\
\midrule
    \addlinespace[2pt]

    $\mathcal{L}_{\mathrm{total}}$
      & {7.68e-1} & {1.66e+0}  & {3.10e-2} & {3.49e-2} & {2.09e-3} & {6.08e-4} \\
    \bottomrule
  \end{tabular}

  \end{threeparttable}

\end{table}

\begin{figure}
	\centering
	\includegraphics[width=.8\textwidth]{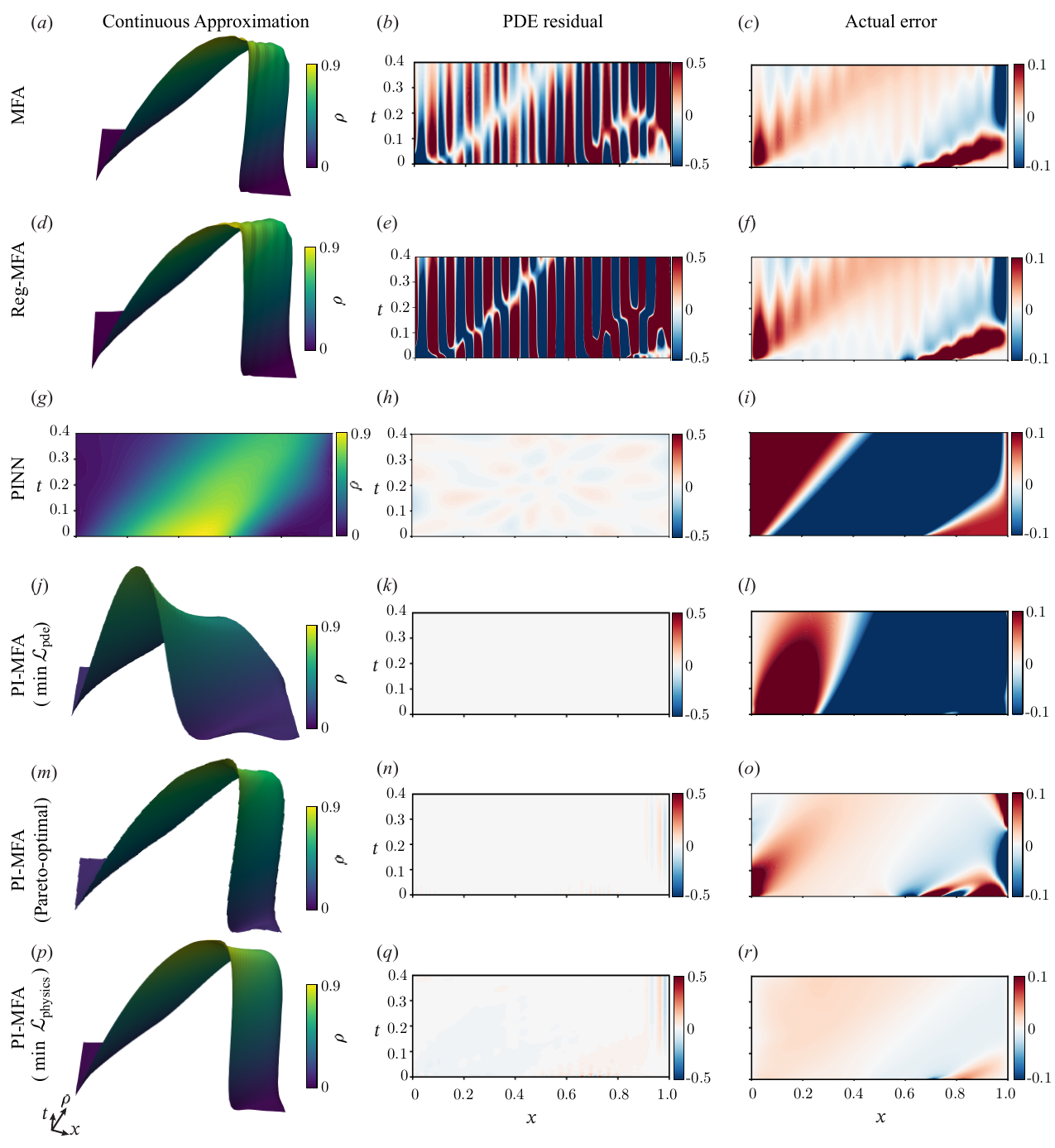}
  \caption{Continuous-field approximations (left column), strong-form PDE residuals (middle column), and actual relative errors with respect to the analytical solution (right column). Rows correspond to standard MFA (a--c), Reg-MFA (d--f), PINN (g--i), and PI-MFA under three representative weight selections: $\min \mathcal{L}_{\mathrm{pde}}$ (j--l), Pareto-optimal (m--o), and $\min \mathcal{L}_{\mathrm{physics}}$ (p--r).}\label{Fig_pointwiseerrorFields}
\end{figure}

Figure~\ref{Fig_pointwiseerrorFields} visually compares the continuous space-time approximations generated by the baselines (MFA, Reg-MFA, and PINN) against the proposed PI-MFA framework evaluated under the three representative weight selections ($\min \mathcal{L}_{\mathrm{pde}}$, Pareto-optimal, and $\min \mathcal{L}_{\mathrm{physics}}$). For each evaluated method, the left column displays the reconstructed continuous field, the middle column visualizes the pointwise strong-form PDE residual, and the right column maps the actual relative error with respect to the exact analytical solution. As anticipated, the purely data-driven standard MFA reconstructs a visually plausible macroscopic field that matches the discrete data accurately but structurally violates the governing equations, producing widespread, highly oscillatory PDE residuals. Adding naive smoothing penalties (Reg-MFA) does little to correct this physical discrepancy. While the PINN baseline noticeably reduces the volumetric PDE residual relative to standard MFA, it still suffers from elevated actual errors near the spatial boundaries. This visual artifact is consistent with its higher $\mathcal{L}_{\mathrm{BC}}$ value of $1.56\times 10^{-2}$ reported in Table~\ref{tab:losses_1d}, highlighting the persistent challenge of cleanly capturing localized, sharp flow dynamics using a continuous global neural network despite utilizing 855 training iterations. In contrast, explicitly embedding hard physical constraints directly into the locally supported B-spline objective via PI-MFA systematically suppresses PDE violations across the entire spatiotemporal domain. The $\min \mathcal{L}_{\mathrm{pde}}$ configuration naturally produces the cleanest, most suppressed residual field; however, its aggressive, unilateral PDE enforcement forces slight deviations from the true initial state, resulting in a somewhat broader actual error distribution. Both the Pareto-optimal and $\min \mathcal{L}_{\mathrm{physics}}$ configurations offer highly balanced, superior reconstructions. They successfully reduce the severe structural residuals present in the standard MFA while maintaining extremely low actual errors relative to the ground-truth analytical solution.

  \begin{figure}
	\centering
	\includegraphics[width=.7\textwidth]{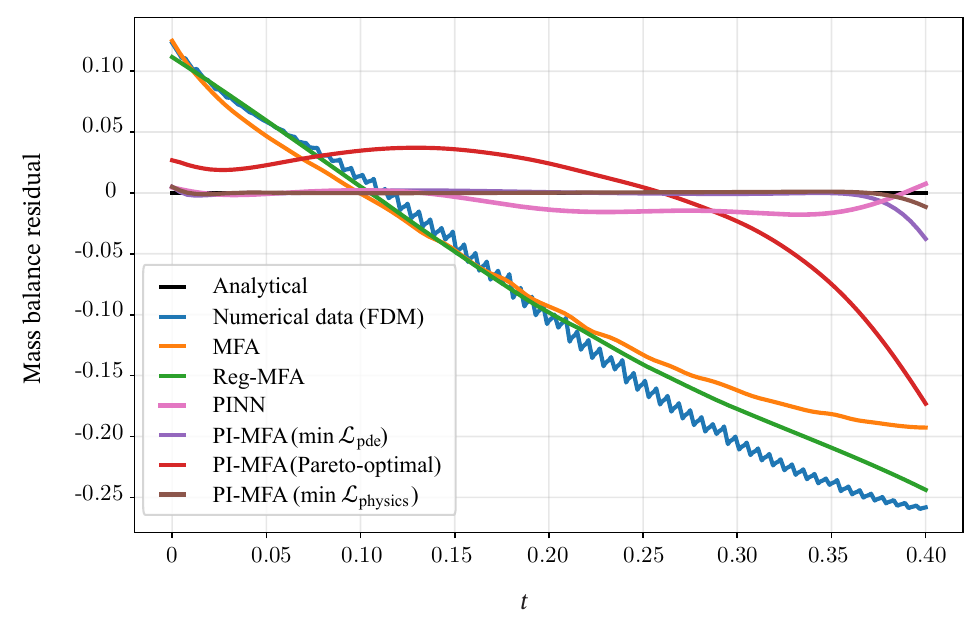}
  \caption{Temporal evolution of the integrated mass-balance (flux) residual for the 1D convection-diffusion problem. The exact analytical solution conserves mass (zero residual). The non-conservative numerical data computed by the finite difference method (FDM) exhibits a severe negative drift, which the purely data-driven MFA and Reg-MFA baselines inherently absorb. The PI-MFA configurations correct this physical discrepancy, with the $\min \mathcal{L}_{\mathrm{physics}}$ model achieving accurate global conservation by enforcing both the PDE and the boundary constraints.}\label{Fig_massbalance_1D}
\end{figure}

We also objectively assess whether the reconstructed continuous fields satisfy the global flux-conservation criterion defined in Eq.~\eqref{eq:1dmassbalance}. Figure~\ref{Fig_massbalance_1D} tracks the temporal evolution of the integrated mass-balance residual for the exact analytical solution, the discrete numerical data (the residual is computed by finite difference), the baseline methods (MFA, Reg-MFA, and PINN), and the proposed PI-MFA framework evaluated under the three representative weight selections ($\min \mathcal{L}_{\mathrm{pde}}$, Pareto-optimal, and $\min \mathcal{L}_{\mathrm{physics}}$). The numerical data was computed by using a non-conservative scheme. Consequently, its evaluated mass balance exhibits a severe, highly oscillatory negative drift away from the exact analytical zero-residual baseline. As expected, the purely data-driven standard MFA and smoothed Reg-MFA successfully filter out the high-frequency numerical noise but passively inherit the underlying macroscopic numerical discrepancy, perfectly tracking the flawed data and failing entirely to conserve mass. While the PINN baseline leverages physical equations to actively flatten this error curve and improve upon the data-only splines, it still suffers from a persistent, fluctuating offset throughout the temporal domain. In contrast, explicitly embedding physical constraints into the PI-MFA objective restores global flux conservation, although the extent of the correction depends heavily on the chosen constraint weights. The $\min \mathcal{L}_{\mathrm{physics}}$ configuration delivers the most robust performance. By penalizing the PDE residual alongside the initial and boundary constraints, it effectively overrides the flawed input data, locking the continuous spline onto a mass-conserving trajectory that closely tracks the exact analytical zero line. The $\min \mathcal{L}_{\mathrm{pde}}$ configuration also dramatically improves mass conservation for the majority of the time domain. However, by unilaterally prioritizing the interior PDE at the expense of the boundary conditions (as previously noted by its higher $\mathcal{L}_{\mathrm{BC}}$ error in Table~\ref{tab:losses_1d}), this configuration allows artificial numerical flux to leak at the domain edges, causing a noticeable drop near the final time $t=0.4$. The Pareto-optimal configuration ($\min \mathcal{L}_{\mathrm{Pareto}}$) exhibits intermediate behavior, visually encapsulating the expected trade-off in the optimization. Sacrificing a degree of enforcement of physics to accommodate non-conservative observational data ultimately yields a long-term global conservation discrepancy.

Overall, the 1D convection-diffusion example illustrates two central features of the proposed PI-MFA framework. First, purely data-driven MFA and mathematical smoothing (Reg-MFA) can achieve extremely low data-fitting errors with moderate spline resolutions. However, the resulting continuous surrogates remain physically oblivious, readily absorbing underlying discretization errors, non-conservative drifts, and large PDE residuals. Second, augmenting the MFA objective with explicitly formulated, physics-informed penalties ($\mathcal{L}_{\mathrm{pde}}$, $\mathcal{L}_{\mathrm{IC}}$, and $\mathcal{L}_{\mathrm{BC}}$) recovers physical consistency. By tuning the multiobjective weights ($\lambda_\ast$), PI-MFA acts as a powerful mathematical filter, preserving the compact, analytically differentiable B-spline representation. This canonical test case therefore provides a controlled benchmark demonstrating how the inherent tension between data fidelity and physical enforcement can be quantified and dynamically tuned. In the next subsection, we demonstrate that these capabilities scale to the more complex, higher-dimensional, nonlinear Burgers equations.

\subsection{Two-dimensional coupled viscous Burgers equations}\label{subsec:2DBurgers}
  \begin{figure}
	\centering
	\includegraphics[width=.9\textwidth]{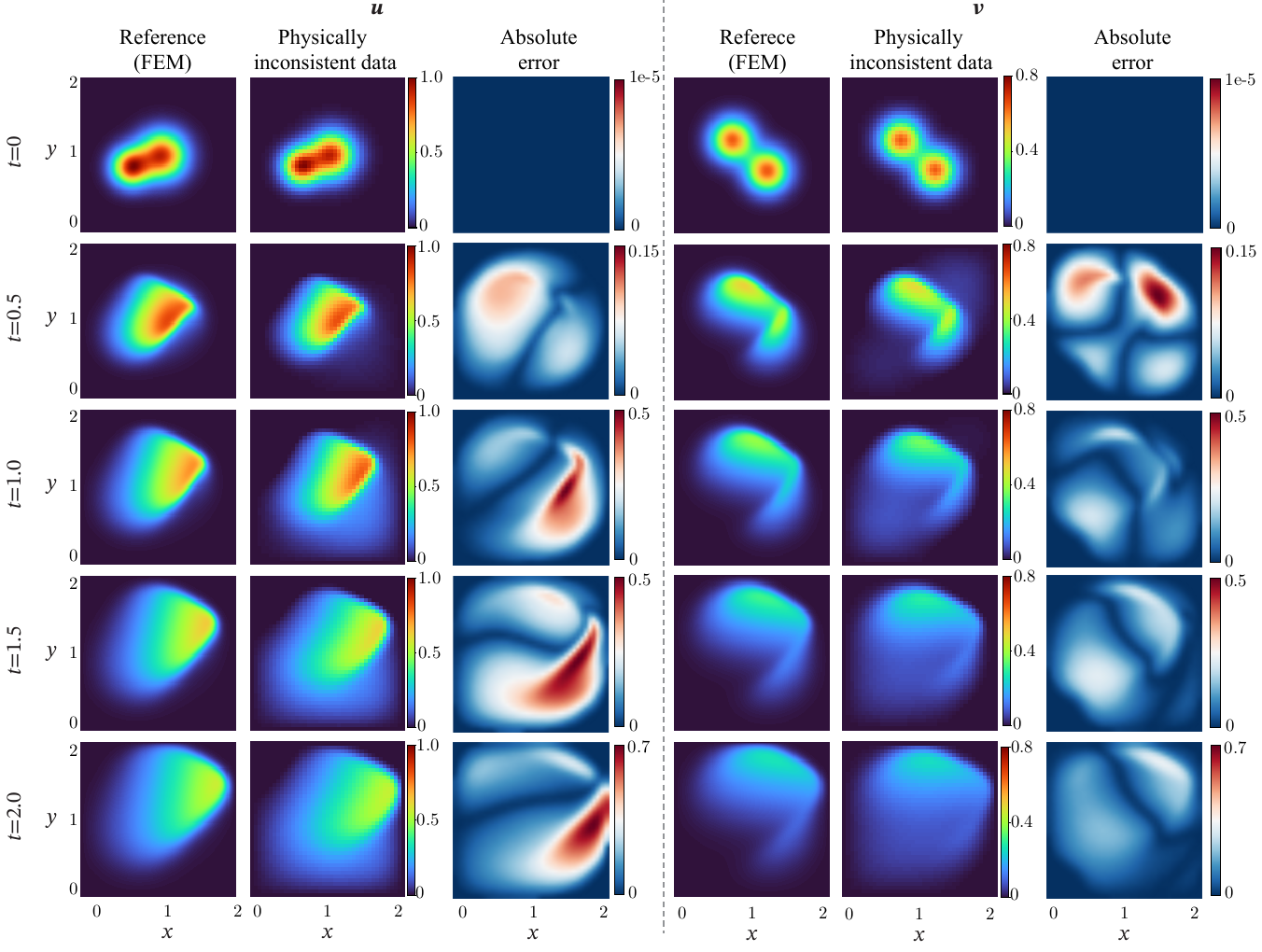}
  \caption{Snapshots of the 2D coupled viscous Burgers solution at $t = 0,\ 0.5,\ 1.0,\ 1.5,$ and $2.0$. The left panels show the $u$ field, and the right panels show the $v$ field at the corresponding times. The FEM solution serves as the high-accuracy reference, while the input data is shown separately as physically inconsistent data. The absolute error with respect to the reference solution is reported in the rightmost column for both the $u$ and $v$ fields.}
  \label{Fig_burgerdata}
\end{figure}

We consider the two-dimensional coupled viscous Burgers system in non-conservative (advective) form on the rectangular domain
\(\Omega=(0,L_x)\times(0,L_y)\) over \(t\in[0,T]\):
\begin{subequations}\label{eq:burgers2d}
\begin{align}
\frac{\partial u}{\partial t} + u \frac{\partial u}{\partial x} + v \frac{\partial u}{\partial y}
&=
\nu \left( \frac{\partial^2 u}{\partial x^2} + \frac{\partial^2 u}{\partial y^2} \right), \\
\frac{\partial v}{\partial t} + u \frac{\partial v}{\partial x} + v \frac{\partial v}{\partial y}
&=
\nu \left( \frac{\partial^2 v}{\partial x^2} + \frac{\partial^2 v}{\partial y^2} \right).
\end{align}
\end{subequations}
As in the one-dimensional convection-diffusion example in Section~\ref{subsec:1d-convdiff}, homogeneous Dirichlet boundary conditions are imposed on \(\partial\Omega\):
\begin{align}
u(x,y,t)=0,\qquad v(x,y,t)=0,\qquad (x,y)\in\partial\Omega,\quad t\ge 0.
\end{align}
Following common practice for Burgers-type benchmark problems \citep{Cole1951,LeVeque2002FVM,ClarkDiLeoni2018BurgersGaussians}, we prescribe smooth initial conditions as superpositions of Gaussian functions:
\begin{subequations}\label{eq:ICburgers2d}
\begin{align}
u_{\mathrm{IC}}(x,y)
&=
A_u \sum_{k}
\exp\!\left(
-\frac{(x-c^{u}_{x,k})^2+(y-c^{u}_{y,k})^2}{2\,s_{u,k}^2}
\right), \\
v_{\mathrm{IC}}(x,y)
&=
A_v \sum_{k}
\exp\!\left(
-\frac{(x-c^{v}_{x,k})^2+(y-c^{v}_{y,k})^2}{2\,s_{v,k}^2}
\right).
\end{align}
\end{subequations}
The normalization constants $A_u$ and $A_v$ are prescribed to enforce the peak initial velocity amplitudes, yielding $\max_{(x,y)} |u_{\mathrm{IC}}| = \mathrm{amp}_u$ and $\max_{(x,y)} |v_{\mathrm{IC}}| = \mathrm{amp}_v$. Throughout the subsequent numerical experiments, unless otherwise noted, the geometric and physical parameters are fixed at $L_x=L_y=2$ and $\nu=0.01$, with the initial amplitudes set to $\mathrm{amp}_u=1.0$ and $\mathrm{amp}_v=0.8$.

As an additional study of physics consistency, we monitor a global integral balance for each component of the two-dimensional Burgers system. To this end, we define the componentwise integral
\begin{equation}
M_q(t)=\int_{\Omega} q(\boldsymbol{x},t)\,\mathrm{d}\Omega,
\qquad q\in\{u,v\}.
\label{eq:app_component_mass}
\end{equation}
Starting from Eq.~\eqref{eq:burgers2d}, we use the divergence theorem and Green's identity~\citep{Evans2010,Borthwick2017}. Under the homogeneous Dirichlet boundary condition $q=0$ on $\partial\Omega$, this yields the componentwise global integral-balance residuals 
\begin{equation}
r_{q,M}(t)
=
\frac{\mathrm{d}M_{q}}{\mathrm{d}t}
-\int_{\Omega} q\,(\nabla\cdot\boldsymbol{U})\,\mathrm{d}\Omega
-\nu\int_{\partial\Omega}\nabla q\cdot\boldsymbol{n}\,\mathrm{d}s,
\qquad q\in\{u,v\}, \qquad \boldsymbol{U}=(u,v)^{\mathsf T},
\label{eq:burgers_mass_residual}
\end{equation}
which vanish for the exact solution of the prescribed PDE, that is, $r_{u,M}(t)=r_{v,M}(t)=0.$
In the numerical experiments, we evaluate $r_{u,M}(t)$ and $r_{v,M}(t)$ directly from the continuous PI-MFA representation. The B-spline model is used to compute $\mathrm{d}M_q/\mathrm{d}t$ and the associated spatial and boundary integrals, and the resulting residuals are reported over time as diagnostics of physics consistency. Additional details are provided in Appendix~\ref{app:componentwise_mass_balance}.

 \subsubsection{Data generation and standard MFA}\label{subsubsec:standardMFA_2DBurgers}
 
To evaluate the PI-MFA framework, we construct a benchmark dataset by introducing a controlled discrepancy into the coarse-grid observations. Specifically, we inject source terms that are absent from the prescribed governing equations. This setup mimics missing or unresolved physics, which is a common source of model-form discrepancy between observed dynamics and analytical models~\citep{KennedyOHagan2001,Hooker2009,MorrisonOliverMoser2018,MasudNasharGoraya2023}. The corrupted "observed" trajectory is generated on a coarse spatial grid ($n_x^{\mathrm{low}} = n_y^{\mathrm{low}} = 40$) by augmenting the right-hand side of the Burgers equations with additive forcing terms $f_u$ and $f_v$. These localized forcings are parameterized as $f_u(x,y,t) = \sum_{j=1}^{M} a_j^{(k)} \phi_j(x,y)$ and $f_v(x,y,t) = \sum_{j=1}^{M} b_j^{(k)} \phi_j(x,y)$ over intervals $t \in [t_k, t_{k+1})$, using spatial modes $\phi_j(x,y) = \sin(m_j\pi x / L_x) \sin(n_j\pi y / L_y)$ with $M=4$. The modal coefficients are activated within a defined temporal window, producing a smooth, localized perturbation that preserves the initial and boundary conditions of the unforced system. This physically inconsistent coarse-grid solution is advanced by using fourth-order spatial finite differences and a fourth-order Runge--Kutta time integrator with a step size of $\Delta t = 10^{-2}$. In order to serve as the ground-truth reference, a high-fidelity solution of the unforced system is computed by using the Firedrake finite element framework~\citep{FiredrakeUserManual} on a fine $200 \times 200$ grid. This reference employs a continuous-Galerkin $P^2/P^1$ discretization with a stringent time step of $\Delta t = 5 \times 10^{-4}$. While continuous-Galerkin FEM formulations satisfy the weak form, evaluating their pointwise strong-form residual inherently yields non-zero discretization errors. However, because the prescribed Burgers system evolves from smooth Gaussian initial conditions rather than harsh wall-driven shear layers, the inherent strong-form residual of the high-fidelity Firedrake solution remains negligible. Therefore, the dominant source of physical inconsistency in the coarse dataset arises deliberately from the unmodeled macroscopic forcing, allowing us to accurately assess whether PI-MFA can successfully suppress the latent perturbation and recover the true unforced dynamics.

Figure~\ref{Fig_burgerdata} illustrates representative snapshots of the velocity components $(u,v)$ at $t\in\{0, 0.5, 1.0, 1.5, 2.0\}$, including the initial condition. Given the viscous and unforced nature of the reference system, its exact solution naturally dissipates and decays toward zero over time. In contrast, the coarse-grid observations incorporate an additive forcing term over the interval $t\in[0.1, 1.8]$, inducing a smooth but pronounced deviation from the reference trajectory. Since the spatial discretization error in this  example is comparatively small, the dominant source of physical inconsistency stems directly from this unmodeled forcing. This discrepancy generates large residuals when the coarse data is evaluated against the prescribed unforced PDE. The resulting dataset comprises $2N_xN_y(n_{\mathrm{out}}+1)$ spatiotemporal samples across both velocity components; for the chosen parameters, this yields $2\times 41\times 41\times 201 = 675,762$ total samples. Leveraging these simulation snapshots, we construct a standard, purely data-driven MFA baseline by representing each state variable with a three-dimensional tensor-product B-spline over $(x,y,t)$. For the $u$-velocity component, this continuous approximation is written as
\begin{equation}\label{eq:3D_tensorproduct_Bspline}
u(x,y,t) \approx
\sum_{i=1}^{P_x}\sum_{j=1}^{P_y}\sum_{k=1}^{P_t}
N_i^{(p)}(x)\,N_j^{(p)}(y)\,N_k^{(p)}(t)\,P^{u}_{ijk},
\end{equation}
where the B-spline degree is $p=3$ in all parametric directions. The $v(x,y,t)$ field is represented analogously by using an independent set of control points $P_{ijk}^{v}$. While the global error is quantified by using the same MSE definition as in the 1D convection-diffusion example, for 3D volumetric visualization we report the pointwise absolute errors $r_u$ and $r_v$. These are evaluated at each reference-data coordinate as the magnitude of the difference between the continuous approximations ($\tilde{u}, \tilde{v}$) and the high-fidelity reference solutions ($u, v$)
\begin{equation}\label{eq:absoluteerror}r_u = \left|\tilde{u}(x,y,t)-u(x,y,t)\right|, \qquad r_v = \left|\tilde{v}(x,y,t)-v(x,y,t)\right|.
\end{equation}
The rendered 3D fields for the data-fitting error, PDE residual, and actual reconstruction error are all visualized by utilizing these pointwise absolute differences.

\begin{table}[t]
\centering
\small
\setlength{\tabcolsep}{5pt}
\renewcommand{\arraystretch}{1.15}
\caption{Contributions of individual loss components as a function of the control-point grid resolution $(P_x \times P_y \times P_t)$ for the 2D coupled Burgers equations. The total loss is defined as the unweighted sum of the data, PDE, IC, and BC losses for both velocity fields: $\mathcal{L}_{\mathrm{total}} = \mathcal{L}^{u}_{\mathrm{data}} + \mathcal{L}^{v}_{\mathrm{data}} + \mathcal{L}^{u}_{\mathrm{pde}} + \mathcal{L}^{v}_{\mathrm{pde}} + \mathcal{L}^{u}_{\mathrm{IC}} + \mathcal{L}^{v}_{\mathrm{IC}} + \mathcal{L}^{u}_{\mathrm{BC}} + \mathcal{L}^{v}_{\mathrm{BC}}$.}
\label{tab:loss_terms_for_plain_MFA2Dburgers}
\begin{tabular}{l
S[table-format=1.3e-1]
S[table-format=1.3e-1]
S[table-format=1.3e-1]
S[table-format=1.3e-1]}
\toprule
& \multicolumn{4}{c}{\textbf{Control-point grid} \(\boldsymbol{P_x \times P_y \times P_t}\)} \\
\cmidrule(lr){2-5}
\textbf{Loss term}
& {\(\mathbf{10\times10\times10}\)}
& {\(\mathbf{20\times20\times20}\)}
& {\(\mathbf{30\times30\times30}\)}
& {\(\mathbf{40\times40\times40}\)} \\
\midrule
\(\mathcal{L}^{u}_{\text{data}}\) & 4.648e-4  & 1.790e-5  & 2.155e-6  & 9.236e-7  \\
\(\mathcal{L}^{v}_{\text{data}}\) & 3.275e-4  & 1.258e-5  & 1.188e-6  & 5.258e-7  \\
\(\mathcal{L}^{u}_{\text{pde}}\)  & 1.919e-2  & 1.677e-2  & 1.464e-2  & 1.607e-2  \\
\(\mathcal{L}^{v}_{\text{pde}}\)  & 1.559e-2  & 1.168e-2  & 1.096e-2  & 1.116e-2  \\
\(\mathcal{L}^{u}_{\text{IC}}\)   & 2.552e-4  & 7.536e-7  & 4.317e-8  & 1.245e-8  \\
\(\mathcal{L}^{v}_{\text{IC}}\)   & 4.419e-4  & 5.440e-7  & 4.838e-8  & 1.093e-8  \\
\(\mathcal{L}^{u}_{\text{BC}}\)   & 4.967e-5  & 3.068e-8  & 1.206e-13 & 3.669e-20 \\
\(\mathcal{L}^{v}_{\text{BC}}\)   & 5.137e-5  & 4.809e-8  & 2.018e-13 & 5.512e-20 \\
\addlinespace[2pt]
\textbf{\(\mathcal{L}_{\text{total}}\)}
& \textbf{3.637e-2}
& \textbf{2.849e-2}
& \textbf{2.560e-2}
& \textbf{2.722e-2} \\
\bottomrule
\end{tabular}

\end{table}

  \begin{figure}
	\centering
	\includegraphics[width=.9\textwidth]{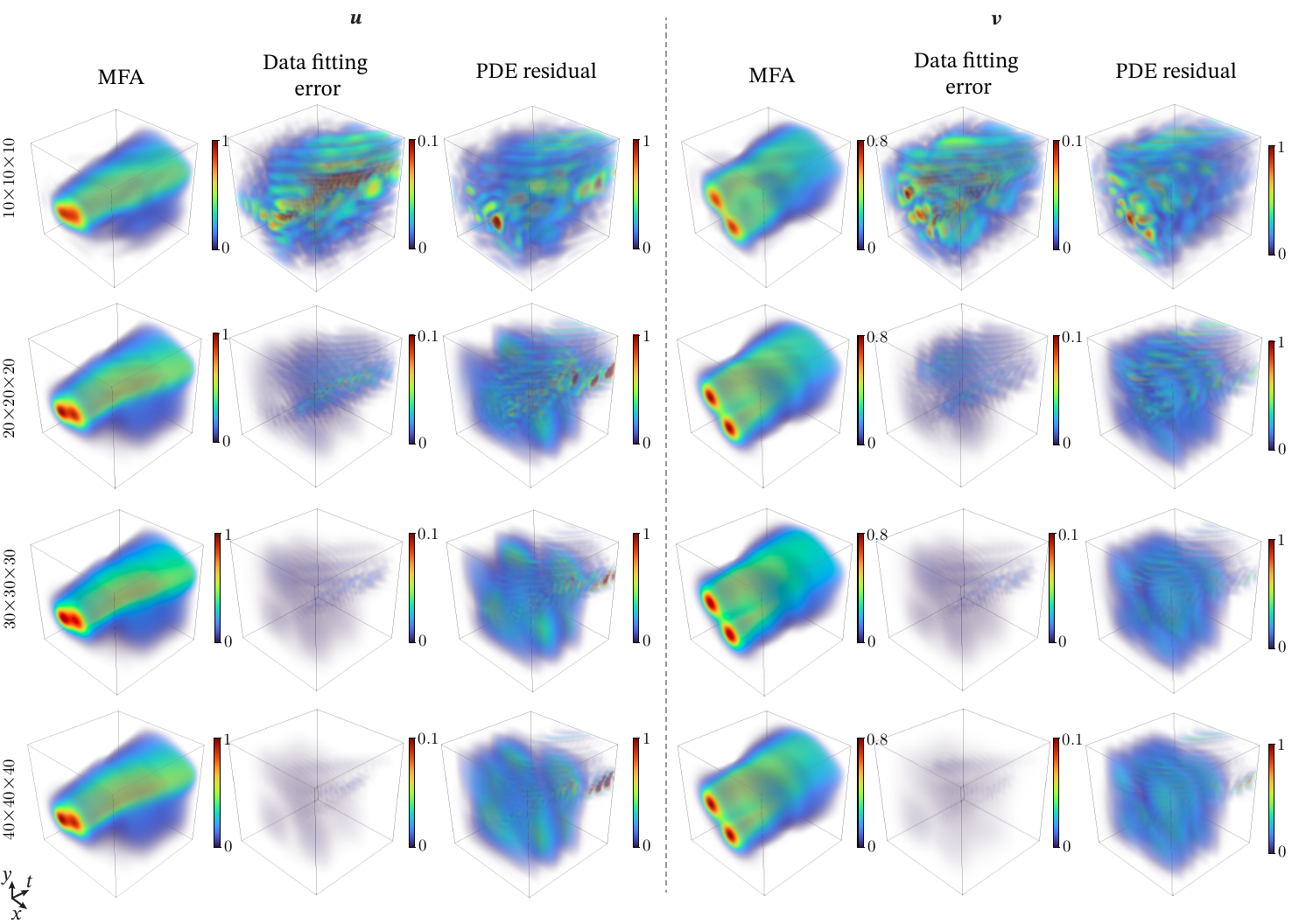}
\caption{Volumetric rendering of the standard MFA reconstructions for the 2D coupled Burgers velocity fields, $u(x,y,t)$ and $v(x,y,t)$, evaluated across four distinct control-point grid resolutions $(P_x, P_y, P_t)$. For each grid density, the pointwise absolute error relative to the simulation data and the corresponding strong-form PDE residuals are displayed.}
  \label{Fig_burger_datafitting}
\end{figure}

In order to establish a baseline,  Table~\ref{tab:loss_terms_for_plain_MFA2Dburgers} and Fig.~\ref{Fig_burger_datafitting} present the performance of a standard, purely data-driven MFA reconstruction across varying control-point grid resolutions. As in the previous 1D example, all loss terms are evaluated by using the MSE formulation defined in Eq.~\eqref{eq:MSE}, extended here to account for the additional spatial dimension $y$ and the science variable $v$. As the space-time grid is refined from $10\times10\times10$ to $40\times40\times40$, the reconstructed MFA fields for both velocity components, $u$ and $v$, become progressively better resolved. This improvement is visually evident in the 3D volumetric renderings in Fig.~\ref{Fig_burger_datafitting}, where the spatial distribution of the data-fitting error visibly diminishes as grid density increases. This visual trend aligns with the quantitative results in Table~\ref{tab:loss_terms_for_plain_MFA2Dburgers}, which demonstrate that the data-misfit terms ($\mathcal{L}^{u}_{\mathrm{data}}$ and $\mathcal{L}^{v}_{\mathrm{data}}$), along with the initial and boundary condition losses, decrease by several orders of magnitude and become essentially negligible on the finer grids. Notably, the bulk of this improvement in interpolation fidelity is already realized at the $30\times30\times30$ resolution. Conversely, a distinct and problematic behavior emerges regarding the physical consistency of the reconstruction. Although the PDE residuals ($\mathcal{L}^{u}_{\mathrm{pde}}$ and $\mathcal{L}^{v}_{\mathrm{pde}}$) decrease moderately during the initial grid refinements, they quickly plateau and remain the dominant contributors to the total loss. As shown in both the quantitative metrics (Table~\ref{tab:loss_terms_for_plain_MFA2Dburgers}) and the volumetric renders (Fig.~\ref{Fig_burger_datafitting}), refining the grid from $30\times30\times30$ to $40\times40\times40$ yields no meaningful improvement in physical accuracy; the PDE residuals for both resolutions are remarkably similar. Consequently, the overall reduction in $\mathcal{L}_{\mathrm{total}}$ stagnates, decreasing from $3.637\times10^{-2}$ on the $10\times10\times10$ grid to $2.560\times10^{-2}$ on the $30\times30\times30$ grid, before shifting marginally to $2.722\times10^{-2}$ on the finest grid. Even at these higher resolutions, the PDE residual fields remain highly structured and exhibit persistent error magnitudes localized around regions with pronounced velocity gradients. These results expose a critical limitation of standard MFA. The additional approximation power provided by a finer parameterization primarily improves the approximation of discrete snapshot data but does not inherently enforce consistency with the governing Burgers equations. Ultimately, a vanishing data-misfit error cannot be regarded as evidence of a physically admissible continuous reconstruction. Balancing this observation, the $30\times30\times30$ grid yields a highly accurate approximation of the data while maintaining a moderate computational footprint. Therefore, we select this resolution as a practical compromise for all subsequent PI-MFA computations.

\subsubsection{PI-MFA for the 2D Burgers system}\label{subsubsec:PIMFA_burgers}

We begin by defining the vectorized control-point arrays for the two velocity components as $\mathbf{p}_u := \operatorname{vec}(P_u) \in \mathbb{R}^{n_c}$ and $\mathbf{p}_v := \operatorname{vec}(P_v) \in \mathbb{R}^{n_c}$, where $n_c = n_x n_y n_t$ is the total number of control points per field. Evaluating the tensor product B-spline basis and its analytic derivatives at the $M_{\mathrm{pde}}$ interior PDE-collocation points (where the number of collocation points is chosen such that $M_{\mathrm{pde}} = M_{\mathrm{data}}$) yields the basis matrices $N, N_t, N_x, N_y, N_{xx}, N_{yy} \in \mathbb{R}^{M_{\mathrm{pde}} \times n_c}$. Consequently, the continuous reconstructed fields and their spatial and temporal derivatives at these collocation points can be expressed algebraically as
\begin{subequations}
\begin{align}
u      &= N \mathbf{p}_u, & u_t    &= N_t \mathbf{p}_u, & u_x    &= N_x \mathbf{p}_u, & u_y    &= N_y \mathbf{p}_u, & u_{xx} &= N_{xx} \mathbf{p}_u, & u_{yy} &= N_{yy} \mathbf{p}_u, \\
v      &= N \mathbf{p}_v, & v_t    &= N_t \mathbf{p}_v, & v_x    &= N_x \mathbf{p}_v, & v_y    &= N_y \mathbf{p}_v, & v_{xx} &= N_{xx} \mathbf{p}_v, & v_{yy} &= N_{yy} \mathbf{p}_v.
\end{align}
\end{subequations}

Similarly, let $N_{\mathrm{data}}$, $N_{\mathrm{IC}}$, and $N_{\mathrm{BC}}$ represent the basis matrices restricted to the spatial and temporal coordinates of the observation data, IC, and BC sets, respectively. The corresponding target value vectors for the $u$ and $v$ fields are denoted by $\mathbf q_u^d$ and $\mathbf q_v^d$ for the observational dataset, $\mathbf q_u^0$ and $\mathbf q_v^0$ for the initial snapshots, and $\mathbf q_u^b$ and $\mathbf q_v^b$ for the prescribed boundary values. Using these discrete operators, we construct the strong-form PDE residuals for the coupled Burgers system  directly in the control-point space:
\begin{subequations}
\begin{align}
r_u &= u_t + u\odot u_x + v\odot u_y - \nu(u_{xx}+u_{yy}),\label{eq:burgers_res_u_clean} \\ 
r_v &= v_t + u\odot v_x + v\odot v_y - \nu(v_{xx}+v_{yy}), \label{eq:burgers_res_v_clean}
\end{align}
\end{subequations}
where the operator $\odot$ denotes the Hadamard (elementwise) product. The global PI-MFA objective function is then formulated as a weighted sum of the  MSE associated with the data misfits and the physics-informed constraint residuals:
\begin{align}
\mathcal{L}_{\mathrm{Burgers}}(\mathbf{p}_u, \mathbf{p}_v) 
&= \frac{\lambda_{\mathrm{data}}}{M_{\mathrm{data}}} \left( \left\| N_{\mathrm{data}} \mathbf{p}_u - \mathbf q_u^d \right\|_2^2 + \left\| N_{\mathrm{data}} \mathbf{p}_v - \mathbf q_v^d \right\|_2^2 \right) \nonumber \\
&\quad + \frac{\lambda_{\mathrm{PDE}}}{M_{\mathrm{pde}}} \left( \left\| r_u \right\|_2^2 + \left\| r_v \right\|_2^2 \right) \nonumber \\
&\quad + \frac{\lambda_{\mathrm{IC}}}{M_{\mathrm{IC}}} \left( \left\| N_{\mathrm{IC}} \mathbf{p}_u - \mathbf q_u^0 \right\|_2^2 + \left\| N_{\mathrm{IC}} \mathbf{p}_v - \mathbf q_v^0 \right\|_2^2 \right) \nonumber \\
&\quad + \frac{\lambda_{\mathrm{BC}}}{M_{\mathrm{BC}}} \left( \left\| N_{\mathrm{BC}} \mathbf{p}_u - \mathbf q_u^b \right\|_2^2 + \left\| N_{\mathrm{BC}} \mathbf{p}_v - \mathbf q_v^b \right\|_2^2 \right).
\label{eq:burgers_objective_clean}
\end{align}

To efficiently minimize this objective using the L-BFGS algorithm, we must derive and supply the exact analytical gradients. We first define the stacked residual vector $r(\mathbf{p}_u,\mathbf{p}_v) = [r_u^\top, r_v^\top]^\top$ and compute its global Jacobian with respect to the control points, $J_r = \partial r / \partial (\mathbf{p}_u,\mathbf{p}_v)$. This global Jacobian is partitioned into four block matrices,
\begin{equation}
J_r =
\begin{bmatrix}
J_{uu} & J_{uv}\\
J_{vu} & J_{vv}
\end{bmatrix},
\end{equation}
which are evaluated as
\begin{subequations}
\begin{align}
J_{uu} &= N_t + \operatorname{diag}(u_x)N + \operatorname{diag}(u)N_x + \operatorname{diag}(v)N_y - \nu(N_{xx}+N_{yy}), \\
J_{uv} &= \operatorname{diag}(u_y)N, \\
J_{vu} &= \operatorname{diag}(v_x)N, \\
J_{vv} &= N_t + \operatorname{diag}(u)N_x + \operatorname{diag}(v_y)N + \operatorname{diag}(v)N_y - \nu(N_{xx}+N_{yy}).
\end{align}
\end{subequations}

The gradients of the PDE-residual penalty, $\mathcal{L}_{\mathrm{PDE}}$, with respect to the control vectors are computed via the chain rule as
\begin{subequations}
\begin{align}
\nabla_{\mathbf{p}_u}\mathcal{L}_{\mathrm{PDE}} &= \frac{2}{M_{\mathrm{pde}}}\left(J_{uu}^\top r_u + J_{vu}^\top r_v\right), \\
\nabla_{\mathbf{p}_v}\mathcal{L}_{\mathrm{PDE}} &= \frac{2}{M_{\mathrm{pde}}}\left(J_{uv}^\top r_u + J_{vv}^\top r_v\right).
\end{align}
\end{subequations}

In order to avoid the explicit construction and associated memory overhead of large diagonal matrices, these gradients are computed more efficiently in practice by using Hadamard (elementwise) products ($\odot$):
\begin{subequations}
\begin{align}
\nabla_{\mathbf{p}_u}\mathcal{L}_{\mathrm{PDE}}
&=
\frac{2}{M_{\mathrm{pde}}}\Big[
N_t^\top r_u
+N^\top(u_x\odot r_u)
+N_x^\top(u\odot r_u)
+N_y^\top(v\odot r_u)
-\nu(N_{xx}^\top+N_{yy}^\top)r_u
+N^\top(v_x\odot r_v)
\Big],
\label{eq:burgers_grad_u_pde_clean}
\\
\nabla_{\mathbf{p}_v}\mathcal{L}_{\mathrm{PDE}}
&=
\frac{2}{M_{\mathrm{pde}}}\Big[
N^\top(u_y\odot r_u)
+N_t^\top r_v
+N_x^\top(u\odot r_v)
+N^\top(v_y\odot r_v)
+N_y^\top(v\odot r_v)
-\nu(N_{xx}^\top+N_{yy}^\top)r_v
\Big].
\label{eq:burgers_grad_v_pde_clean}
\end{align}
\end{subequations}

Assembling the contributions from the data, initial, and boundary constraints yields the full analytical gradients of the optimization objective:
\begin{subequations}
\begin{align}
\nabla_{\mathbf{p}_u}\mathcal{L}_{\mathrm{Burgers}}
&=
\frac{2\lambda_{\mathrm{data}}}{M_{\mathrm{data}}}N_{\mathrm{data}}^\top(N_{\mathrm{data}}\mathbf{p}_u-\mathbf q_u^d)
+
\lambda_{\mathrm{PDE}}\nabla_{\mathbf{p}_u}\mathcal{L}_{\mathrm{PDE}}
\nonumber\\
&\quad
+
\frac{2\lambda_{\mathrm{IC}}}{M_{\mathrm{IC}}}N_{\mathrm{IC}}^\top(N_{\mathrm{IC}}\mathbf{p}_u-\mathbf q_u^0)
+
\frac{2\lambda_{\mathrm{BC}}}{M_{\mathrm{BC}}}N_{\mathrm{BC}}^\top(N_{\mathrm{BC}}\mathbf{p}_u-\mathbf q_u^b),
\\
\nabla_{\mathbf{p}_v}\mathcal{L}_{\mathrm{Burgers}}
&=
\frac{2\lambda_{\mathrm{data}}}{M_{\mathrm{data}}}N_{\mathrm{data}}^\top(N_{\mathrm{data}}\mathbf{p}_v-\mathbf q_v^d)
+
\lambda_{\mathrm{PDE}}\nabla_{\mathbf{p}_v}\mathcal{L}_{\mathrm{PDE}}
\nonumber\\
&\quad
+
\frac{2\lambda_{\mathrm{IC}}}{M_{\mathrm{IC}}}N_{\mathrm{IC}}^\top(N_{\mathrm{IC}}\mathbf{p}_v-\mathbf q_v^0)
+
\frac{2\lambda_{\mathrm{BC}}}{M_{\mathrm{BC}}}N_{\mathrm{BC}}^\top(N_{\mathrm{BC}}\mathbf{p}_v-\mathbf q_v^b).
\end{align}
\end{subequations}

\subsubsection{Recovery from physically inconsistent data}
\begin{table}[t]
  \centering
  \footnotesize
  \setlength{\tabcolsep}{11pt} 
  \renewcommand{\arraystretch}{1.12}
  \caption{Quantitative comparison of the final loss components for the 2D coupled Burgers' equations across purely data-driven MFA, regularized MFA (Reg-MFA), a physics-informed neural network (PINN) baseline, and four representative PI-MFA configurations. The data-misfit, PDE-residual, initial condition (IC), and boundary condition (BC) terms are evaluated separately for the $u$ and $v$ velocity fields with the unweighted total composite loss.}
  \label{tab:loss_PIMFA_2dburgers}
  \begin{threeparttable}
  \begin{tabular}{@{} l ccccccc @{}}
    \toprule
      & \multicolumn{1}{c}{\textbf{MFA}}
      & \multicolumn{1}{c}{\textbf{Reg-MFA}}
      & \multicolumn{1}{c}{\textbf{PINN}}
      & \multicolumn{4}{c}{\textbf{PI-MFA}} \\
    \cmidrule(lr){2-2}\cmidrule(lr){3-3}\cmidrule(lr){4-4}\cmidrule(l){5-8}
      & \begin{tabular}{@{}c@{}}\scriptsize(data only)\end{tabular}
      & \begin{tabular}{@{}c@{}}\scriptsize(smoothing)\end{tabular}
      & \begin{tabular}{@{}c@{}}\scriptsize(PDE/IC)\end{tabular}
      & \begin{tabular}{@{}c@{}}Case 1 \\[-0.5ex] \scriptsize(naive)\end{tabular}
      & \begin{tabular}{@{}c@{}}Case 2 \\[-0.5ex] \scriptsize(physics)\end{tabular}
      & \begin{tabular}{@{}c@{}}Case 3 \\[-0.5ex] \scriptsize(data/PDE)\end{tabular}
      & \begin{tabular}{@{}c@{}}Case 4 \\[-0.5ex] \scriptsize(PDE/IC)\end{tabular} \\
    \midrule

    Iteration
      & 0 & 0 & -- & 4 & 131 & 101 & 145 \\
      \midrule
    $\lambda_{\mathrm{data}}$
      & 1 & 1 & $10^{-3}$ & 1 & $10^{-2}$ & $10$ & $10^{-3}$ \\

    $\lambda_{\mathrm{pde}}$
      & 0 & 0 & $10^{2}$ & 1 & $10^{2}$ & $10^{2}$ & $10^{2}$ \\

    $\lambda_{\mathrm{IC}}$
      & 0 & 0 & $10^{2}$ & 1 & $10^{2}$ & $10^{-3}$ & $10^{2}$ \\

    $\lambda_{\mathrm{BC}}$
      & 0 & 0 & $10^{-3}$ & 1 & $10^{2}$ & $10^{-3}$ & $10^{-3}$ \\
    \midrule

    $\mathcal{L}^{u}_{\mathrm{data}}$
      & 2.155e-6 & 9.296e-6 & 2.821e-2 & 4.663e-5 & 2.465e-3 & 1.312e-3 & 2.682e-3 \\
    $\mathcal{L}^{v}_{\mathrm{data}}$
      & 1.188e-5 & 5.092e-6 & 8.196e-3 & 4.902e-5 & 7.650e-4 & 7.811e-4 & 7.821e-4 \\
    \addlinespace[3pt]

    $\mathcal{L}^{u}_{\mathrm{pde}}$
      & 1.464e-2 & 1.184e-1 & 2.063e-3 & 4.223e-3 & 2.549e-4 & 5.135e-5 & 2.485e-4 \\
    $\mathcal{L}^{v}_{\mathrm{pde}}$
      & 1.096e-2 & 7.740e-2 & 1.926e-3 & 1.913e-3 & 2.066e-4 & 3.124e-5 & 2.037e-4 \\
    \addlinespace[3pt]

    $\mathcal{L}^{u}_{\mathrm{IC}}$
      & 4.317e-8 & 1.487e-7 & 2.087e-3 & 1.107e-6 & 3.923e-5 & 1.984e-3 & 3.770e-5 \\
    $\mathcal{L}^{v}_{\mathrm{IC}}$
      & 4.838e-8 & 1.868e-7 & 3.347e-3 & 3.224e-7 & 6.009e-5 & 2.186e-3 & 5.519e-5 \\
    \addlinespace[3pt]

    $\mathcal{L}^{u}_{\mathrm{BC}}$
      & 1.206e-13 & 5.275e-14 & 9.302e-4 & 3.694e-7 & 3.819e-7 & 5.534e-5 & 2.557e-4 \\
    $\mathcal{L}^{v}_{\mathrm{BC}}$
      & 2.018e-13 & 9.680e-14 & 9.294e-4 & 5.690e-7 & 5.975e-7 & 1.011e-5 & 5.061e-5 \\
    \midrule

    \textbf{$\mathcal{L}_{\mathrm{total}}$}
      & \textbf{2.560e-2} & \textbf{1.958e-1} & \textbf{4.769e-2} & \textbf{6.234e-3} & \textbf{3.791e-3} & \textbf{6.411e-3} & \textbf{4.315e-3} \\
    \bottomrule
  \end{tabular}

  \end{threeparttable}

\end{table}

For the two-dimensional coupled Burgers system, the global loss function comprises four distinct penalty categories: data fidelity, PDE residual, IC, and  BC enforcement. While the system governs two separate velocity components ($u$ and $v$), which could theoretically introduce eight independent hyperparameters, we assign identical weights to the $u$- and $v$-contributions within each category. This strategy reduces the hyperparameter space to four independent weights: $\lambda_{\mathrm{data}}$, $\lambda_{\mathrm{pde}}$, $\lambda_{\mathrm{IC}}$, and $\lambda_{\mathrm{BC}}$. Given the increased computational expense of L-BFGS optimization in three space-time dimensions $(x,y,t)$, conducting an exhaustive 4D parameter sweep is impractical. Instead, we select representative weight configurations informed by both the insights gained from the previous 1D example and the error magnitudes observed in the purely data-driven MFA baseline (Table~\ref{tab:loss_terms_for_plain_MFA2Dburgers}). Our primary objective in this 2D example is to suppress the unmodeled, localized forcing embedded within the coarse observation data, thereby recovering a continuous trajectory that is more consistent with the prescribed, unforced Burgers equations.

To evaluate the flexibility of the proposed framework, we analyze four representative weight configurations. Case 1 ($\lambda_{*}=1$) serves as a naive baseline, applying equal importance to all constraints. In Case 2 (physics-dominated), the data weight $\lambda_{\mathrm{data}}$ is intentionally heavily reduced ($10^{-2}$) to acknowledge that the input observations are inherently polluted by unmodeled physics. This configuration forces the optimizer to prioritize the governing equations, alongside the IC and BC constraints. Case 3 (data- and PDE-dominated) maintains a strong emphasis on data interpolation ($\lambda_{\mathrm{data}}=10$) and PDE consistency ($\lambda_{\mathrm{pde}}=10^2$), while dramatically relaxing the boundary and initial constraints ($\lambda_{\mathrm{IC}}, \lambda_{\mathrm{BC}}=10^{-3}$). This choice is motivated by the observation that standard MFA naturally satisfies the IC and BC constraints with high precision even without explicit enforcement (Table~\ref{tab:loss_terms_for_plain_MFA2Dburgers}).   Case 4 (PDE- and IC-dominated) combines a suppressed data weight with enforcement of the PDE and initial conditions, providing an alternative strategy for filtering the internal volumetric discrepancy while locking the starting state.

The performance metrics for these four PI-MFA configurations, along with standard MFA, Regularized MFA (Reg-MFA), and a PINN baseline, are summarized in Table~\ref{tab:loss_PIMFA_2dburgers}. As expected, the purely data-driven standard MFA yields the smallest data-misfit errors but incurs massive PDE residuals, confirming its inability to filter out the unmodeled forcing. The Reg-MFA baseline, which applies smoothing penalties without physical knowledge, exacerbates this issue. It degrades the data fit while simultaneously increasing the PDE residuals to the highest levels observed across all methods ($\mathcal{L}_{\mathrm{total}}=1.958\times10^{-1}$). The PINN baseline succeeds in lowering the PDE residuals compared with standard and regularized MFA but struggles significantly with boundary and initial condition enforcement, resulting in a higher total loss ($\mathcal{L}_{\mathrm{total}}=4.769\times10^{-2}$) than any of the PI-MFA configurations.

In contrast, all four PI-MFA configurations successfully balance constraint enforcement, reducing the unweighted total composite loss relative to the baselines. While enforcing physical constraints inherently causes a minor degradation in data approximation fidelity compared with the data-only MFA, the PDE residuals drop by several orders of magnitude. Among the tested configurations, Case 2 achieves the lowest total loss ($\mathcal{L}_{\mathrm{total}}=3.791\times10^{-3}$), demonstrating that down-weighting the corrupted data while enforcing the PDE and boundary conditions provides the most robust filtration of the unmodeled source term. Case 3 successfully drives the PDE residuals to their absolute minimum values but incurs noticeably larger IC and BC errors due to its relaxed constraint weights. Case 4 performs similarly to Case 2, yielding excellent physics consistency but slightly elevated data-misfit values. Furthermore, in comparison with the PINN baseline, PI-MFA benefits intrinsically from the local support of the B-spline basis. This mathematical structure allows the framework to anchor physical constraints locally and produce exceptionally smooth gradient fields, even when operating on coarse discretizations. Ultimately, these findings demonstrate that intelligently down-weighting physically inconsistent observations empowers the PI-MFA framework to reject latent perturbations and accurately recover the prescribed governing dynamics.

  \begin{figure}
	\centering
	\includegraphics[width=.95\textwidth]{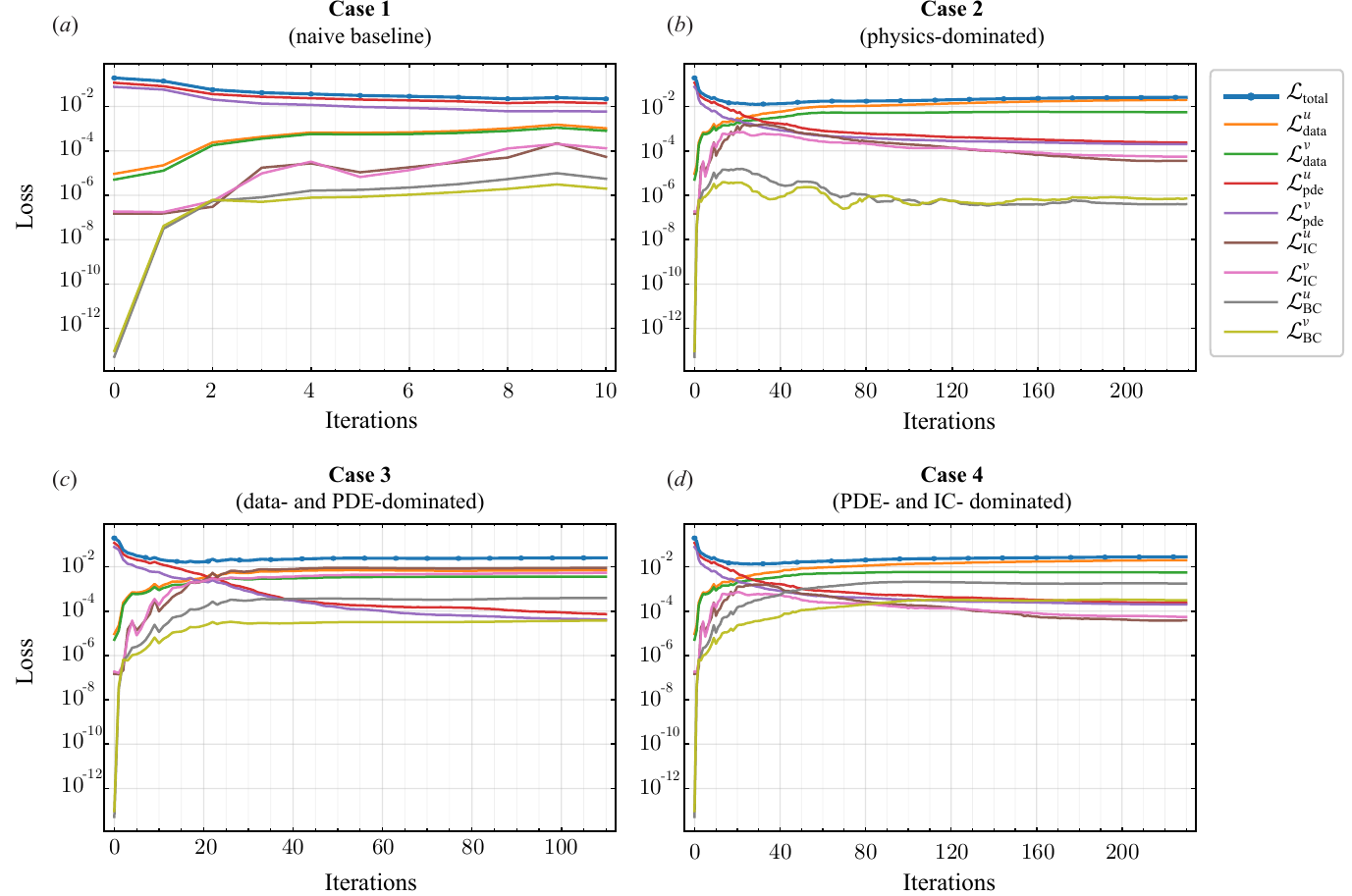}
  \caption{Convergence histories of the unweighted loss components during L-BFGS optimization for the four representative PI-MFA configurations applied to the 2D coupled Burgers problem. Each panel tracks the evolution of the total loss alongside the individual data-misfit, PDE-residual, initial-condition, and boundary-condition contributions for both the $u$ and $v$ velocity fields.}\label{Fig_Burgers_iter}
\end{figure}


Figure~\ref{Fig_Burgers_iter} complements the quantitative final losses in Table~\ref{tab:loss_PIMFA_2dburgers} by detailing the full L-BFGS convergence histories for the four PI-MFA configurations. These optimization trajectories clearly illustrate how weight selection fundamentally alters the dynamic trade-off among data interpolation, PDE residual minimization, and IC/BC constraint satisfaction. For instance, the naive baseline (Case 1, Fig.~\ref{Fig_Burgers_iter}(\textit{a})) terminates after only a few iterations. The optimizer quickly fits the physically inconsistent observation data, causing the unweighted PDE residuals to permanently stagnate at unacceptably high levels. Conversely, configurations utilizing a reduced data weight, such as Case 2 (Fig.~\ref{Fig_Burgers_iter}(\textit{b})) and Case 4 (Fig.~\ref{Fig_Burgers_iter}(\textit{d})), require significantly more iterations to converge. In these physics-dominated regimes, the algorithm actively works to filter the unmodeled forcing, systematically driving the PDE residuals down by several orders of magnitude until they prominently cross below the data-misfit errors. Meanwhile, Case 3 (Fig.~\ref{Fig_Burgers_iter}(\textit{c})) demonstrates the direct consequence of relaxing the boundary and initial constraints. While the PDE residuals are successfully driven to their absolute minimum, the $\mathcal{L}_{\mathrm{IC}}$ and $\mathcal{L}_{\mathrm{BC}}$ terms experience a rapid, pronounced surge early in the optimization process before stabilizing. Ultimately, these convergence histories visually confirm that an appropriate hyperparameter balance is crucial; down-weighting corrupted data prevents premature overfitting and allows the L-BFGS algorithm to actively prioritize physical consistency throughout the optimization trajectory.

  \begin{figure}
	\centering
	\includegraphics[width=.85\textwidth]{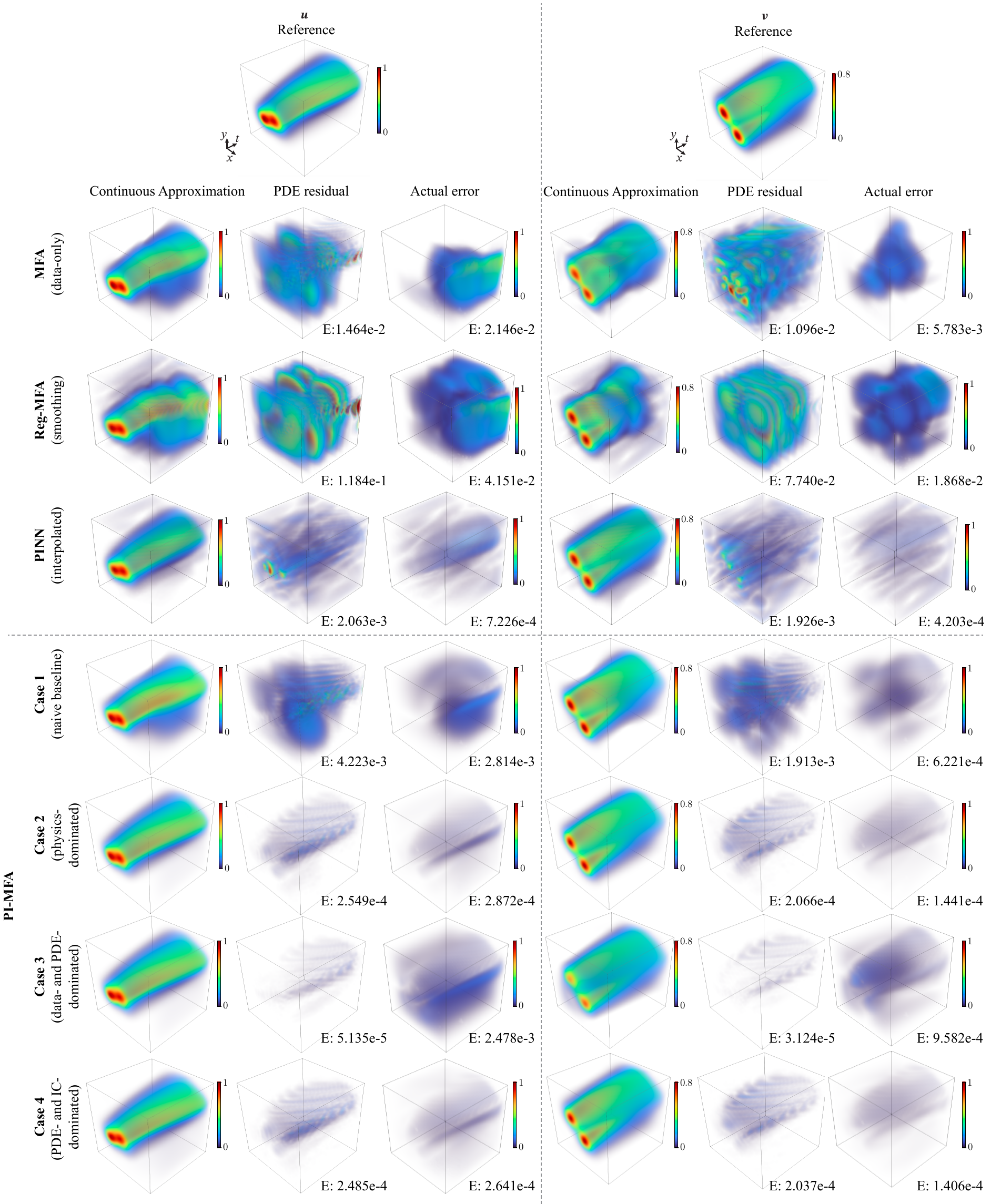}
\caption{Volumetric comparison of the reconstructed 2D Burgers velocity fields ($u$ and $v$) across standard MFA, regularized MFA (Reg-MFA), a PINN baseline, and four PI-MFA configurations. For each velocity component, the panels display the continuous approximation, the corresponding strong-form PDE residual, and the pointwise actual error relative to the high-fidelity FEM reference solution. The aggregated MSE for the PDE residual and actual error is reported beneath each respective volume.}\label{Fig_Burgers_PIMFA_actErr_uv}
\end{figure}


Figure~\ref{Fig_Burgers_PIMFA_actErr_uv} visually compares the reconstructed velocity components ($u$ and $v$) across standard MFA, Reg-MFA, a PINN, and the four PI-MFA configurations. For each method, the continuous approximation is plotted alongside its corresponding strong-form PDE residual and its actual pointwise error relative to the unforced FEM reference solution. In the first row, standard MFA captures the macroscopic space-time flow patterns but exhibits broad, spatially distributed residuals and errors. This situation occurs because the purely data-driven spline fits the unmodeled forcing present in the input data (simulation data). The Reg-MFA baseline attempts to mitigate noise via blind mathematical smoothing. Because it lacks physical awareness, however, it oversmooths the true gradients of the target model. This exacerbates the physical discrepancy, resulting in a highly blurred approximation and increases in both the PDE residual and actual error. The PINN baseline incorporates the governing equations but struggles with the continuous spatial representation. As seen in the actual error and PDE residual volumes, the PINN reconstruction suffers from noticeable diagonal streaking and high-frequency artifacts, indicating an inability to capture smooth, stable gradients natively.

  \begin{figure}
	\centering
	\includegraphics[width=.9\textwidth]{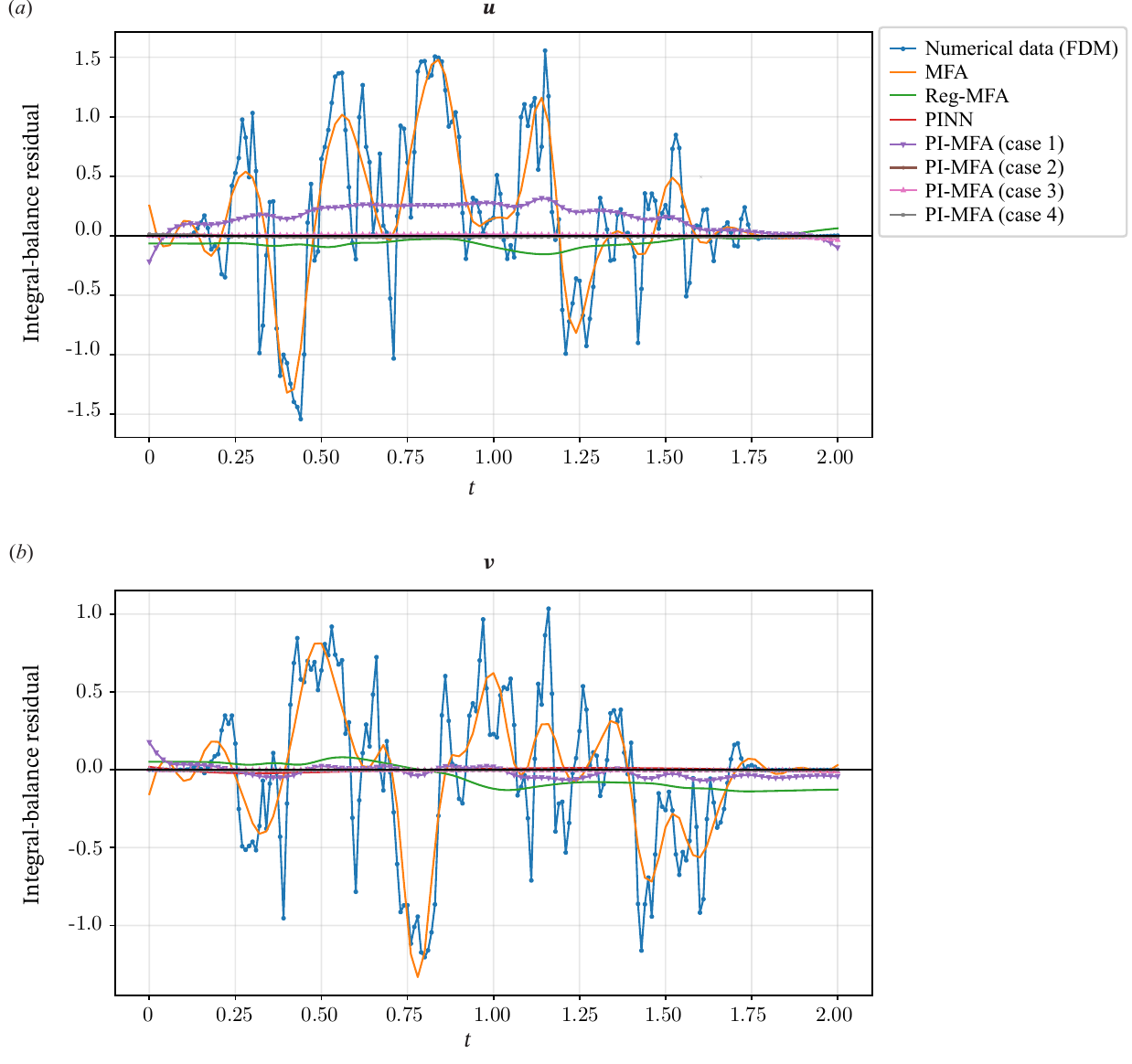}
\caption{Temporal evolution of the global integral-balance residuals, (a) $r_{u,M}(t)$ and (b) $r_{v,M}(t)$, for the two-dimensional coupled Burgers equations (see Eq.~\eqref{eq:burgers_mass_residual}).}\label{Fig_Burgers_massbalance}
\end{figure}


In contrast, applying the physics-informed MFA framework (PI-MFA) systematically drives the reconstructed fields closer to the exact FEM reference. While the continuous approximations of all four PI-MFA cases appear visually reasonable, their residual and actual-error panels reveal profound, quantifiable improvements over the baselines, especially standard MFA and Reg-MFA. For the $u$ velocity component, the actual error decreases from $2.146\times10^{-2}$ (standard MFA) to $2.814\times10^{-3}$, $2.872\times10^{-4}$, $2.478\times10^{-3}$, and $2.641\times10^{-4}$ for PI-MFA Cases 1 through 4, respectively. A similar magnitude of improvement is observed for the $v$ component, where the actual error drops from $5.783\times10^{-3}$ (MFA) down to $1.406\times10^{-4}$ (Case 4). Simultaneously, the PDE residual volumes are heavily suppressed and tightly localized compared with the standard MFA and Reg-MFA reconstruction, confirming a restoration of physical consistency. To compare with PINN reconstruction, the PINN approach needs another postprocessing to have a continuous field from the simulation data. From a computational-cost perspective, using a PINN as an intermediate model and then postprocessing it into another continuous representation is less efficient for this workflow, while the actual error can be significantly dropped. Among the four variants, Cases 2 and 4 yield the best overall performance, achieving the smallest residuals and the lowest actual errors for both velocity components. Ultimately, these visual and quantitative results demonstrate that embedding the governing equations directly into the MFA objective allows the continuous representation to reject corrupted observation data and reconstruct the true underlying physics far more effectively than the previous pure data-fitting and smoothing MFA framework.

Figure~\ref{Fig_Burgers_massbalance} reports the time histories of the componentwise global integral-balance residuals, $r_{u,M}(t)$ and $r_{v,M}(t)$ as defined in Eq.~\eqref{eq:burgers_mass_residual}, for the various reconstruction methods. As a baseline, the residuals computed directly from the discrete observation data using the finite difference method  are included. For the exact solution of the prescribed, unforced Burgers system, the componentwise integral-balance residuals should be zero. These time histories provide a diagnostic of macroscopic physical consistency that extends beyond the localized loss metrics summarized in Table~\ref{tab:loss_PIMFA_2dburgers}. The FDM evaluation of the corrupted observation data exhibits high-frequency oscillations. While the purely data-driven standard MFA leverages its continuous spline representation to smooth out the harshest numerical artifacts, its trajectory still broadly tracks the corrupted data, failing entirely to satisfy the componentwise integral-balance diagnostic. Similarly, Reg-MFA applies mathematical smoothing to further dampen the integral variance; but lacking physical constraints, it still drifts significantly from the true zero-line. In contrast, integrating physics directly into the optimization fundamentally corrects this behavior. The PINN baseline and the physics-dominated PI-MFA configurations (Cases 2, 3, and 4) successfully collapse the balance residuals tightly around zero. Case 1 (the naive baseline) improves upon standard MFA but still exhibits noticeable fluctuations due to its equal weighting, which allows the corrupted data to exert too much influence. Among the tuned variants, Cases 2 and 4 yield the tightest adherence to zero across the entire time domain, aligning perfectly with their strong emphasis on the PDE constraints. Case 3, while producing exceptionally small PDE residuals locally, exhibits marginally larger excursions in the global balance at certain time steps, reflecting its weaker enforcement of the initial and boundary conditions. Overall, Fig.~\ref{Fig_Burgers_massbalance} confirms that appropriate parameter reweighting in the PI-MFA framework not only minimizes local optimization losses but effectively guarantees the global integral-balance consistency of the prescribed 2D Burgers system. An exhaustive parametric study for larger-scale systems falls outside the scope of the present article and will be addressed in future investigations.

\subsection{Two-dimensional incompressible Navier-Stokes equation}\label{subsec:2DNS_cavity}
We consider the two-dimensional, unsteady, incompressible Navier-Stokes equations
in a lid-driven cavity on the unit-square domain $\Omega=(0,1)\times(0,1)$ over
the time interval $t\in[0,T]$. Let $\boldsymbol{u}=(u,v)^{\mathsf T}$ denote the
velocity field and $p$ the kinematic pressure. The governing equations in strong
form are
\begin{subequations}\label{eq:NS_strong}
\begin{align}
\frac{\partial u}{\partial t}
+ u\frac{\partial u}{\partial x}
+ v\frac{\partial u}{\partial y}
+ \frac{\partial p}{\partial x}
- \nu\left(
\frac{\partial^2 u}{\partial x^2}
+ \frac{\partial^2 u}{\partial y^2}
\right)
&= 0,
\qquad (x,y)\in\Omega,\ t>0, \label{eq:NS_strong_mom_x}\\
\frac{\partial v}{\partial t}
+ u\frac{\partial v}{\partial x}
+ v\frac{\partial v}{\partial y}
+ \frac{\partial p}{\partial y}
- \nu\left(
\frac{\partial^2 v}{\partial x^2}
+ \frac{\partial^2 v}{\partial y^2}
\right)
&= 0,
\qquad (x,y)\in\Omega,\ t>0, \label{eq:NS_strong_mom_y}\\
\frac{\partial u}{\partial x}
+
\frac{\partial v}{\partial y}
&= 0,
\qquad (x,y)\in\Omega,\ t>0. \label{eq:NS_strong_cont}
\end{align}
\end{subequations}

Here $\nu$ denotes the kinematic viscosity. The flow is driven by the motion of the top lid. In the classical lid-driven cavity benchmark~\citep{GHIA1982387}, the top boundary is assigned a uniform tangential velocity. However, this creates physical discontinuities at the top corners, $(0,1)$ and $(1,1)$, where the moving lid meets the stationary side walls. These velocity discontinuities lead to non-physical pressure singularities that can degrade the convergence order of high-fidelity numerical schemes and present significant challenges. To regularize the flow and ensure a bounded, globally smooth pressure and velocity field, we adopt the regularized lid-driven cavity formulation, which is widely used to evaluate high-order continuous numerical methods \citep{shen1991hopf}. We replace the discontinuous top boundary condition with a smooth velocity profile that tapers to zero at the corners. Specifically, we prescribe a fourth-order polynomial velocity profile on the top lid ($y=1$):
\begin{equation}
u(x, 1, t) = 16x^2(1 - x)^2, \quad v(x, 1, t) = 0, \quad \text{for } x \in [0,1], ; t \ge 0.
\end{equation}
This specific polynomial is chosen because it satisfies the no-slip condition ($u = 0$) at the corners ($x = 0$ and $x = 1$), effectively preventing the singularity, while maintaining a maximum tangential velocity of $U = 1$ at the center of the lid ($x = 0.5$). Standard no-slip boundary conditions are imposed on the remaining three stationary walls:
\begin{equation}u(x, y, t) = 0, \quad v(x, y, t) = 0, \quad \text{for } x=0, ; x=1, \text{ or } y=0, ; t \ge 0.
\end{equation}
At time $t = 0$, the interior velocity field and kinematic pressure are initialized from rest:
\begin{equation}u(x, y, 0) = 0, \quad v(x, y, 0) = 0, \quad p(x, y, 0) = 0, \quad \text{for } (x, y) \in \Omega.
\end{equation}
The cavity-length-based Reynolds number is set to $Re = U L/\nu = 1000$, where $U=1$ is the maximum lid velocity and $L=1$ is the cavity length.

\subsubsection{Dataset generation and standard MFA} \label{subsubsec:lid_datageneration_and_MFA}

  \begin{figure}
	\centering
	\includegraphics[width=.65\textwidth]{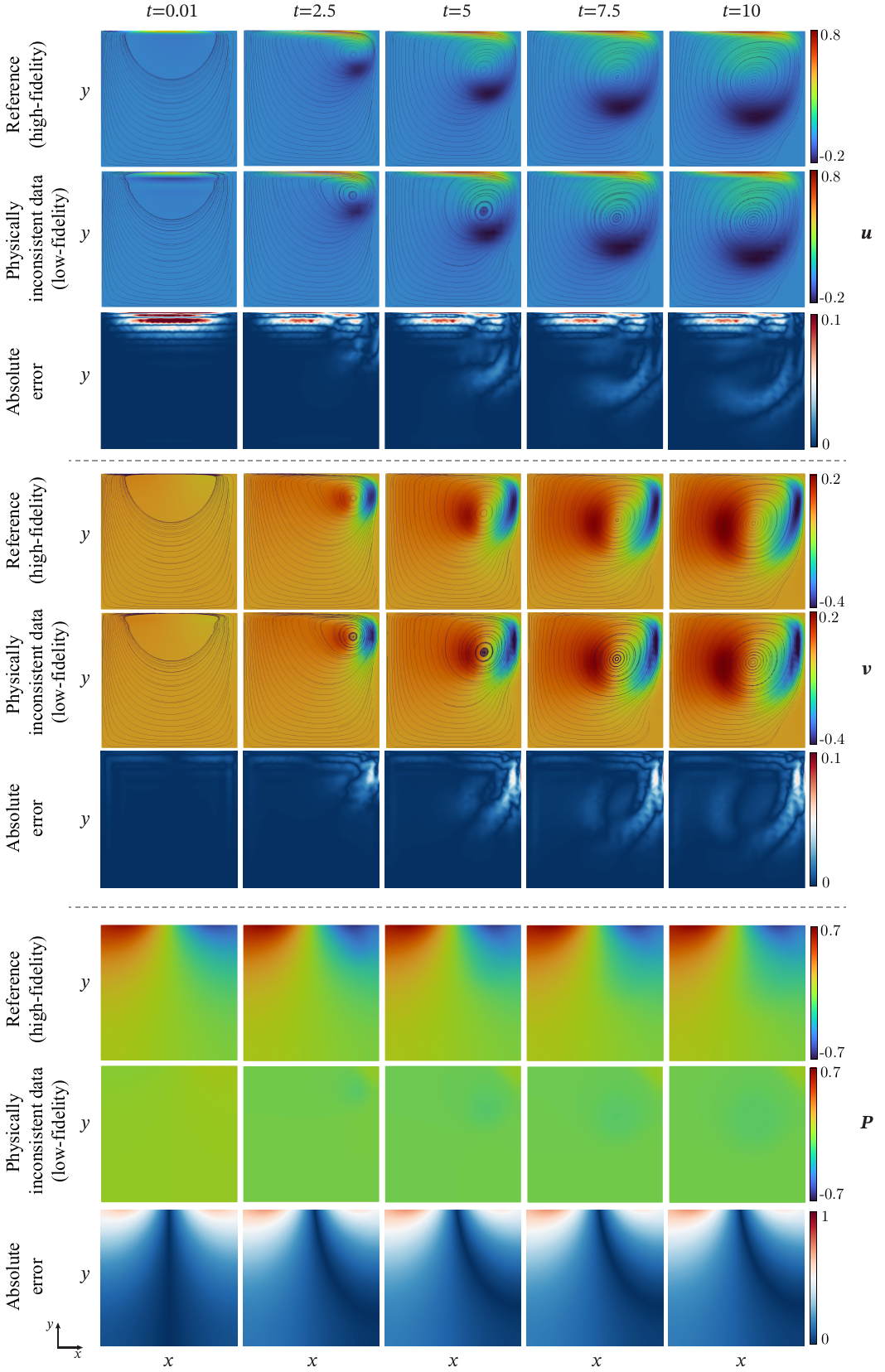}
\caption{Snapshots of the horizontal velocity ($u$), vertical velocity ($v$), and kinematic pressure ($p$) for the regularized 2D lid-driven cavity flow at $t = 0.01,\ 2.5,\ 5,\ 7.5,$ and $10$. For each flow variable, the top row displays the exact high-fidelity FEM reference solution, the middle row shows the physically inconsistent low-fidelity observation data (degraded by spatial underresolution and numerical diffusion), and the bottom row plots the pointwise absolute error. The overlaid solid black lines indicate contours of constant vorticity. In all plots, $x\in[0,1]$ and $y\in[0,1]$.}
\label{Fig_lid_dataset_u}
\end{figure}

To generate the benchmark dataset, we solve the regularized incompressible Navier-Stokes equations using the Firedrake finite element framework \citep{FiredrakeUserManual}. A high-fidelity reference trajectory is computed on a fine $100 \times 100$ uniform spatial mesh. To satisfy the LBB (inf-sup) stability condition, we employ a high-order Taylor--Hood element pair, utilizing continuous Galerkin elements of degree four ($P^4$) for the velocity field and degree three ($P^3$) for the kinematic pressure. Time integration is performed by using the second-order, non-dissipative Crank--Nicolson scheme with a refined time step of $\Delta t_{\mathrm{ref}} = 10^{-3}$, integrated up to a final time of $T = 10.0$. Notably, the sharp corners inherent to this flow configuration induce distinct behavior in the PDE residuals. Given that the finite element method satisfies the weak form of the equations, evaluating the strong-form pointwise residual naturally yields non-zero values clustered heavily around these geometric singularities.

To emulate physically inconsistent observations, we intentionally degrade the simulation fidelity using a coarse grid and lower-order numerical schemes. Filtering discretization and numerical errors in such data is highly relevant, as many emerging AI/ML-driven frameworks rely on cheaper, low-resolution simulations to accelerate computational workflows \citep{Kochkov2021Machine, Pathak2020Using, Li2021Fourier}. Specifically, this corrupted dataset is generated on a much coarser $20 \times 20$ spatial mesh using lower-order $P^2/P^1$ Taylor--Hood elements. Furthermore, the temporal integration is downgraded to the unconditionally stable but highly dissipative first-order backward Euler scheme with a larger time step of $\Delta t = 10^{-2}$. This combination of spatial underresolution and first-order temporal integration deliberately introduces significant artificial numerical diffusion and truncation errors, rendering the resulting discrete data physically inconsistent with the exact continuous Navier-Stokes dynamics.

Figure~\ref{Fig_lid_dataset_u} visualizes the impact of this numerical degradation across the velocity and pressure fields. While the low-fidelity observation data broadly captures the primary macroscopic vortex driven by the sliding lid, the pointwise absolute error plots reveal substantial discrepancies. The $u$ and $v$ velocity fields suffer from noticeable numerical smearing, particularly near the high-gradient region of the moving lid and the primary vortex core. More critically, the pressure field in the low-fidelity dataset is severely degraded, failing almost entirely to capture the sharp gradients at the singular top corners.

To evaluate the capabilities of the PI-MFA framework, we investigate its performance on this purely discretization-induced inconsistency, which is a common source of data discrepancy in scientific visualization. In this regime, the target data violates the continuous governing equations due to numerical diffusion. While PI-MFA successfully drives the continuous momentum and divergence residuals to near zero, producing a highly smooth, mass-conserving velocity field ($\nabla \cdot \mathbf{u} = 0$) suitable for reliable downstream visualization, the absolute error relative to the high-fidelity reference remains bounded. This behavior is mathematically expected. PI-MFA accurately projects the corrupted data onto a physically admissible macroscopic state, but macroscopic PDE constraints alone cannot spontaneously reconstruct subgrid flow features that were irreversibly damped by the coarse solver. However, since the PI-MFA objective inherently embeds the full Navier-Stokes momentum equations, it unlocks a powerful secondary capability---the ability to simultaneously infer and recover hidden or severely corrupted physical variables, such as the continuous pressure field, directly from the observed velocity data.

Similar to the approaches detailed in Sections \ref{subsubsec:1D_datageneration_and_MFA} and \ref{subsubsec:standardMFA_2DBurgers}, we begin by establishing a purely data-driven MFA baseline. This involves reconstructing continuous representations of the two velocity components, $u(x,y,t)$ and $v(x,y,t)$, and the kinematic pressure, $p(x,y,t)$, directly from the physically inconsistent lid-driven cavity observation data. Each physical field is parameterized by an independent, three-dimensional tensor product B-spline over the space-time domain $(x,y,t)\in\Omega\times[0,T]$, as defined in Eq.~\eqref{eq:3D_tensorproduct_Bspline}.

To determine the optimal complexity for the spline space, we perform a resolution study by varying the number of control points, $P_x$, $P_y$, and $P_t$, across the domain. The resulting loss contributions are summarized in Table~\ref{tab:loss_terms_for_2DNS}. Given that the dataset encompasses 1,000 discrete time steps, we allocate relatively denser control-point resolution to the temporal axis. As expected, the results demonstrate a strict trade-off between data fitting fidelity and physical consistency. Increasing the spatial resolution from $10 \times 10$ to $20 \times 20$ improves the data-fitting accuracy, driving the velocity approximation errors down from $\sim10^{-4}$ to $\sim10^{-8}$. However, this tight conformity to the corrupted observation data comes at a steep physical cost. The strong-form momentum and divergence residuals increase by over an order of magnitude. Pushing the spatial resolution further to $30 \times 30$ causes these PDE errors to explode severely ($\mathcal{L}_{\mathrm{total}} \approx 3.6 \times 10^2$). In this highly refined regime, the B-spline basis possesses sufficient flexibility to entirely overfit the artificial numerical diffusion and truncation errors embedded in the low-fidelity observations, completely abandoning the underlying fluid dynamics.

\begin{figure}
	\centering
	\includegraphics[width=.9\textwidth]{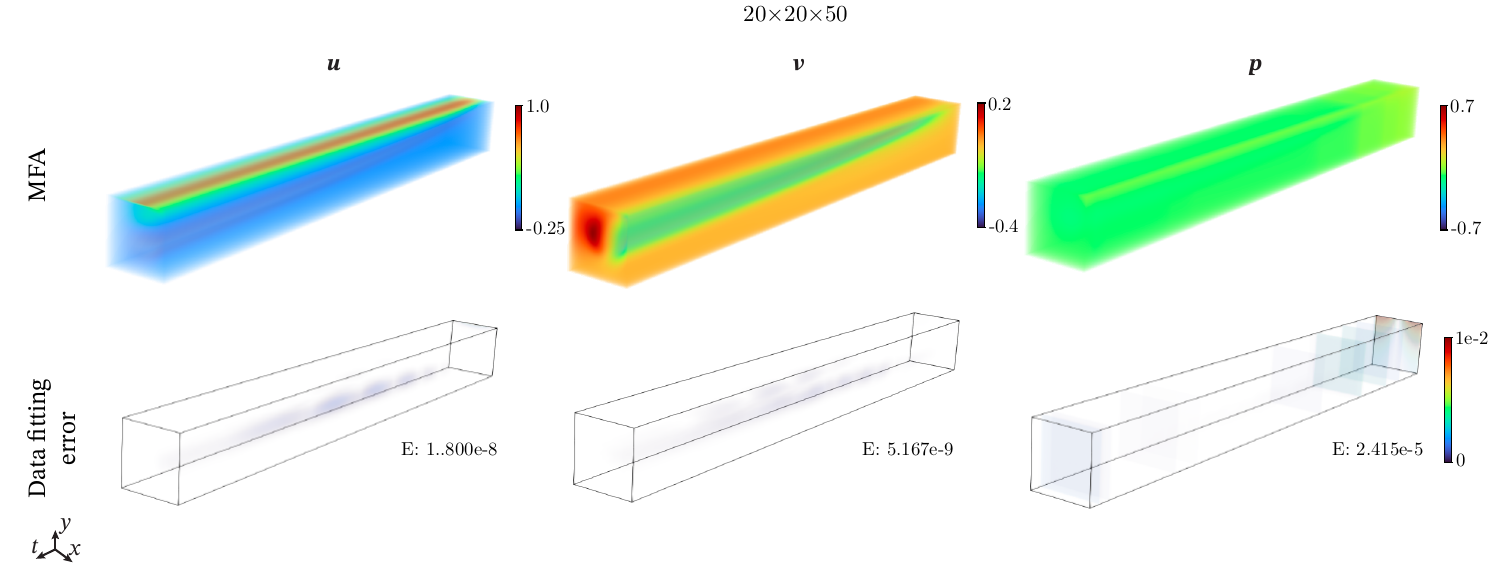}
\caption{Standard, purely data-driven MFA reconstruction of the horizontal velocity ($u$), vertical velocity ($v$), and kinematic pressure ($p$) space-time fields for the 2D lid-driven cavity problem. The volumetric plots depict the continuous approximations evaluated on the selected $20 \times 20 \times 50$ control-point grid (top row) alongside their corresponding pointwise data-fitting errors relative to the corrupted observation data (bottom row). The aggregate mean squared error ($E$) for data interpolation is reported beneath each respective error volume.}
  \label{Fig_lid_datafitting}
\end{figure}

Based on this analysis, we select the $20 \times 20 \times 50$ control-point grid for all subsequent PI-MFA evaluations. As visualized in Fig.~\ref{Fig_lid_datafitting}, this intermediate configuration provides an ideal balance. It provides an adequate geometric fit to the input data while preventing the catastrophic degradation of the PDE residuals observed at higher grid resolutions.

\begin{table*}[t]
  \centering
  \footnotesize 
  \setlength{\tabcolsep}{5.5pt} 
  \renewcommand{\arraystretch}{1.15}
  \caption{Loss contributions for varying control-point grids \((P_x \times P_y \times P_t)\) in the 2D Navier-Stokes lid-driven cavity problem. The total loss reflects the sum of all data, PDE, IC, and BC penalties for the three physical fields (\(u\), \(v\), \(p\)).}  \label{tab:loss_terms_for_2DNS}
  \begin{tabular}{@{} l ccc ccc ccc @{}}
    \toprule
    \multirow{2}{*}{\textbf{Loss term}} & 
    \multicolumn{3}{c}{\textbf{Grid} \(\mathbf{10\times10}\)} & 
    \multicolumn{3}{c}{\textbf{Grid} \(\mathbf{20\times20}\)} & 
    \multicolumn{3}{c}{\textbf{Grid} \(\mathbf{30\times30}\)} \\
    \cmidrule(lr){2-4} \cmidrule(lr){5-7} \cmidrule(l){8-10}
    & \(P_t = 10\) & \(P_t = 50\) & \(P_t = 100\) 
    & \(P_t = 20\) & \(P_t = 50\) & \(P_t = 100\) 
    & \(P_t = 30\) & \(P_t = 50\) & \(P_t = 100\) \\
    \midrule
    
    \(\mathcal{L}^{u}_{\mathrm{data}}\) & 2.487e-4 & 2.455e-4 & 2.455e-4 & 1.966e-7 & 1.800e-8 & 1.575e-8 & 1.849e-8 & 9.666e-9 & 7.419e-9 \\
    \(\mathcal{L}^{v}_{\mathrm{data}}\) & 4.728e-5 & 4.668e-5 & 4.668e-5 & 1.637e-8 & 5.167e-9 & 5.111e-9 & 5.356e-10& 1.654e-10& 1.099e-10 \\
    \(\mathcal{L}^{p}_{\mathrm{data}}\) & 7.120e-5 & 2.530e-5 & 1.206e-5 & 5.114e-5 & 2.415e-5 & 1.091e-5 & 3.872e-5 & 2.415e-5 & 1.091e-5 \\
    \addlinespace[3pt]
    
    \(\mathcal{L}^{x}_{\mathrm{mom}}\)  & 3.025e-1 & 3.031e-1 & 3.032e-1 & 4.103e+0 & 4.103e+0 & 4.104e+0 & 1.581e+2 & 1.583e+2 & 1.584e+2 \\
    \(\mathcal{L}^{y}_{\mathrm{mom}}\)  & 8.698e-2 & 8.744e-2 & 8.759e-2 & 2.633e-1 & 2.637e-1 & 2.638e-1 & 1.836e+2 & 1.837e+2 & 1.839e+2 \\
    \(\mathcal{L}_{\mathrm{div}}\)      & 1.019e-1 & 1.024e-1 & 1.024e-1 & 2.799e-1 & 2.807e-1 & 2.807e-1 & 1.765e+1 & 1.765e+1 & 1.765e+1 \\
    \addlinespace[3pt]
    
    \(\mathcal{L}^{u}_{\mathrm{IC}}\)   & 6.627e-4 & 4.023e-4 & 3.995e-4 & 8.653e-6 & 5.634e-6 & 2.675e-6 & 2.387e-6 & 5.634e-6 & 2.675e-6 \\
    \(\mathcal{L}^{v}_{\mathrm{IC}}\)   & 1.404e-5 & 4.978e-7 & 4.407e-7 & 2.924e-7 & 9.839e-8 & 3.963e-8 & 9.233e-8 & 9.839e-8 & 3.963e-8 \\
    \addlinespace[3pt]
    
    \(\mathcal{L}^{u}_{\mathrm{BC}}\)   & 1.974e-1 & 1.941e-1 & 1.886e-1 & 1.937e-1 & 1.974e-1 & 1.928e-1 & 1.928e-1 & 1.928e-1 & 1.928e-1 \\
    \(\mathcal{L}^{v}_{\mathrm{BC}}\)   & 1.310e-2 & 8.900e-5 & 8.900e-5 & 7.275e-5 & 4.646e-3 & 7.278e-5 & 7.278e-5 & 7.278e-5 & 7.278e-5 \\
    \midrule
    
    \textbf{\(\mathcal{L}_{\mathrm{total}}\)}
    & \textbf{7.029e-1} & \textbf{6.879e-1} & \textbf{6.826e-1}
    & \textbf{4.840e+0} & \textbf{4.850e+0} & \textbf{4.841e+0}
    & \textbf{3.595e+2} & \textbf{3.596e+2} & \textbf{3.599e+2} \\
    \bottomrule
  \end{tabular}
\end{table*}

\subsubsection{PI-MFA formulation for the Navier-Stokes system}\label{subsubsec:PIMFA_NS}

We define the vectorized control-point arrays for the two velocity components and the kinematic pressure as $\mathbf{p}_u := \operatorname{vec}(P_u) \in \mathbb{R}^{n_c}$, $\mathbf{p}_v := \operatorname{vec}(P_v) \in \mathbb{R}^{n_c}$, and $\mathbf{p}_p := \operatorname{vec}(P_p) \in \mathbb{R}^{n_c}$, where $n_c$ denotes the total number of control points per physical field. All three fields are parameterized over the exact same tensor product B-spline space. Consequently, evaluating the basis functions and their analytic derivatives at the $M_{\mathrm{pde}}$ interior PDE-collocation points yields the shared basis matrices $N, N_t, N_x, N_y, N_{xx}, \text{ and } N_{yy}$. The continuously reconstructed fields and their relevant spatial and temporal derivatives at these collocation points can thus be expressed algebraically as
\begin{subequations}
\begin{align}
u      &= N \mathbf{p}_u, & u_t    &= N_t \mathbf{p}_u, & u_x    &= N_x \mathbf{p}_u, & u_y    &= N_y \mathbf{p}_u, & u_{xx} &= N_{xx}\mathbf{p}_u, & u_{yy} &= N_{yy}\mathbf{p}_u, \\
v      &= N \mathbf{p}_v, & v_t    &= N_t \mathbf{p}_v, & v_x    &= N_x \mathbf{p}_v, & v_y    &= N_y \mathbf{p}_v, & v_{xx} &= N_{xx}\mathbf{p}_v, & v_{yy} &= N_{yy}\mathbf{p}_v, \\
p      &= N \mathbf{p}_p, & p_x    &= N_x \mathbf{p}_p, & p_y    &= N_y \mathbf{p}_p.
\end{align}
\end{subequations}

Similarly, let \(N_{\mathrm{data}}\), \(N_{\mathrm{IC}}\), and
\(N_{\mathrm{BC}}\) represent the shared basis matrices restricted to the
coordinates of the observation data, IC, and BC sets, respectively. Since the
incompressible Navier--Stokes equations determine pressure only up to a
spatially constant, time-dependent offset, an additional pressure gauge is
required to fix the pressure level. In this work, we use a pointwise gauge. At a fixed spatial reference location \((x_g,y_g)\) the reconstructed pressure
is constrained to satisfy
\[
p_h(x_g,y_g,t_k)=p_g, \qquad k=1,\ldots,M_{\mathrm{gauge}},
\]
where \(t_k\) denotes the sampled time levels used for the gauge condition. Let \(N_{\mathrm{gauge}}\) denote
the pressure-basis matrix evaluated at the gauge points
\((x_g,y_g,t_k)\), and let
\(\mathbf q_p^{\mathrm{gauge}}=p_g\mathbf 1\) denote the corresponding
reference-pressure vector.

Using these discrete operators, we construct the strong-form residuals for the momentum and incompressibility (divergence) equations directly in the control-point space:
\begin{subequations}
\begin{align}
r_x &= u_t + u\odot u_x + v\odot u_y + p_x - \nu(u_{xx}+u_{yy}), \label{eq:NS_res_x_clean} \\
r_y &= v_t + u\odot v_x + v\odot v_y + p_y - \nu(v_{xx}+v_{yy}), \label{eq:NS_res_y_clean} \\
r_c &= u_x + v_y, \label{eq:NS_res_c_clean}
\end{align}
\end{subequations}
where the operator $\odot$ denotes the Hadamard (elementwise) product. 

To systematically balance the physical constraints, we group the volumetric residuals into a combined PDE-residual penalty, $\mathcal{L}_{\mathrm{PDE}}$
\begin{equation}
\mathcal{L}_{\mathrm{PDE}}(\mathbf{p}_u,\mathbf{p}_v,\mathbf{p}_p) = \frac{\lambda_{\mathrm{mom}}^x}{M_{\mathrm{pde}}}\left\|r_x\right\|_2^2 + \frac{\lambda_{\mathrm{mom}}^y}{M_{\mathrm{pde}}}\left\|r_y\right\|_2^2 + \frac{\lambda_{\mathrm{div}}}{M_{\mathrm{pde}}}\left\|r_c\right\|_2^2.
\end{equation}

The global PI-MFA optimization objective for the Navier--Stokes system, \(\mathcal{L}_{\mathrm{NS}}\), is formulated as the weighted sum of the
data-fidelity, PDE-residual, IC, BC, and pressure-gauge penalties as
\begin{align}
\mathcal{L}_{\mathrm{NS}}(\mathbf{p}_u,\mathbf{p}_v,\mathbf{p}_p)
&= \frac{\lambda_{\mathrm{data}}^u}{M_{\mathrm{data}}}\left\|N_{\mathrm{data}}\mathbf{p}_u-\mathbf q_u^d\right\|_2^2
+ \frac{\lambda_{\mathrm{data}}^v}{M_{\mathrm{data}}}\left\|N_{\mathrm{data}}\mathbf{p}_v-\mathbf q_v^d\right\|_2^2
+ \frac{\lambda_{\mathrm{data}}^p}{M_{\mathrm{data}}}\left\|N_{\mathrm{data}}\mathbf{p}_p-\mathbf q_p^d\right\|_2^2 \nonumber\\
&\quad + \mathcal{L}_{\mathrm{PDE}}(\mathbf{p}_u,\mathbf{p}_v,\mathbf{p}_p) \nonumber\\
&\quad + \frac{\lambda_{\mathrm{IC}}^u}{M_{\mathrm{IC}}}\left\|N_{\mathrm{IC}}\mathbf{p}_u-\mathbf q_u^0\right\|_2^2
+ \frac{\lambda_{\mathrm{IC}}^v}{M_{\mathrm{IC}}}\left\|N_{\mathrm{IC}}\mathbf{p}_v-\mathbf q_v^0\right\|_2^2 \nonumber\\
&\quad + \frac{\lambda_{\mathrm{BC}}^u}{M_{\mathrm{BC}}}\left\|N_{\mathrm{BC}}\mathbf{p}_u-\mathbf q_u^b\right\|_2^2
+ \frac{\lambda_{\mathrm{BC}}^v}{M_{\mathrm{BC}}}\left\|N_{\mathrm{BC}}\mathbf{p}_v-\mathbf q_v^b\right\|_2^2 \nonumber\\
&\quad + \frac{\lambda_{\mathrm{gauge}}}{M_{\mathrm{gauge}}} \left\| N_{\mathrm{gauge}}\mathbf p_p-\mathbf q_p^{\mathrm{gauge}} \right\|_2^2.
\label{eq:NS_objective_clean}
\end{align}
Explicit IC and BC penalties are omitted for pressure, since pressure in incompressible flow is determined through the momentum equations up to an additive gauge. The pressure level is therefore anchored through the gauge penalty. 

To efficiently minimize this objective using the L-BFGS algorithm, we derive the exact analytical gradients. The global Jacobian blocks for the momentum equations with respect to the control variables are evaluated as
\begin{subequations}
\begin{align}
J_{xu} &= N_t + \operatorname{diag}(u_x)N + \operatorname{diag}(u)N_x + \operatorname{diag}(v)N_y - \nu(N_{xx}+N_{yy}), \\
J_{xv} &= \operatorname{diag}(u_y)N, \\
J_{xp} &= N_x, \\
J_{yu} &= \operatorname{diag}(v_x)N, \\
J_{yv} &= N_t + \operatorname{diag}(u)N_x + \operatorname{diag}(v_y)N + \operatorname{diag}(v)N_y - \nu(N_{xx}+N_{yy}), \\
J_{yp} &= N_y.
\end{align}
\end{subequations}
For the divergence constraint, the corresponding Jacobian blocks are simply defined as
\begin{equation}
J_{cu} = N_x, \quad J_{cv} = N_y, \quad J_{cp} = \mathbf{0}.
\end{equation}

Using the chain rule, we compute the gradients of the combined PDE-residual penalty, $\mathcal{L}_{\mathrm{PDE}}$, with respect to the control vectors as
\begin{subequations}
\begin{align}
\nabla_{\mathbf{p}_u}\mathcal{L}_{\mathrm{PDE}} &= \frac{2\lambda_{\mathrm{mom}}^x}{M_{\mathrm{pde}}}J_{xu}^\top r_x + \frac{2\lambda_{\mathrm{mom}}^y}{M_{\mathrm{pde}}}J_{yu}^\top r_y + \frac{2\lambda_{\mathrm{div}}}{M_{\mathrm{pde}}}J_{cu}^\top r_c, \\
\nabla_{\mathbf{p}_v}\mathcal{L}_{\mathrm{PDE}} &= \frac{2\lambda_{\mathrm{mom}}^x}{M_{\mathrm{pde}}}J_{xv}^\top r_x + \frac{2\lambda_{\mathrm{mom}}^y}{M_{\mathrm{pde}}}J_{yv}^\top r_y + \frac{2\lambda_{\mathrm{div}}}{M_{\mathrm{pde}}}J_{cv}^\top r_c, \\
\nabla_{\mathbf{p}_p}\mathcal{L}_{\mathrm{PDE}} &= \frac{2\lambda_{\mathrm{mom}}^x}{M_{\mathrm{pde}}}J_{xp}^\top r_x + \frac{2\lambda_{\mathrm{mom}}^y}{M_{\mathrm{pde}}}J_{yp}^\top r_y.
\end{align}
\end{subequations}

In order to avoid the explicit construction and associated memory overhead of large diagonal matrices, these gradients are implemented in a highly efficient matrix-free form using Hadamard (elementwise) products ($\odot$)
\begin{subequations}
\begin{align}
\nabla_{\mathbf{p}_u}\mathcal{L}_{\mathrm{PDE}}
&= \frac{2\lambda_{\mathrm{mom}}^x}{M_{\mathrm{pde}}}\Big[ N_t^\top r_x + N^\top(u_x\odot r_x) + N_x^\top(u\odot r_x) + N_y^\top(v\odot r_x) - \nu(N_{xx}^\top+N_{yy}^\top)r_x \Big] \nonumber\\
&\quad + \frac{2\lambda_{\mathrm{mom}}^y}{M_{\mathrm{pde}}}\Big[ N^\top(v_x\odot r_y) \Big] + \frac{2\lambda_{\mathrm{div}}}{M_{\mathrm{pde}}}N_x^\top r_c, \\
\nabla_{\mathbf{p}_v}\mathcal{L}_{\mathrm{PDE}}
&= \frac{2\lambda_{\mathrm{mom}}^x}{M_{\mathrm{pde}}}\Big[ N^\top(u_y\odot r_x) \Big] \nonumber\\
&\quad + \frac{2\lambda_{\mathrm{mom}}^y}{M_{\mathrm{pde}}}\Big[ N_t^\top r_y + N_x^\top(u\odot r_y) + N^\top(v_y\odot r_y) + N_y^\top(v\odot r_y) - \nu(N_{xx}^\top+N_{yy}^\top)r_y \Big] \nonumber\\
&\quad + \frac{2\lambda_{\mathrm{div}}}{M_{\mathrm{pde}}}N_y^\top r_c, \\
\nabla_{\mathbf{p}_p}\mathcal{L}_{\mathrm{PDE}}
&= \frac{2\lambda_{\mathrm{mom}}^x}{M_{\mathrm{pde}}}N_x^\top r_x + \frac{2\lambda_{\mathrm{mom}}^y}{M_{\mathrm{pde}}}N_y^\top r_y.
\end{align}
\end{subequations}

Assembling the contributions from the data, initial, boundary, and gauge constraints yields the full analytical gradients of the optimization objective
\begin{subequations}
\begin{align}
\nabla_{\mathbf{p}_u}\mathcal{L}_{\mathrm{NS}} &= \frac{2\lambda_{\mathrm{data}}^u}{M_{\mathrm{data}}}N_{\mathrm{data}}^\top(N_{\mathrm{data}}\mathbf{p}_u-\mathbf q_u^d) + \nabla_{\mathbf{p}_u}\mathcal{L}_{\mathrm{PDE}} \nonumber\\
&\quad + \frac{2\lambda_{\mathrm{IC}}^u}{M_{\mathrm{IC}}}N_{\mathrm{IC}}^\top(N_{\mathrm{IC}}\mathbf{p}_u-\mathbf q_u^0) + \frac{2\lambda_{\mathrm{BC}}^u}{M_{\mathrm{BC}}}N_{\mathrm{BC}}^\top(N_{\mathrm{BC}}\mathbf{p}_u-\mathbf q_u^b), \\
\nabla_{\mathbf{p}_v}\mathcal{L}_{\mathrm{NS}} &= \frac{2\lambda_{\mathrm{data}}^v}{M_{\mathrm{data}}}N_{\mathrm{data}}^\top(N_{\mathrm{data}}\mathbf{p}_v-\mathbf q_v^d) + \nabla_{\mathbf{p}_v}\mathcal{L}_{\mathrm{PDE}} \nonumber\\
&\quad + \frac{2\lambda_{\mathrm{IC}}^v}{M_{\mathrm{IC}}}N_{\mathrm{IC}}^\top(N_{\mathrm{IC}}\mathbf{p}_v-\mathbf q_v^0) + \frac{2\lambda_{\mathrm{BC}}^v}{M_{\mathrm{BC}}}N_{\mathrm{BC}}^\top(N_{\mathrm{BC}}\mathbf{p}_v-\mathbf q_v^b), \\
\nabla_{\mathbf{p}_p}\mathcal{L}_{\mathrm{NS}} &= \frac{2\lambda_{\mathrm{data}}^p}{M_{\mathrm{data}}}N_{\mathrm{data}}^\top(N_{\mathrm{data}}\mathbf{p}_p-\mathbf q_p^d) + \nabla_{\mathbf{p}_p}\mathcal{L}_{\mathrm{PDE}} \nonumber\\
&\quad +\frac{2\lambda_{\mathrm{gauge}}}{M_{\mathrm{gauge}}}N_{\mathrm{gauge}}^{\top}\left(N_{\mathrm{gauge}}\mathbf p_p-\mathbf q_p^{\mathrm{gauge}} \right).
\end{align}
\end{subequations}

The final term in \(\nabla_{\mathbf p_p}\mathcal{L}_{\mathrm{NS}}\) comes from the pressure-gauge penalty and removes the arbitrary pressure offset in the penalized reconstruction. When \(\lambda^p_{\mathrm{data}}=0\), this gauge condition is the only direct pressure-value constraint. The remaining pressure information enters through the momentum residuals.

\subsubsection{Recovery of physical consistency from low-fidelity data}\label{subsubsec:lid_recovery}

  \begin{figure}
	\centering
	\includegraphics[width=.95\textwidth]{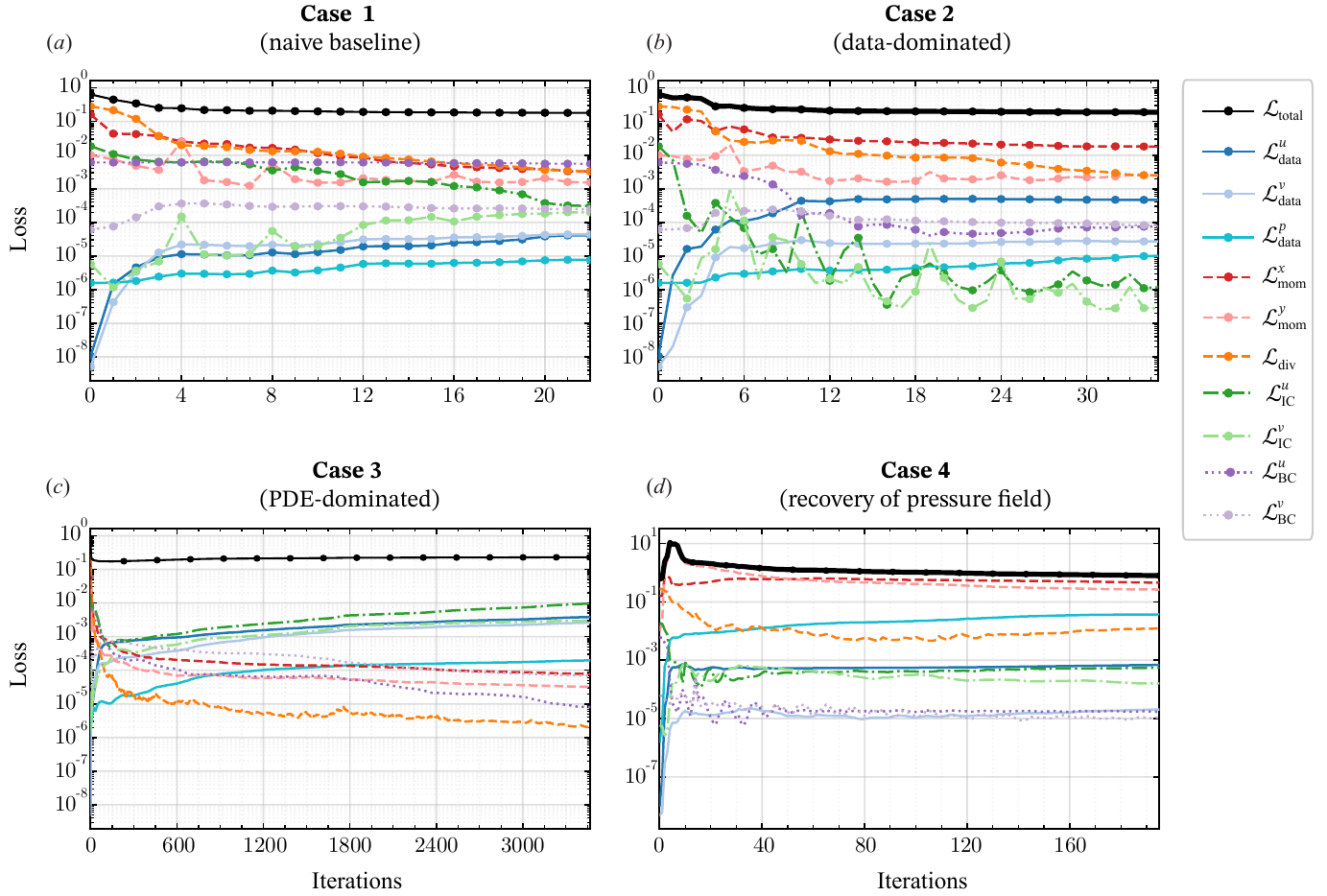}
  \caption{Convergence histories of the PI-MFA composite loss components for the 2D lid-driven cavity problem. All configurations are warm-started from the data-only MFA solution. Cases 1 and 2 rapidly stagnate as they overfit the coarse input data, leaving the physical residuals (dashed lines) elevated. Case 3 successfully drives the PDE residuals toward zero over a prolonged optimization by relaxing the data constraints. Case 4 demonstrates hidden state inference; the pressure field undergoes an initial transient reorganization to balance the momentum equations from velocity kinematics.}
  \label{Fig_lid_PIMFAresult_loss}
\end{figure}


  \begin{figure}
	\centering
	\includegraphics[width=.8\textwidth]{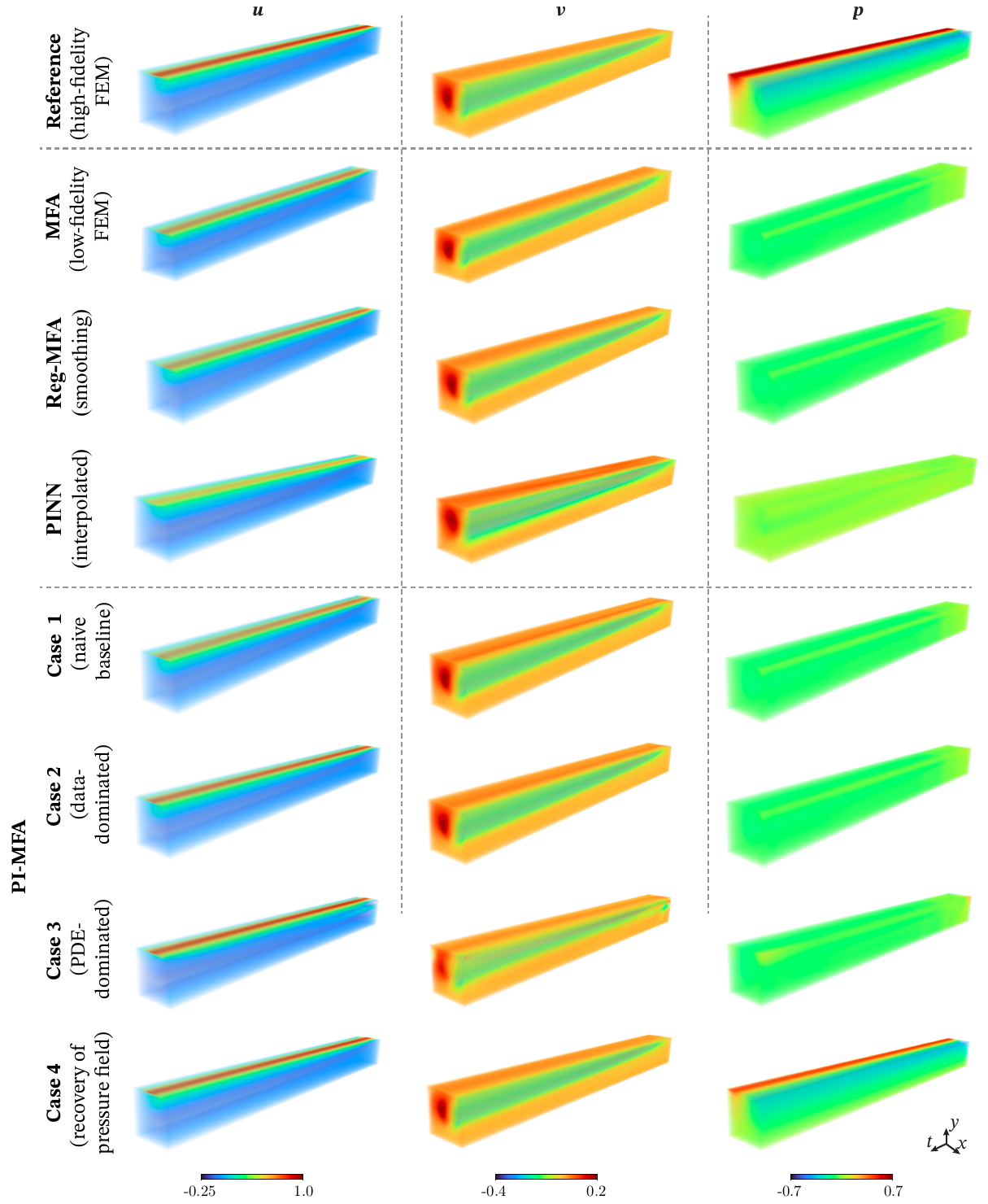}
  \caption{Continuous space-time volumetric rendering ($x, y, t$) of the $u$-velocity, $v$-velocity, and pressure ($p$) fields for the lid-driven cavity problem. The high-fidelity FEM serves as the benchmark reference. The data-driven MFA accurately interpolates the noisy low-fidelity data but inherits its corrupted, smeared pressure field. The Reg-MFA and PINN baselines introduce excessive mathematical blurring and high-frequency spatial artifacts, respectively.}
  \label{Fig_lid_PIMFAresult_uvp}
\end{figure}

  \begin{figure}
	\centering
	\includegraphics[width=.8\textwidth]{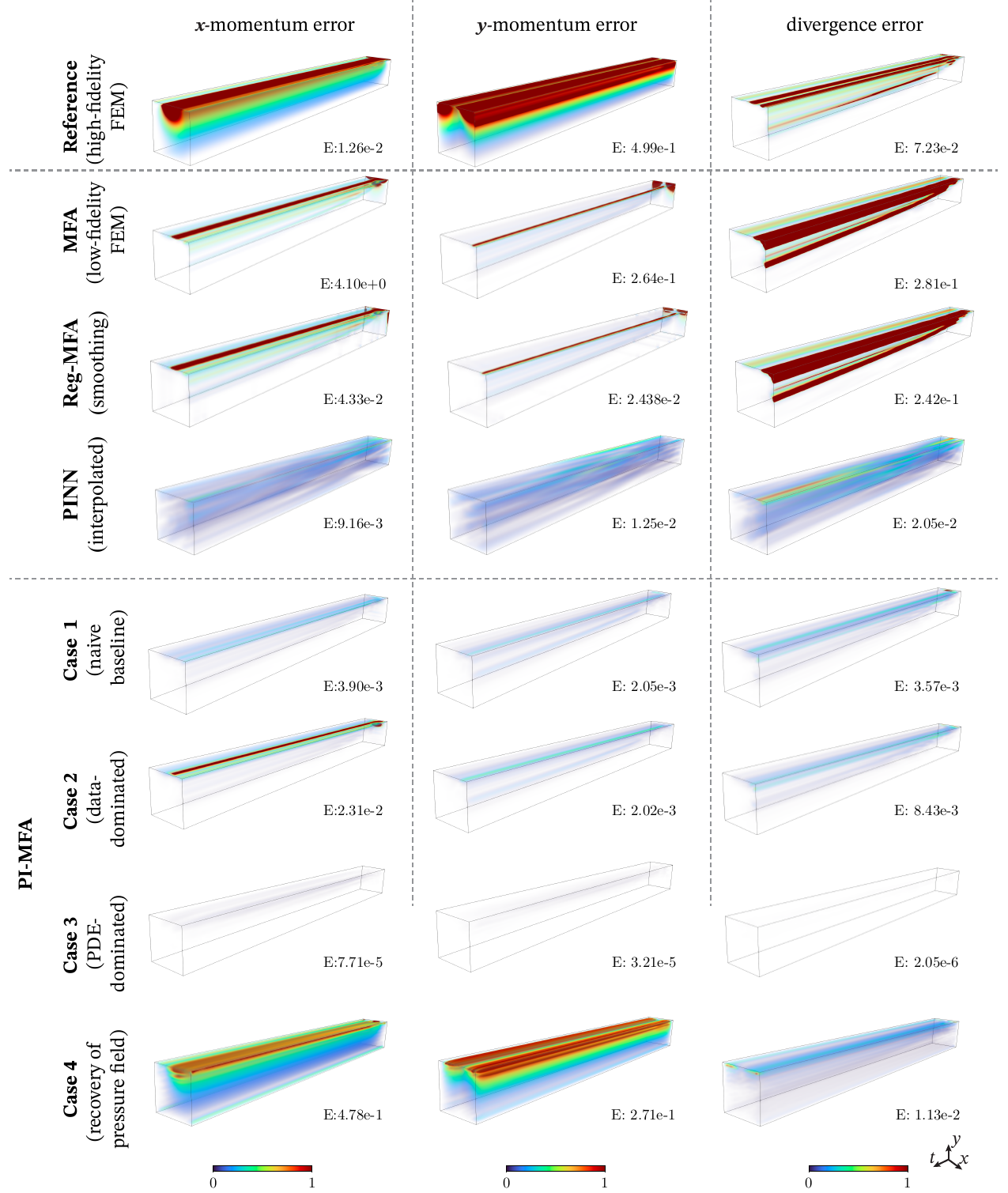}
  \caption{Volumetric rendering of the strong-form PDE residual errors ($x$-momentum, $y$-momentum, and divergence) over the space-time domain based on the metric in Eq.~\eqref{eq:MSE}.}
  \label{Fig_lid_PIMFAresult_physicsErr}
\end{figure}
In this scenario we investigate the capability of the proposed PI-MFA framework to restore physical consistency and reconstruct high-fidelity continuous fields from coarse, low-fidelity FEM simulation data. A fundamental challenge in physics-informed modeling using FEM data is that discrete continuous-Galerkin FEM solutions satisfy the governing equations in a discrete weak (Galerkin) sense rather than pointwise in a strong form \citep{hughes1987finite,ern2004theory}. Furthermore, low-fidelity simulations suffer from pronounced numerical diffusion and truncation errors, $\mathcal{O}(h^p)$. When standard, purely data-driven MFA is applied to this low-fidelity data, the continuous B-spline representation inadvertently approximates these discretization artifacts. Consequently, evaluating exact continuous spatial derivatives on the MFA representation yields significant, non-physical spikes in the strong-form momentum and divergence residuals. By penalizing these strong-form residuals during the control-point optimization, PI-MFA acts as a physics-based regularizer, smoothing out numerical noise and coercing the representation onto a physically admissible manifold.

To systematically evaluate the trade-off between fitting the flawed low-fidelity data and enforcing the continuous governing equations, we investigate four weighting strategies ($\lambda$) for the PI-MFA composite loss function, similar to the 2D Burgers problem in Section~\ref{subsec:2DBurgers}. First, Case 1 (naive baseline) sets all weights to unity ($\lambda_{*} = 1$), providing an equal penalty across data fidelity and physics constraints. Second, Case 2 (data-dominated) assigns high weights to the data fidelity terms ($\lambda_{\mathrm{data}} = 10^2$), while the PDE terms are weakly penalized ($\lambda_{\mathrm{mom}} = \lambda_{\mathrm{div}} = 1$). Conversely, Case 3 (PDE-dominated) heavily penalizes the optimization for violating the continuous physics ($\lambda_{\mathrm{mom}} = \lambda_{\mathrm{div}} = 10^3$) while relaxing the data fidelity constraints ($\lambda_{\mathrm{data}} = 10^{-2}$). This forces the B-spline representation to aggressively correct the underlying simulation data to satisfy the continuous fluid dynamics. Case 4 (recovery of pressure field) demonstrates the framework's capability for hidden state inference \citep{raissi2020hidden}. In this configuration, the data weight for the pressure field is completely removed ($\lambda_{\mathrm{data}}^p = 0$). Instead, the pressure gradients are inferred from the momentum equations while the pressure level is fixed by the reference condition.

\begin{table*}[!t]
  \centering
  \footnotesize
  \setlength{\tabcolsep}{5pt}
  \renewcommand{\arraystretch}{1.12}

  \caption{Comparison of MFA and PI-MFA loss values for the two-dimensional incompressible Navier-Stokes lid-driven cavity problem, alongside the Reg-MFA baseline and PINN. For each case, the data-misfit, PDE-residual, IC, and BC contributions are reported separately, followed by the total unweighted composite loss ($\mathcal{L}_{\mathrm{total}}$). The true pointwise errors ($\mathrm{MSE}$) computed against the high-fidelity reference data and their combined sum $\mathrm{MSE}_{\mathrm{total}}$ are provided at the bottom of the table to quantify the objective accuracy of the reconstructed fields.}
  \label{tab:loss_PIMFA_NSlid}
  \begin{threeparttable}
  \begin{tabular}{l ccccccc}
    \toprule
      & \multicolumn{1}{c}{\textbf{MFA}}
      & \multicolumn{1}{c}{\textbf{Reg-MFA}}
      & \multicolumn{1}{c}{\textbf{PINN}}
      & \multicolumn{4}{c}{\textbf{PI-MFA}} \\
    \cmidrule(lr){2-2}\cmidrule(lr){3-3}\cmidrule(lr){4-4}\cmidrule(lr){5-8}
      & \multicolumn{1}{c}{\begin{tabular}{@{}c@{}}(data only)\end{tabular}}
      & \multicolumn{1}{c}{\begin{tabular}{@{}c@{}}(smoothing)\end{tabular}}
      & \multicolumn{1}{c}{\begin{tabular}{@{}c@{}}(case 4)\end{tabular}}
      & \multicolumn{1}{c}{\begin{tabular}{@{}c@{}}Case 1 \\ \scriptsize(naive baseline)\end{tabular}}
      & \multicolumn{1}{c}{\begin{tabular}{@{}c@{}}Case 2 \\ \scriptsize(data-dominated)\end{tabular}}
      & \multicolumn{1}{c}{\begin{tabular}{@{}c@{}}Case 3 \\ \scriptsize(PDE-dominated)\end{tabular}}
      & \multicolumn{1}{c}{\begin{tabular}{@{}c@{}}Case 4 \\ \scriptsize(recovery of pressure field)\end{tabular}} \\
    \midrule

    Iteration
      & $0$ & $0$ & -- & $22$ & $35$ & $3460$ & $180$ \\
    \midrule
    $\lambda^{u}_{\mathrm{data}}$
      & $1$ & $1$ & $5\times 10^{2}$ & $1$ & $10^{2}$ & $10^{-2}$ & $5\times 10^{2}$ \\
    $\lambda^{v}_{\mathrm{data}}$
      & $1$ & $1$ & $10^{3}$ & $1$ & $10^{2}$ & $10^{-2}$ & $10^{3}$ \\
    $\lambda^{p}_{\mathrm{data}}$
      & $1$ & $1$ & $0$ & $1$ & $10^{2}$ & $0$ & $0$ \\
    \addlinespace[3pt]

    $\lambda^{x}_{\mathrm{mom}}$
      & $0$ & $0$ & $1$ & $1$ & $1$ & $10^{3}$ & $1$ \\
    $\lambda^{y}_{\mathrm{mom}}$
      & $0$ & $0$ & $1$ & $1$ & $1$ & $10^{3}$ & $1$ \\
    $\lambda_{\mathrm{div}}$
      & $0$ & $0$ & $1$ & $1$ & $1$ & $10^{3}$ & $1$ \\
    \addlinespace[3pt]
      
    $\lambda^{u}_{\mathrm{IC}}$
      & $0$ & $0$ & $1$ & $1$ & $10^{2}$ & $1$ & $1$ \\
    $\lambda^{v}_{\mathrm{IC}}$
      & $0$ & $0$ & $1$ & $1$ & $10^{2}$ & $1$ & $1$ \\
    \addlinespace[3pt]

    $\lambda^{u}_{\mathrm{BC}}$
      & $0$ & $0$ & $10^{3}$ & $1$ & $10^{2}$ & $10^{3}$ & $10^{3}$ \\
    $\lambda^{v}_{\mathrm{BC}}$
      & $0$ & $0$ & $10^{3}$ & $1$ & $10^{2}$ & $10^{3}$ & $10^{3}$ \\
    \midrule

    $\mathcal{L}^{u}_{\mathrm{data}}$
      & 1.800e-8 & 1.361e-8 & 4.913e-4 & 3.806e-5 & 4.984e-4 & 3.805e-3 & 6.635e-4 \\
    $\mathcal{L}^{v}_{\mathrm{data}}$
      & 5.167e-9 & 2.566e-8 & 2.546e-4 & 4.266e-5 & 2.326e-5 & 2.586e-3 & 1.872e-5 \\
    $\mathcal{L}^{p}_{\mathrm{data}}$
      & 2.415e-5 & 1.362e-6 & 4.012e-2 & 7.465e-6 & 4.612e-6 & 1.963e-4 & 3.696e-2 \\
    \addlinespace[3pt]

    $\mathcal{L}^{x}_{\mathrm{mom}}$
      & 4.103e+0 & 4.329e-2 & 9.161e-3 & 3.905e-3 & 2.313e-2 & 7.716e-5 & 4.789e-1 \\
    $\mathcal{L}^{y}_{\mathrm{mom}}$
      & 2.637e-1 & 2.425e-2 & 1.246e-2 & 2.056e-3 & 2.024e-3 & 3.208e-5 & 2.714e-1 \\
    $\mathcal{L}_{\mathrm{div}}$
      & 2.807e-1 & 2.419e-1 & 2.054e-2 & 3.574e-3 & 8.433e-3 & 2.050e-6 & 1.137e-2 \\
    \addlinespace[3pt]

    $\mathcal{L}^{u}_{\mathrm{IC}}$
      & 5.634e-6 & 6.314e-3 & 7.112e-3 & 3.758e-4 & 2.995e-6 & 9.561e-3 & 5.317e-4 \\
    $\mathcal{L}^{v}_{\mathrm{IC}}$
      & 9.839e-8 & 1.198e-5 & 1.884e-4 & 1.805e-4 & 2.252e-6 & 2.987e-3 & 1.743e-4 \\
    \addlinespace[3pt]

    $\mathcal{L}^{u}_{\mathrm{BC}}$
      & 1.974e-1 & 5.949e-3 & 8.163e-5 & 5.586e-3 & 5.142e-5 & 8.110e-6 & 1.772e-5 \\
    $\mathcal{L}^{v}_{\mathrm{BC}}$
      & 4.646e-5 & 6.096e-5 & 6.635e-5 & 2.514e-4 & 1.074e-4 & 6.865e-5 & 9.935e-6 \\
    \midrule
    
    \textbf{$\mathcal{L}_{\mathrm{total}}$}
      & \textbf{4.850e+0} & \textbf{3.218e-1} & \textbf{9.047e-2} & \textbf{1.602e-2} & \textbf{3.428e-2} & \textbf{1.932e-2} & \textbf{8.000e-1} \\
    \midrule

    $\mathrm{MSE}_{u}$
      & 6.596e-4 & 6.599e-4 & 2.656e-4 & 6.334e-4 & 1.155e-4 & 3.152e-3 & 2.282e-4 \\
    $\mathrm{MSE}_{v}$
      & 7.972e-5 & 7.956e-5 & 1.222e-4 & 1.011e-4 & 8.353e-5 & 2.506e-3 & 8.393e-5 \\
    $\mathrm{MSE}_{p}$
      & 5.438e-2 & 5.366e-2 & 4.778e-2 & 5.386e-2 & 5.393e-2 & 5.487e-2 & 3.811e-3 \\
    \midrule
    
    \textbf{$\mathrm{MSE}_{\mathrm{total}}$}
      & \textbf{5.511e-2} & \textbf{5.439e-2} & \textbf{4.827e-2} & \textbf{5.460e-2} & \textbf{5.413e-2} & \textbf{6.053e-2} & \textbf{4.123e-3} \\
    \bottomrule
  \end{tabular}

  \end{threeparttable}

\end{table*}

The optimization dynamics for these four cases are illustrated in Fig.~\ref{Fig_lid_PIMFAresult_loss}. For computational efficiency, all PI-MFA optimizations are warm-started using the control points from the data-only MFA solution. Given that the optimizer initializes near the data-fit minimum, the naive baseline (Case 1) and data-dominated model (Case 2) satisfy their optimization criteria rapidly, within 20 iterations. Because they remain heavily anchored to the flawed data, however, their physical PDE residuals (dashed lines) stagnate at elevated levels. In contrast, the PDE-dominated configuration (Case 3) exhibits a prolonged optimization trajectory (3,460 iterations). Here, the optimizer intentionally trades off initial data fidelity, evidenced by a slight early increase in the data loss terms, to systematically drive the momentum and divergence residuals down by several orders of magnitude. In Case 4, where the pressure field is decoupled from the data, the PI-MFA undergoes a massive initial transient reorganization to establish a valid pressure field that balances the velocity kinematics, eventually converging to a stable, physics-compliant state.

The qualitative impact of these weighting strategies, alongside the Reg-MFA and PINN baselines, is presented in Fig.~\ref{Fig_lid_PIMFAresult_uvp}, which displays the space-time volumetric rendering ($x, y, t$) of the $u, v,$ and $p$ fields. While the data-only MFA manages to capture the macroscale velocity features of the low-fidelity data, it propagates numerical diffusion into the pressure field, yielding a highly corrupted contour. The Reg-MFA baseline attempts to mitigate this via mathematical smoothing but ultimately fails to sharpen the pressure gradients, instead blurring the kinematic features. The PINN baseline incorporates physical constraints but struggles to resolve the sharp gradients near the sliding lid, introducing visible high-frequency streaking and spatial artifacts throughout the continuous domain. PI-MFA Cases 1 through 3 attempt to balance the low-fidelity data with the governing PDEs but ultimately struggle to reconstruct the true pressure dynamics. In contrast, Case 4 sets \(\lambda^p_{\mathrm{data}}=0\), so no pressure observations are used in the data-misfit term. The pressure field is instead
inferred from the velocity observations through the momentum and incompressibility residuals, together with the pointwise pressure gauge \(p_h(x_g,y_g,t_k)\). This case demonstrates pressure recovery from velocity observations and a pressure gauge, but it does not simultaneously minimize the Navier–Stokes momentum residuals.

The true advantage of the physics-informed spatial regularization is further isolated in Fig.~\ref{Fig_lid_PIMFAresult_physicsErr}, which visualizes the spatial distribution of the strong-form $x$-momentum, $y$-momentum, and divergence residual errors. We note that even the high-fidelity reference FEM solution exhibits non-zero strong-form errors near the moving lid (e.g., $x$-momentum error $E = 1.26\times 10^{-2}$). This is an expected artifact of evaluating exact continuous analytical derivatives on discretely integrated solutions with continuous finite elements near singular boundary conditions \citep{cottrell2009isogeometric}. When the low-fidelity FEM data is encoded by using the pure data-driven MFA, the lack of spatial resolution translates into massive structural violations of the divergence-free condition ($E=2.81 \times 10^{-1}$). The Reg-MFA baseline retains these massive boundary errors, while the PINN baseline, although reducing the error magnitudes, exhibits highly oscillatory residual bands that permeate the entire volume. In contrast, the PDE-dominated PI-MFA configuration (Case 3) acts as a mathematical smoother. The residuals transition from wide, noisy bands to near-zero continuous fields, proving that the B-spline control-point optimization has successfully coerced the low-fidelity dynamics onto a physically consistent, mass-conserving continuum (divergence error reduced to $E=2.05 \times 10^{-6}$). Since Case 4 is designed to recover the high-fidelity pressure field from the low-fidelity solution, the errors in the x- and y-momentum and in the divergence behave similarly to those of the high-fidelity solution, although they do not match perfectly.

To objectively quantify the reconstruction accuracy, we summarize in Table~\ref{tab:loss_PIMFA_NSlid} the final optimization loss components and the MSE computed against the highly resolved, grid-converged FEM reference simulation. As expected, the purely data-driven MFA achieves a low data-misfit loss against the coarse input but exhibits unacceptably large PDE residuals. The Reg-MFA baseline marginally worsens the total actual error ($\mathrm{MSE}_{\mathrm{total}} = 5.439 \times 10^{-2}$) compared with standard MFA ($\mathrm{MSE}_{\mathrm{total}} = 5.511 \times 10^{-2}$), confirming that the smoothing degrades physical fidelity. While the PINN baseline improves upon standard MFA, it remains an order of magnitude less accurate than PI-MFA (Case 4). PI-MFA in Cases 1, 2, and 3 also systematically reduces the physical errors. Notably, Case 4 achieves the lowest overall error relative to the ground truth, with ($\mathrm{MSE}_{\mathrm{total}} = 4.123 \times 10^{-3}$). This reduction is primarily driven by a more accurate pressure field reconstruction ($\mathrm{MSE}_p = 3.811 \times 10^{-3}$) despite the absence of pressure data during optimization, indicating that the framework effectively avoids the discretization errors of the low-fidelity baseline.

\subsection{Computational benefits and comparison with baselines}
\label{subsec:computational_benefits}

\begin{table*}[t]
\centering
\footnotesize
\setlength{\tabcolsep}{14pt}
\renewcommand{\arraystretch}{1.15}
\caption{Representation cost, compression, and fit/train time. Raw-data DOFs count all stored scalar samples across all reconstructed fields. Model DOFs count all spline control points or all trainable network parameters. MFA, Reg-MFA, and PI-MFA use the same spline control lattice.}
\label{tab:efficiency_summary}
\begin{threeparttable}
\begin{tabular}{@{} l r r r r @{}}
\toprule
\textbf{Method} &
\textbf{Model DOFs} &
\textbf{Compression ratio} &
\textbf{Storage reduction [\%]} &
\textbf{Fit/train time [s]} \\
\midrule
  \multicolumn{5}{@{}l}{\textbf{1D Convection-Diffusion} \quad \textit{(Raw data DOFs: 24,000)}} \\
  \addlinespace[2pt]
  MFA            & 1,600  & 15.00 & 93.33 & 2 \\
  Reg-MFA        & 1,600  & 15.00 & 93.33 & 3 \\
  PI-MFA         & 1,600  & 15.00 & 93.33 & 14 \\
  PINN & 16,897 & 1.42  & 29.60 & 45 \\
  \addlinespace[8pt]
   
  \multicolumn{5}{@{}l}{\textbf{2D Burgers} \quad \textit{(Raw data DOFs: 675,762)}} \\
  \addlinespace[2pt]
  MFA            & 54,000 & 12.51 & 92.01 & 30 \\
  Reg-MFA        & 54,000 & 12.51 & 92.01 & 35 \\
  PI-MFA         & 54,000 & 12.51 & 92.01 & 301 \\
  PINN & 83,330 & 8.11  & 87.67 & 502 \\
  \addlinespace[8pt]
  
  \multicolumn{5}{@{}l}{\textbf{2D Navier-Stokes} \quad \textit{(Raw data DOFs: 1,324,323)}} \\
  \addlinespace[2pt]
  MFA            & 60,000 & 22.07  & 95.47 & 31 \\
  Reg-MFA        & 60,000 & 22.07  & 95.47 & 32 \\
  PI-MFA         & 60,000 & 22.07  & 95.47 & 602 \\
  PINN & 83,459  & 15.87 & 93.70 & 2,520 \\
  \bottomrule
\end{tabular}

\begin{tablenotes}[flushleft]
  \item \textit{Note:} Burgers raw/model DOFs include both reconstructed fields $(u,v)$. Navier-Stokes raw/model DOFs include all three reconstructed fields $(u,v,p)$.
\end{tablenotes}
\end{threeparttable}
\end{table*}

We next evaluate the computational benefits of PI-MFA and compare it with the baseline reconstruction methods. Since PI-MFA is a post hoc reconstruction framework rather than a standalone forward PDE solver, its efficiency should be assessed in terms of representation cost, compression, and fitting/training time. This perspective is consistent with the objective of MFA to construct compact, continuous, and differentiable representations of already generated scientific data for downstream analysis and visualization.

For reproducibility, all comparisons reported in this section use the same setting and parameters established in the preceding numerical examples. The spline-based methods use the same tensor-product B-spline degree, knot construction, physical-to-parametric coordinate mapping, and control lattice within each benchmark, so that MFA, Reg-MFA, and PI-MFA differ only in their optimization objectives. The selected control-point grids are \(40 \times 40\) for the 1D convection--diffusion problem, \(30 \times 30 \times 30\) per velocity component for the 2D Burgers problem, and \(20 \times 20 \times 50\) per field for the 2D Navier--Stokes problem. In all PI-MFA runs, the data-only MFA solution is used as the initial guess for the L-BFGS optimization. The PDE, IC, BC, and gauge collocation sets are fixed during each optimization rather than resampled dynamically. The loss values reported in the numerical tables are unweighted diagnostic MSE components evaluated on the specified data or collocation sets, whereas the \(\lambda\)-values define the weighted objective minimized during training. This distinction is important because different weight choices can lead to different trade-offs between data fidelity and physical residual reduction.

The PINN baselines are used as post hoc reconstruction baselines rather than as parametric forward solvers. They are trained on the same observational datasets and use PDE, IC, and BC penalties analogous to those used in PI-MFA. The network architectures, number of trainable parameters, sampling counts, optimizer schedule, and learning-rate settings are summarized in Appendix~\ref{app:pinn_baseline_settings}. The timing results in Table~\ref{tab:efficiency_summary} should therefore be interpreted as implementation-level comparisons for the selected spline lattices, PINN architectures, loss weights, and optimization schedules, rather than as hardware-independent complexity bounds.

Assuming a predefined spline basis, the storage footprint of the spline reconstruction is governed primarily by the number of stored control-point values. As detailed in Table~\ref{tab:efficiency_summary}, MFA, Reg-MFA, and PI-MFA have identical representation costs for a fixed benchmark since they use the same spline control lattice. For the 1D convection--diffusion problem, PI-MFA stores \(1{,}600\) control-point values for \(24{,}000\) raw scalar samples, corresponding to a compression ratio of \(15.00\). For the 2D Burgers problem, it stores \(54{,}000\) control-point values for \(675{,}762\) raw scalar samples, corresponding to a compression ratio of \(12.51\). For the 2D Navier--Stokes problem, it stores \(60{,}000\) control-point values for \(1{,}324{,}323\) raw scalar samples, corresponding to a compression ratio of \(22.07\) and a \(95.47\%\) reduction relative to the raw data. These counts include all reconstructed physical fields: \((u,v)\) for Burgers and \((u,v,p)\) for Navier--Stokes. Small representation metadata, such as spline degree, knot vectors, field names, and coordinate-domain information, are not included in the degree-of-freedom count, but they are negligible compared with the raw scalar field storage for the tested datasets.

The tested PINN baselines also provide compact continuous representations, with compression ratios of \(1.42\), \(8.11\), and \(15.87\) for the three examples. For the selected architectures, PI-MFA uses fewer model degrees of freedom than the corresponding PINN baselines in all three cases, with the largest relative advantage in the 1D example. However, the more important distinction is not only the number of stored parameters. PI-MFA stores a locally supported spline representation, so derivative evaluations, localized updates, residual mapping, and balance-law diagnostics can be performed directly from analytic B-spline basis derivatives. The PINN representation is also differentiable, but it relies on automatic differentiation of a global neural parametrization.

All fitting and training times were measured on an Apple M4 Mac using CPU execution only. Embedding physical residuals into the optimization objective increases the computational cost of PI-MFA relative to standard MFA and Reg-MFA. This cost increase is expected because PI-MFA repeatedly evaluates PDE, IC, BC, and auxiliary residuals and their analytical gradients during L-BFGS optimization. Nevertheless, for the benchmark configurations considered here, PI-MFA remained faster than training the implemented PINN baselines. As reported in Table~\ref{tab:efficiency_summary}, the PI-MFA fit times for the 1D convection--diffusion, 2D Burgers, and 2D Navier--Stokes problems were \(14~\mathrm{s}\), \(301~\mathrm{s}\), and \(602~\mathrm{s}\), respectively. The corresponding PINN training times were \(45~\mathrm{s}\), \(502~\mathrm{s}\), and \(2{,}520~\mathrm{s}\). Thus, under the selected architectures, sampling strategies, and optimization schedules, PI-MFA occupies a pragmatic operational regime: it is more expensive than data-only spline fitting, but it provides stronger physics-based control over the reconstructed continuum at a lower fitting cost than the tested PINN baselines.

Ultimately, the primary computational advantage of PI-MFA is not that it universally minimizes parameter counts or guarantees exact satisfaction of the governing equations. Rather, its advantage is that it unifies compact data representation, local spline support, exact analytical differentiation, and penalty-based PDE/IC/BC residual reduction within a single post hoc reconstruction objective. This combination is tailored to downstream visualization and scientific data interrogation, where continuous field evaluation, residual mapping, balance-law diagnostics, and repeated derivative computations must be performed efficiently and reliably.

\section{Conclusion}\label{conclusion}

This work introduces a physics-informed extension of multivariate functional approximation (PI-MFA) for the post hoc reconstruction of spatiotemporal scientific fields. By embedding governing PDEs, along with boundary and initial conditions, directly into the optimization of tensor-product B-spline control points, the proposed framework preserves the compactness, local support, and exact analytical differentiability of standard MFA~\citep{peterka2018mfa} while constraining the reconstructed continuous fields to be physically consistent. Leveraging exact B-spline derivatives, the formulation enables the matrix-free assembly of strong-form residuals, block Jacobians, and objective gradients, allowing the nonlinear PDE-constrained optimization problem to be solved via L-BFGS.

Across three canonical fluid benchmark problems, numerical evaluations demonstrated that PI-MFA systematically outperforms purely data-driven baselines. In the 1D convection-diffusion system, PI-MFA reduced the PDE residuals and improved global flux-conservation behavior, illustrating the tunable trade-off between data-only approximation and physical consistency. For the 2D coupled Burgers equations, PI-MFA successfully rejected latent physical inconsistencies caused by unmodeled macroscopic forcing, achieving significantly lower actual errors and tighter global integral-balance conservation than both standard data-driven MFA and regularized MFA. In the 2D incompressible Navier-Stokes lid-driven cavity problem, PI-MFA acted as a powerful physics-based filter for coarse, numerically dissipative FEM data. By enforcing the divergence-free and momentum constraints, the framework not only filtered out numerical noise but also successfully inferred a pressure field from velocity observations, momentum constraints, incompressibility, and a pointwise pressure gauge, producing a pressure reconstruction with lower error than the low-fidelity pressure observations. Furthermore, relative to the tested PINN baselines, PI-MFA achieved favorable accuracy--cost trade-offs while retaining a compact, locally supported spline representation, with up to 95.47$\%$ storage reduction relative to the raw data in the Navier--Stokes example. Taken together, these results establish that PI-MFA is an effective and efficient physics-constrained continuous representation explicitly tailored for downstream scientific data interrogation, derivative evaluation, and residual-based diagnostics. 

Several directions for future work naturally follow from these observations. An immediate priority is the development of dynamic or adaptive weighting strategies to autonomously balance the competing data and physics loss terms during L-BFGS optimization. Additionally, investigating weak-form or mixed strong/weak physics constraints within the B-spline objective could prove highly beneficial, particularly for data generated by variational or finite-volume discretizations. A particularly compelling direction involves applying the proposed PI-MFA framework to data generated by emerging AI/ML-coupled simulations, such as hybrid physics-ML simulations accelerated via machine learning. As these data-driven accelerators can inadvertently introduce non-physical artifacts or numerical imbalances into the rapid-output data, PI-MFA is positioned to serve as a post hoc physics filter, restoring physical consistency to these computationally accelerated surrogates. Extending PI-MFA to accommodate highly sparse experimental observations, multiphysics systems, and explicit discrepancy-learning formulations will further solidify its role as a critical bridge between computational mechanics, surrogate modeling, and advanced scientific visualization.

\section*{Acknowledgments} This work was supported by the U.S. Department of Energy, Office of Science, Office of Advanced Scientific Computing Research (ASCR), through the Scientific Discovery through Advanced Computing (SciDAC) FASTMath Institute and the Competitive Portfolios Project on Energy Efficient Computing: A Holistic Methodology, under Contract No. DE-AC02-06CH11357. JJ also acknowledges support from an Argonne Leadership Computing Facility (ALCF) postdoctoral fellowship and the Laboratory Directed Research and Development (LDRD) program.

\section*{Declaration of competing interest} The authors declare that they have no known competing financial interests or personal relationships that could have appeared
to influence the work reported in this paper.



\section*{Data availability} 
Data will be made available on request.

\section*{Author ORCIDs}
\orcidlink{https://orcid.org} Junoh Jung \href{https://orcid.org/0000-0003-0962-3127}{https://orcid.org/0000-0003-0962-3127};\\
\orcidlink{https://orcid.org} David Lenz \href{https://orcid.org/0000-0002-2587-2783}{https://orcid.org/0000-0002-2587-2783};\\
\orcidlink{https://orcid.org} Emil Constantinescu \href{https://orcid.org/0000-0002-7003-6899}{https://orcid.org/0000-0002-7003-6899};\\
\orcidlink{https://orcid.org} Tom Peterka \href{https://orcid.org/0000-0002-0525-3205}{https://orcid.org/0000-0002-0525-3205}.

\begin{appendices}

\section{PINN baseline settings}
\label{app:pinn_baseline_settings}

This appendix summarizes the PINN baselines used in the numerical comparisons in this work. The PINNs are deployed as post hoc reconstruction baselines rather than as parametric forward solvers. For each benchmark, the network is trained on the exact same observation data used by MFA, Reg-MFA, and PI-MFA, while the governing equations and IC/BCs are imposed through the penalty terms. The pressure gauge used for the PINN baseline is the same pointwise gauge used
for PI-MFA at each sampled time level \(t_k\).

All PINNs are constructed as fully connected multilayer perceptrons with \(\tanh\) activation functions. Input coordinates are affinely mapped to \([-1,1]^d\), with \(d=2\) for the one-dimensional space--time problem and \(d=3\) for the two-dimensional space--time problems. Network weights are initialized with Xavier normal initialization using the \(\tanh\) gain, and biases are initialized to zero. Spatial and temporal derivatives appearing in the physics losses are computed by automatic differentiation. Both Adam and L-BFGS (history size 50, strong-Wolfe line search) optimizers are utilized sequentially for training.

A comprehensive summary of the architectures, sampling sets, and optimization schedules for the three benchmarks is shown in Table~\ref{tab:app_pinn_settings}. Under the selected architectures, sampling, and training schedules, PI-MFA gave better accuracy/cost trade-offs than the implemented PINN baselines. 
\begin{table}[htbp]
\centering
\small
\renewcommand{\arraystretch}{1.2}
\begin{threeparttable}
\caption{Hyperparameter settings for the physics-informed neural network (PINN) reconstruction baselines.}
\label{tab:app_pinn_settings}
\begin{tabular}{@{}lccc@{}}
\toprule
\textbf{Parameter} & \textbf{1D Conv--Diff} & \textbf{2D Burgers} & \textbf{2D Navier--Stokes} \\
\midrule
\multicolumn{4}{@{}l}{\textit{Network Architecture}} \\
Input $\mapsto$ Output map & \((x,t)\mapsto \rho_\theta\) & \((x,y,t)\mapsto (u_\theta,v_\theta)\) & \((x,y,t)\mapsto (u_\theta,v_\theta,p_\theta)\) \\
Hidden layers $\times$ width & $5 \times 64$ & $6 \times 128$ & $6 \times 128$ \\
Trainable parameters & 16,897 & 83,330 & 83,459 \\
\midrule
\multicolumn{4}{@{}l}{\textit{Sampling Points}} \\
Observation data ($N_d$) & 8,000 & 4,096 / 20,000 & 8,192 / 20,000 \\
PDE collocation ($N_r$)  & 12,000 & 8,192 / 40,000 & 16,384 / 40,000 \\
Boundary conditions ($N_{\mathrm{BC}}$) & All available & All available & 8,192 / 20,000 \\
\midrule
\multicolumn{4}{@{}l}{\textit{Optimization Schedule}} \\
Adam iterations & 500 & 8,000 & 4,000 \\
Adam learning rate & $10^{-3}$ & $10^{-3}$ & $10^{-3}$ \\
Step decay (rate $\gamma=0.5$) & Every 1,250 iters & Every 2,000 iters & Every 1,000 iters \\
L-BFGS max iterations & 400 & 100 & 100 \\
\bottomrule
\end{tabular}

\end{threeparttable}
\end{table}

\section{Additional experiments for the one-dimensional convection-diffusion example}
\label{app:1Dconv_differentICs}
  \begin{figure}

	\centering
	\includegraphics[width=.8\textwidth]{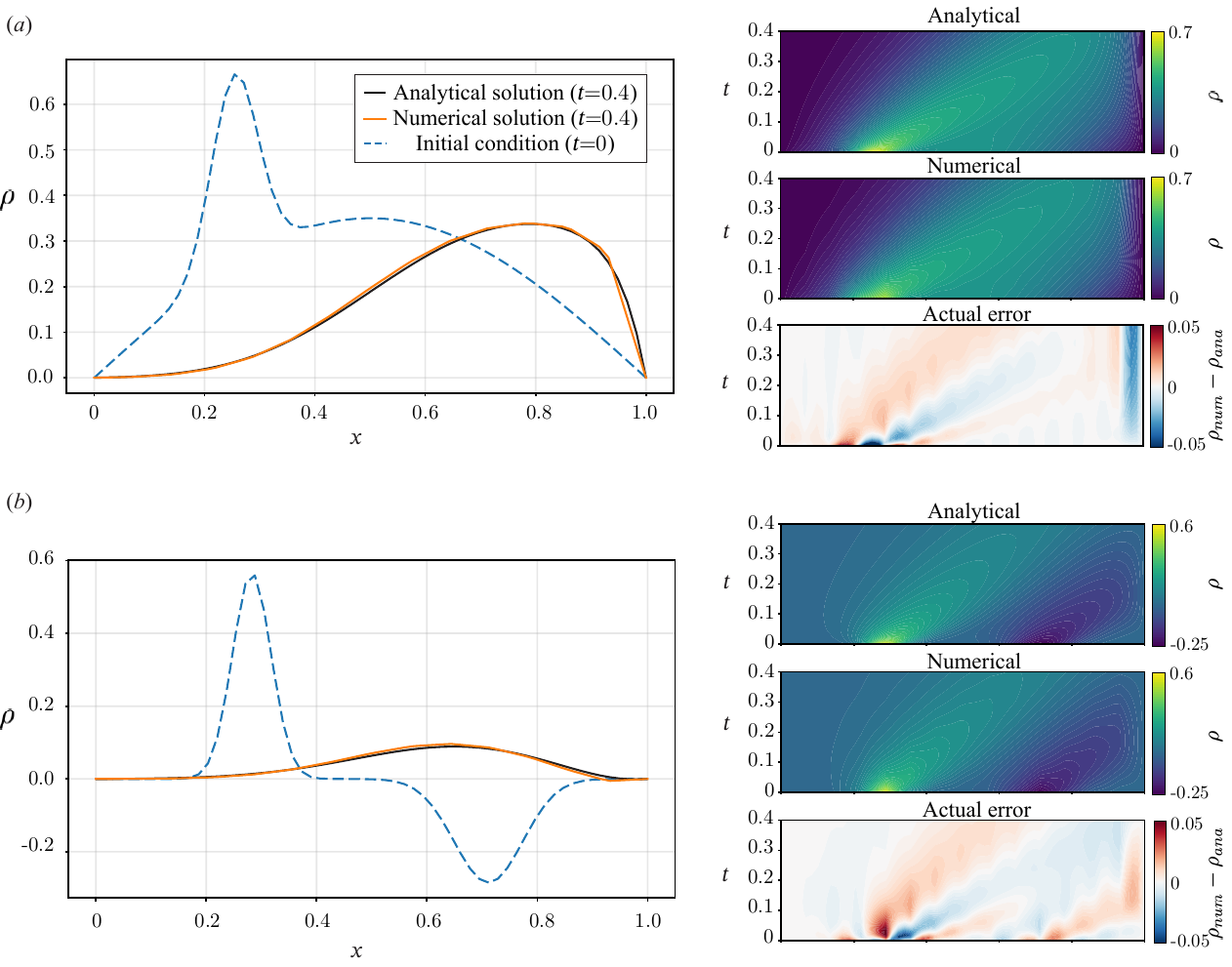}
    \caption{One-dimensional convection-diffusion benchmarks Case 2 and Case 3 with homogeneous Dirichlet boundary conditions.
  Left: initial condition at $t=0$ (blue dashed) and solutions at $T=0.4$ from a non-conservative finite difference simulation (colored) and the analytical reference (black).
  Right: spatiotemporal fields of the analytical solution, the numerical solution, and the pointwise error $e(x,t)=\rho_{\mathrm{num}}(x,t)-\rho_{\mathrm{ana}}(x,t)$.}
  \label{fig:1dconvdiff_case23}
\end{figure}

\begin{figure}

	\centering
	\includegraphics[width=0.8\textwidth]{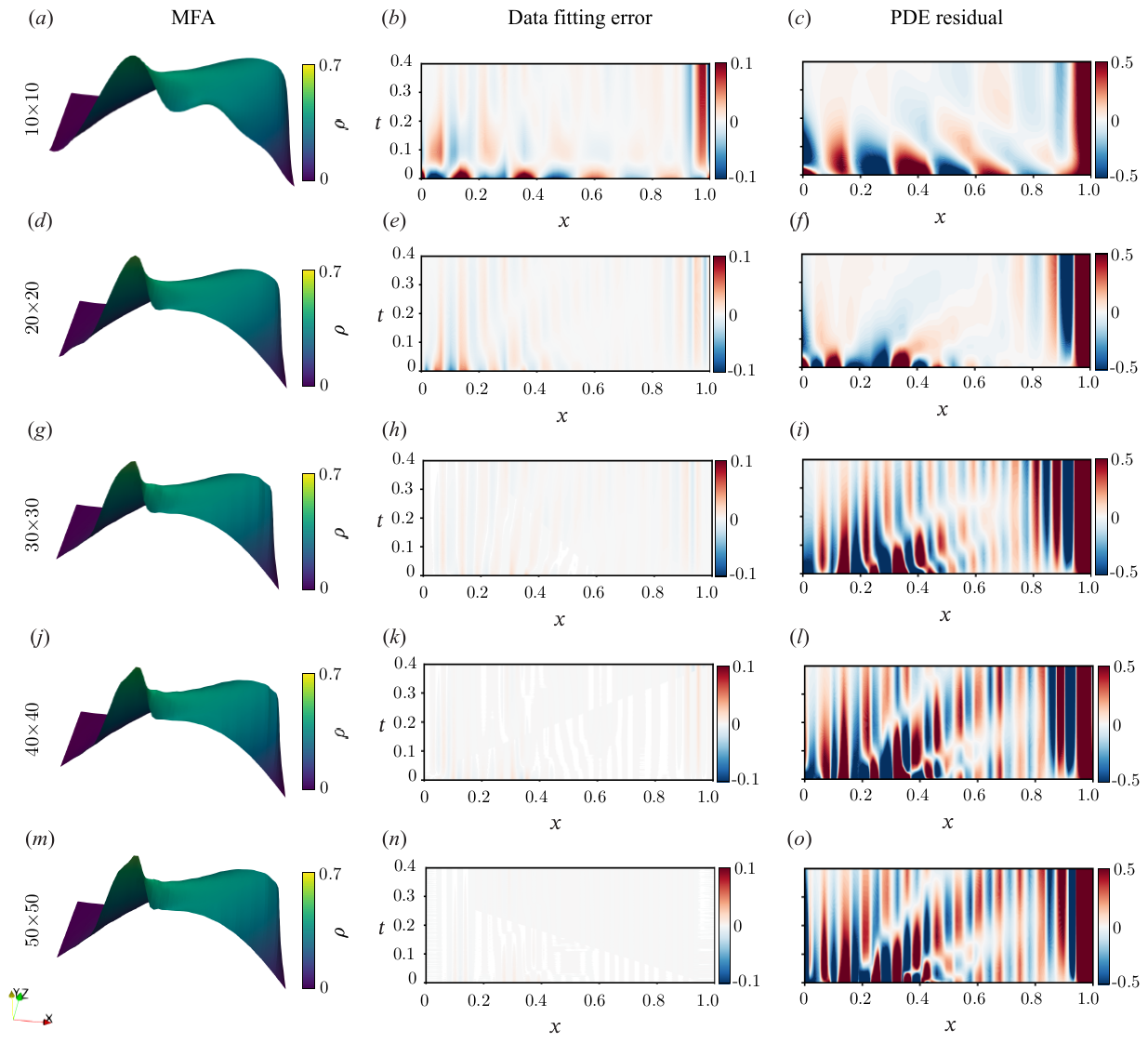}
    \caption{Results of standard MFA data fitting for Case 2, shown as a function of the number of control points.}\label{Fig_1DConvDiff_standardMFA_case2}
\end{figure}

\begin{figure}

	\centering
	\includegraphics[width=0.8\textwidth]{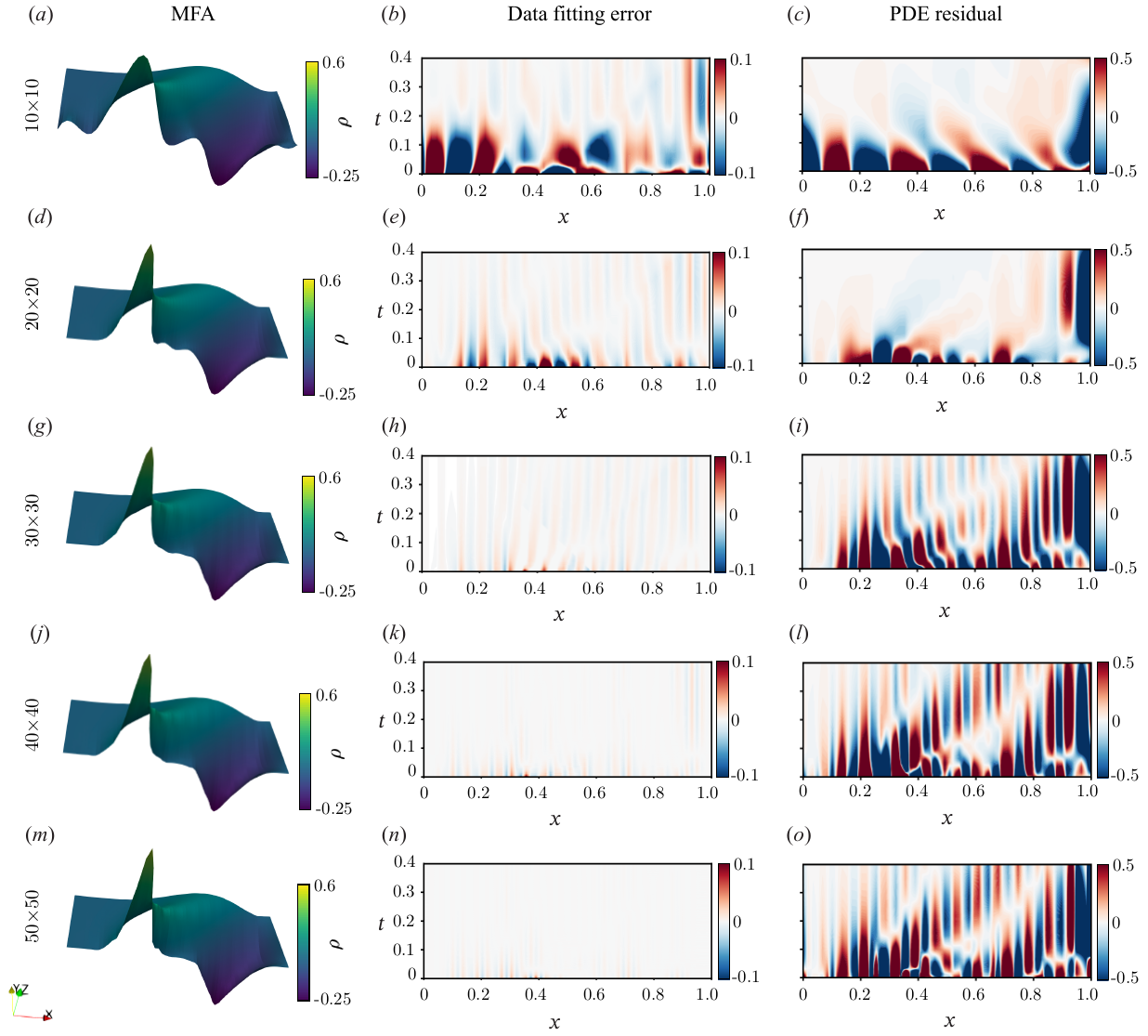}
        \caption{Results of standard MFA data fitting for Case 3, shown as a function of the number of control points.}\label{Fig_1DConvDiff_standardMFA_case3}
\end{figure}

To probe different solution regimes while maintaining compatibility with the homogeneous Dirichlet boundary conditions, we consider the following initial conditions in addition to the case in Eq.~\eqref{eq:initial-condition}.
All profiles are multiplied by $\phi(x)=4x(1-x)$ to enforce $\rho(0,0)=\rho(1,0)=0$
\begin{subequations}\label{eq:IC_T1T2}
\begin{align}
\text{(Case 2)}\;\;\rho(x,0) &=
A_{\sin}\sin(\pi x)
+ A_{\mathrm{b}}\;\phi(x)\exp\!\left(-\frac{(x-x_0)^2}{\sigma^2}\right),
\label{eq:IC_T1}\\
\text{(Case 3)}\;\;\rho(x,0) &=
\phi(x)\left(
A_1\exp\!\left(-\frac{(x-x_1)^2}{s_1^2}\right)
+ A_2\exp\!\left(-\frac{(x-x_2)^2}{s_2^2}\right)
\right).
\label{eq:IC_T2}
\end{align}
\end{subequations}
The parameter values used in the supplementary experiments are
\[
\begin{aligned}
\text{Case 2: } & A_{\sin}=0.35,\; A_{\mathrm{b}}=0.55,\; x_0=0.25,\; \sigma=0.06,\\
\text{Case 3: } & A_1=0.7,\; A_2=-0.35,\; x_1=0.28,\; x_2=0.72,\; s_1=0.05,\; s_2=0.09.
\end{aligned}
\]
Fig.~\ref{fig:1dconvdiff_case23} presents the analytical and numerical solutions for Cases 2 and 3. The left panels show the solutions at the initial condition and at the final time step, whereas the right panels display the corresponding spatiotemporal fields of the analytical and numerical solutions, together with the actual error. Fig.~\ref{Fig_1DConvDiff_standardMFA_case2} and~\ref{Fig_1DConvDiff_standardMFA_case3} present the control-point resolution study, based on which a $40\times40$ resolution is selected, consistent with the choice made for Case 1. Table~\ref{tab:losses_1d_case23} summarizes the final loss values for MFA, Reg-MFA, PINN, and PI-MFA, as well as the corresponding results for the other three weight choices. 
\begin{table}[ht]
\centering
\small
\setlength{\tabcolsep}{6pt}
\renewcommand{\arraystretch}{1.15}
\caption{Loss terms vs.\ control-point resolution (1D convection-diffusion, data-only MFA) for two additional initial condition cases.}
\begin{tabular}{l S S S S S}
\toprule
& \multicolumn{5}{c}{\textbf{Control points per dimension} $N_x\times N_t$} \\
\cmidrule(lr){2-6}
\textbf{Loss term}
& \multicolumn{1}{c}{$10\times10$}
& \multicolumn{1}{c}{$20\times20$}
& \multicolumn{1}{c}{$30\times30$}
& \multicolumn{1}{c}{$40\times40$}
& \multicolumn{1}{c}{$50\times50$} \\
\midrule

\addlinespace[2pt]
\multicolumn{6}{l}{\textbf{Case 2}} \\
$\mathcal{L}_{\text{data}}$ & \num{3.63e-5}  & \num{3.93e-6}  & \num{1.35e-6}  & \num{7.77e-7}  & \num{2.06e-7} \\
$\mathcal{L}_{\text{pde}}$  & \num{1.55e-1}  & \num{1.04e-0}  & \num{7.66e-1}  & \num{8.42e-1}  & \num{7.20e-0} \\
$\mathcal{L}_{\text{IC}}$   & \num{8.35e-4}  & \num{3.54e-5}  & \num{9.08e-6}  & \num{5.99e-6}  & \num{2.76e-6} \\
$\mathcal{L}_{\text{BC}}$   & \num{4.46e-5}  & \num{4.90e-7}  & \num{2.20e-10} & \num{5.15e-11} & \num{3.91e-20} \\
\addlinespace[2pt]
\textbf{$\mathcal{L}_{\text{total}}$} & \num{1.56e-1} & \num{1.04} & \num{7.66e-1} & \num{8.42e-1} & \num{7.20e-0} \\
\midrule
\addlinespace[6pt]
\multicolumn{6}{l}{\textbf{Case 3}} \\
$\mathcal{L}_{\text{data}}$ & \num{8.58e-5}  & \num{7.78e-6}  & \num{2.44e-6}  & \num{1.54e-6}  & \num{7.64e-7} \\
$\mathcal{L}_{\text{pde}}$  & \num{3.46e-1}  & \num{3.35e-1}  & \num{5.70e-1}  & \num{7.68e-1}  & \num{2.16e-0} \\
$\mathcal{L}_{\text{IC}}$   & \num{2.02e-3}  & \num{1.11e-4}  & \num{4.36e-5}  & \num{3.72e-5}  & \num{1.76e-5} \\
$\mathcal{L}_{\text{BC}}$   & \num{5.14e-5}  & \num{6.29e-8}  & \num{5.59e-11} & \num{9.43e-12} & \num{2.57e-20} \\
\addlinespace[2pt]
\textbf{$\mathcal{L}_{\text{total}}$} & \num{3.48e-1} & \num{3.35e-1} & \num{5.70e-1} & \num{7.68e-1} & \num{2.16e-0} \\

\bottomrule
\end{tabular}

\label{tab:1DConvDiff_loss_terms_T1T3}
\end{table}
\begin{table}[t]
  \centering
  \footnotesize
  \setlength{\tabcolsep}{4.5pt}
  \renewcommand{\arraystretch}{1.15}

  \sisetup{
    input-exponent-markers = eE,
    output-exponent-marker = \mathrm{e},
    group-digits = false,
    detect-weight = true,
    detect-family = true
  }

  \caption{Final loss components for the data-only MFA baseline,
  regularized MFA (Reg-MFA), physics-informed neural network (PINN),
  and representative PI-MFA runs for two one-dimensional
  convection--diffusion cases. The PI-MFA configurations correspond to
  the minimum PDE loss \((\min \mathcal{L}_{\mathrm{pde}})\), a balanced
  Pareto-optimal selection, and the minimum total physics loss
  \((\min \mathcal{L}_{\mathrm{physics}})\), where
  \(\mathcal{L}_{\mathrm{physics}}
  =
  \mathcal{L}_{\mathrm{pde}}
  +
  \mathcal{L}_{\mathrm{IC}}
  +
  \mathcal{L}_{\mathrm{BC}}\).
  For each case, the PINN baseline is trained using the corresponding
  \(\min \mathcal{L}_{\mathrm{physics}}\) weights. Reported losses are
  evaluated at the final optimizer iterate.}
  \label{tab:losses_1d_case23}

  \begin{threeparttable}
  \begin{tabular}{lcccccc}
    \toprule
      & \multicolumn{1}{c}{\textbf{MFA}}
      & \multicolumn{1}{c}{\textbf{Reg-MFA}}
      & \multicolumn{1}{c}{\textbf{PINN}}
      & \multicolumn{3}{c}{\textbf{PI-MFA}} \\
    \cmidrule(lr){2-2}\cmidrule(lr){3-3}\cmidrule(lr){4-4}\cmidrule(lr){5-7}
      & \multicolumn{1}{c}{(data only)}
      & \multicolumn{1}{c}{(smoother)}
      & \multicolumn{1}{c}{(physics-informed)}
      & \multicolumn{1}{c}{min $\mathcal{L}_{\mathrm{pde}}$}
      & \multicolumn{1}{c}{Pareto-optimal}
      & \multicolumn{1}{c}{min $\mathcal{L}_{\mathrm{physics}}$} \\
    \midrule

    \multicolumn{7}{l}{\textbf{Case 2}} \\
    \midrule

    Iterations
      & -- & -- & 808 & 4122 & 1446 & 1713 \\

    \midrule

    $\mathcal{L}_{\mathrm{data}}$
      & {7.77e-7} & 7.12e-7 & {1.88e-3} & {2.28e-2} & {7.59e-5} & {6.21e-5} \\

    $\mathcal{L}_{\mathrm{pde}}$
      & {8.43e-1} & 8.35e-1 & {2.87e-3} & {2.13e-8} & {8.65e-5} & {2.06e-4} \\

    $\mathcal{L}_{\mathrm{IC}}$
      & {7.81e-4} & 1.52e-4 & {3.98e-3} & {4.13e-2} & {1.01e-3} & {7.12e-5} \\

    $\mathcal{L}_{\mathrm{BC}}$
      & {5.15e-11} & 1.85e-11 & {5.92e-3} & {5.60e-4} & {4.17e-4} & {4.63e-6} \\

    \midrule

    $\mathcal{L}_{\mathrm{total}}$
      & {8.43e-1} & 8.35e-1 & {1.47e-2} & {6.46e-2} & {1.59e-3} & {3.44e-4} \\

    \midrule
    \addlinespace[2pt]

    \multicolumn{7}{l}{\textbf{Case 3}} \\
    \midrule

    Iterations
      & -- & -- & 812 & 4122 & 1485 & 1650 \\

    \midrule

    $\mathcal{L}_{\mathrm{data}}$
      & {1.51e-6} & 1.34e-6 & {5.44e-4} & {3.10e-3} & {7.08e-5} & {6.23e-5} \\

    $\mathcal{L}_{\mathrm{pde}}$
      & {7.28e-1} & 7.22e-1 & {9.22e-4} & {6.84e-9} & {3.23e-5} & {1.83e-4} \\

    $\mathcal{L}_{\mathrm{IC}}$
      & {3.43e-4} & 3.27e-4 & {6.08e-3} & {1.81e-2} & {1.88e-3} & {1.10e-4} \\

    $\mathcal{L}_{\mathrm{BC}}$
      & {9.43e-12} & 8.98e-11 & {8.28e-4} & {8.03e-5} & {7.76e-5} & {4.98e-7} \\

    \midrule

    $\mathcal{L}_{\mathrm{total}}$
      & {7.28e-1} & 7.22e-1 & {8.38e-3} & {2.13e-2} & {2.06e-3} & {3.56e-4} \\

    \bottomrule
  \end{tabular}
  \end{threeparttable}
\end{table}

\section{Detailed derivation of the componentwise integral balance for the 2D Burgers equation}
\label{app:componentwise_mass_balance}

Let $\Omega\subset\mathbb{R}^2$ be a bounded domain with boundary $\partial\Omega$ and outward unit normal $\boldsymbol{n}$. The two-dimensional Burgers system can be written in vector form as
\begin{equation}
\partial_t \boldsymbol{U} + (\boldsymbol{U}\cdot\nabla)\boldsymbol{U} - \nu \Delta \boldsymbol{U} = \boldsymbol{0}
\qquad \text{in }\Omega\times(0,T],
\label{eq:app_burgers_vector}
\end{equation}
with
\begin{equation}
\boldsymbol{U}=(u,v)^{\mathsf T},
\qquad
\boldsymbol{U}=\boldsymbol{0}
\qquad \text{on }\partial\Omega\times(0,T].
\label{eq:app_burgers_bc}
\end{equation}
For either component $q\in\{u,v\}$, Eq.~\eqref{eq:app_burgers_vector} yields the scalar relation
\begin{equation}
\partial_t q + \boldsymbol{U}\cdot\nabla q - \nu \Delta q = 0
\qquad \text{in }\Omega\times(0,T].
\label{eq:app_burgers_component}
\end{equation}

We define the componentwise integral quantity
\begin{equation}
M_q(t)=\int_{\Omega} q(\boldsymbol{x},t)\,\mathrm{d}\Omega,
\qquad q\in\{u,v\}.
\label{eq:app_component_mass}
\end{equation}
Assuming sufficient regularity of $q$, integration of Eq.~\eqref{eq:app_burgers_component} over $\Omega$ and differentiation under the integral sign give
\begin{align}
0
&=
\int_{\Omega}
\left(
\partial_t q + \boldsymbol{U}\cdot\nabla q - \nu \Delta q
\right)\,\mathrm{d}\Omega
\nonumber\\
&=
\frac{\mathrm{d}}{\mathrm{d}t}\int_{\Omega} q\,\mathrm{d}\Omega
+
\int_{\Omega}\boldsymbol{U}\cdot\nabla q\,\mathrm{d}\Omega
-
\nu\int_{\Omega}\Delta q\,\mathrm{d}\Omega.
\label{eq:app_mass_step1}
\end{align}

Next, we use the product-rule identity
\begin{equation}
\boldsymbol{U}\cdot\nabla q
=
\nabla\cdot(q\boldsymbol{U}) - q\,(\nabla\cdot\boldsymbol{U}),
\qquad
\nabla\cdot\boldsymbol{U} = \partial_x u + \partial_y v.
\label{eq:app_transport_identity}
\end{equation}
Substituting Eq.~\eqref{eq:app_transport_identity} into Eq.~\eqref{eq:app_mass_step1} yields
\begin{align}
0
&=
\frac{\mathrm{d}}{\mathrm{d}t}\int_{\Omega} q\,\mathrm{d}\Omega
+
\int_{\Omega}\nabla\cdot(q\boldsymbol{U})\,\mathrm{d}\Omega
-
\int_{\Omega} q\,(\nabla\cdot\boldsymbol{U})\,\mathrm{d}\Omega
-
\nu\int_{\Omega}\Delta q\,\mathrm{d}\Omega.
\label{eq:app_mass_step2}
\end{align}

Applying the divergence theorem and Green's identity \citep{Evans2010,Borthwick2017}
\begin{equation}
\int_{\Omega}\nabla\cdot(q\boldsymbol{U})\,\mathrm{d}\Omega
=
\int_{\partial\Omega} q\,(\boldsymbol{U}\cdot\boldsymbol{n})\,\mathrm{d}s,
\qquad
\int_{\Omega}\Delta q\,\mathrm{d}\Omega
=
\int_{\partial\Omega}\nabla q\cdot\boldsymbol{n}\,\mathrm{d}s,
\label{eq:app_div_green}
\end{equation}
we obtain
\begin{equation}
\frac{\mathrm{d}}{\mathrm{d}t}\int_{\Omega} q\,\mathrm{d}\Omega
+
\int_{\partial\Omega} q\,(\boldsymbol{U}\cdot\boldsymbol{n})\,\mathrm{d}s
-
\int_{\Omega} q\,(\nabla\cdot\boldsymbol{U})\,\mathrm{d}\Omega
-
\nu\int_{\partial\Omega}\nabla q\cdot\boldsymbol{n}\,\mathrm{d}s
=0,
\qquad q\in\{u,v\}.
\label{eq:app_mass_balance_general}
\end{equation}
Under the homogeneous Dirichlet boundary condition, one has $q=0$ on $\partial\Omega$, and therefore
\begin{equation}
\int_{\partial\Omega} q\,(\boldsymbol{U}\cdot\boldsymbol{n})\,\mathrm{d}s = 0.
\label{eq:app_advective_flux_zero}
\end{equation}
Hence, Eq.~\eqref{eq:app_mass_balance_general} reduces to
\begin{equation}
\frac{\mathrm{d}}{\mathrm{d}t}\int_{\Omega} q\,\mathrm{d}\Omega
=
\int_{\Omega} q\,(\nabla\cdot\boldsymbol{U})\,\mathrm{d}\Omega
+
\nu\int_{\partial\Omega}\nabla q\cdot\boldsymbol{n}\,\mathrm{d}s,
\qquad q\in\{u,v\},
\label{eq:app_mass_balance_dirichlet}
\end{equation}
which corresponds to Eq.~\eqref{eq:burgers_mass_residual}. This identity shows that the componentwise integrals are generally not conserved, and their evolution is governed by the volumetric contribution involving $\nabla\cdot\boldsymbol{U}$ and the diffusive boundary flux.

Accordingly, the componentwise integral-balance residual is defined by
\begin{equation}
r_{q,M}(t)
=
\frac{\mathrm{d}}{\mathrm{d}t}\int_{\Omega} q\,\mathrm{d}\Omega
-
\int_{\Omega} q\,(\nabla\cdot\boldsymbol{U})\,\mathrm{d}\Omega
-
\nu\int_{\partial\Omega}\nabla q\cdot\boldsymbol{n}\,\mathrm{d}s,
\qquad q\in\{u,v\}.
\label{eq:app_mass_residual}
\end{equation}
For the exact PDE solution, one has
\begin{equation}
r_{q,M}(t)\equiv 0,
\qquad q\in\{u,v\}.
\label{eq:app_mass_residual_zero}
\end{equation}

In the numerical experiments, $r_{u,M}(t)$ and $r_{v,M}(t)$ are evaluated from the continuous PIMFA approximation by using the B-spline representation to compute $\mathrm{d}M_q/\mathrm{d}t$ together with the associated volume and boundary integrals. Their temporal evolution is then reported as a diagnostic of physics consistency.

\end{appendices}


\bibliographystyle{elsarticle-num}
\bibliography{arXiv_template}

@book{piegl1997nurbs,
  author    = {Piegl, Les and Tiller, Wayne},
  title     = {The {NURBS} Book},
  edition   = {2},
  publisher = {Springer},
  address   = {Berlin, Heidelberg},
  year      = {1997},
  doi       = {10.1007/978-3-642-59223-2}
}

@book{deboor2001spline,
  author    = {de Boor, Carl},
  title     = {A Practical Guide to Splines},
  series    = {Applied Mathematical Sciences},
  volume    = {27},
  edition   = {Revised},
  publisher = {Springer},
  address   = {New York},
  year      = {2001},
  doi       = {10.1007/978-1-4612-6333-3}
}

@book{schumaker2007spline,
  author    = {Schumaker, Larry L.},
  title     = {Spline Functions: Basic Theory},
  series    = {Cambridge Mathematical Library},
  edition   = {3},
  publisher = {Cambridge University Press},
  address   = {Cambridge},
  year      = {2007},
  doi       = {10.1017/CBO9780511618994}
}

@article{Eilers1996,
  author  = {Eilers, Paul H. C. and Marx, Brian D.},
  title   = {Flexible smoothing with {B}-splines and penalties},
  journal = {Statistical Science},
  volume  = {11},
  number  = {2},
  pages   = {89--121},
  year    = {1996},
  doi     = {10.1214/ss/1038425655}
}

@article{johnson2009scattered,
  author  = {Johnson, Michael J. and Shen, Zuowei and Xu, Yuhong},
  title   = {Scattered data reconstruction by regularization in {B}-spline and associated wavelet spaces},
  journal = {Journal of Approximation Theory},
  volume  = {159},
  number  = {2},
  pages   = {197--223},
  year    = {2009},
  doi     = {10.1016/j.jat.2009.02.005}
}

@article{merchel2022fast,
  author  = {Merchel, Stefan and J{\"u}ttler, Bert and Mokri{\v{s}}, David and Pan, Mingjun},
  title   = {Fast formation of matrices for least-squares fitting by tensor-product spline surfaces},
  journal = {Computer-Aided Design},
  volume  = {150},
  pages   = {103307},
  year    = {2022},
  doi     = {10.1016/j.cad.2022.103307}
}

@article{hughes2005isogeometric,
  author  = {Hughes, Thomas J. R. and Cottrell, John A. and Bazilevs, Yuri},
  title   = {Isogeometric analysis: {CAD}, finite elements, {NURBS}, exact geometry and mesh refinement},
  journal = {Computer Methods in Applied Mechanics and Engineering},
  volume  = {194},
  number  = {39--41},
  pages   = {4135--4195},
  year    = {2005},
  doi     = {10.1016/j.cma.2004.10.008}
}

@article{bazilevs2006isogeometric,
  author  = {Bazilevs, Yuri and Beir{\~a}o da Veiga, Louren{\c{c}}o and Cottrell, John A. and Hughes, Thomas J. R. and Sangalli, Giancarlo},
  title   = {Isogeometric analysis: approximation, stability and error estimates for {$h$}-refined meshes},
  journal = {Mathematical Models and Methods in Applied Sciences},
  volume  = {16},
  number  = {7},
  pages   = {1031--1090},
  year    = {2006},
  doi     = {10.1142/S0218202506001455}
}

@article{cottrell2007studies,
  author  = {Cottrell, J. Austin and Hughes, Thomas J. R. and Reali, Alessandro},
  title   = {Studies of refinement and continuity in isogeometric structural analysis},
  journal = {Computer Methods in Applied Mechanics and Engineering},
  volume  = {196},
  number  = {41--44},
  pages   = {4160--4183},
  year    = {2007},
  doi     = {10.1016/j.cma.2007.04.007}
}

@article{auricchio2010isogeometric,
  author  = {Auricchio, Ferdinando and Beir{\~a}o da Veiga, Louren{\c{c}}o and Hughes, Thomas J. R. and Reali, Alessandro and Sangalli, Giancarlo},
  title   = {Isogeometric collocation methods},
  journal = {Mathematical Models and Methods in Applied Sciences},
  volume  = {20},
  number  = {11},
  pages   = {2075--2107},
  year    = {2010},
  doi     = {10.1142/S0218202510004878}
}

@article{auricchio2012isogeometric,
  author  = {Auricchio, Ferdinando and Beir{\~a}o da Veiga, Louren{\c{c}}o and Hughes, Thomas J. R. and Reali, Alessandro and Sangalli, Giancarlo},
  title   = {Isogeometric collocation for elastostatics and explicit dynamics},
  journal = {Computer Methods in Applied Mechanics and Engineering},
  volume  = {249--252},
  pages   = {2--14},
  year    = {2012},
  doi     = {10.1016/j.cma.2012.03.026}
}

@article{schillinger2013isogeometric,
  author  = {Schillinger, Dominik and Evans, John A. and Reali, Alessandro and Scott, Michael A. and Hughes, Thomas J. R.},
  title   = {Isogeometric collocation: Cost comparison with {Galerkin} methods and extension to adaptive hierarchical {NURBS} discretizations},
  journal = {Computer Methods in Applied Mechanics and Engineering},
  volume  = {267},
  pages   = {170--232},
  year    = {2013},
  doi     = {10.1016/j.cma.2013.07.017}
}

@article{montardini2017optimal,
  author  = {Montardini, Matteo and Sangalli, Giancarlo and Tamellini, Lorenzo},
  title   = {Optimal-order isogeometric collocation at {Galerkin} superconvergent points},
  journal = {Computer Methods in Applied Mechanics and Engineering},
  volume  = {316},
  pages   = {741--757},
  year    = {2017},
  doi     = {10.1016/j.cma.2016.09.043}
}

@inproceedings{peterka2018mfa,
  author    = {Peterka, Tom and Nashed, Youssef S. G. and Grindeanu, Iulian R. and Mahadevan, Vijay S. and Yeh, Raine and Tricoche, Xavier},
  title     = {Foundations of Multivariate Functional Approximation for Scientific Data},
  booktitle = {2018 IEEE 8th Symposium on Large Data Analysis and Visualization (LDAV)},
  pages     = {61--71},
  year      = {2018},
  doi       = {10.1109/LDAV.2018.8739195}
}

@article{lenz_adaptive_2023,
  author  = {Lenz, David and Yeh, Raine and Mahadevan, Vijay and Grindeanu, Iulian and Peterka, Tom},
  title   = {Customizable adaptive regularization techniques for {B}-spline modeling},
  journal = {Journal of Computational Science},
  volume  = {71},
  pages   = {102037},
  year    = {2023},
  doi     = {10.1016/j.jocs.2023.102037}
}

@article{sun_scalable_2023,
  author  = {Sun, Jianxin and Lenz, David and Yu, Hongfeng and Peterka, Tom},
  title   = {Scalable volume visualization for big scientific data modeled by functional approximation},
  journal = {IEEE Transactions on Visualization and Computer Graphics},
  volume  = {30},
  number  = {12},
  pages   = {8637--8651},
  year    = {2024},
  doi     = {10.1109/TVCG.2024.3353594}
}

@article{sun_mfa-dvr_2024,
  author  = {Sun, Jianxin and Lenz, David and Yu, Hongfeng and Peterka, Tom},
  title   = {{MFA-DVR}: Direct volume rendering of {MFA} models},
  journal = {Journal of Visualization},
  volume  = {27},
  pages   = {109--126},
  year    = {2024},
  doi     = {10.1007/s12650-023-00946-y}
}

@inproceedings{peterka2023adaptive,
  author    = {Peterka, Tom and Lenz, David and Grindeanu, Iulian and Mahadevan, Vijay S.},
  title     = {Towards Adaptive Refinement for Multivariate Functional Approximation of Scientific Data},
  booktitle = {2023 IEEE 13th Symposium on Large Data Analysis and Visualization (LDAV)},
  pages     = {32--41},
  year      = {2023},
  doi       = {10.1109/LDAV60332.2023.00011}
}

@inproceedings{grindeanu_cluster19,
  author    = {Grindeanu, Iulian and Peterka, Tom and Mahadevan, Vijay S. and Nashed, Youssef S. G.},
  title     = {Scalable, High-Order Continuity Across Block Boundaries of Functional Approximations Computed in Parallel},
  booktitle = {2019 IEEE International Conference on Cluster Computing (CLUSTER)},
  pages     = {1--9},
  year      = {2019},
  doi       = {10.1109/CLUSTER.2019.8891018}
}

@article{mahadevan_jcs24,
  author  = {Mahadevan, Vijay S. and Lenz, David and Grindeanu, Iulian and Peterka, Thomas},
  title   = {Accelerating multivariate functional approximation computation with domain decomposition techniques},
  journal = {Journal of Computational Science},
  volume  = {78},
  pages   = {102268},
  year    = {2024},
  doi     = {10.1016/j.jocs.2024.102268}
}

@inproceedings{ma_topoinvis24,
  author    = {Ma, Guanqun and Lenz, David and Peterka, Tom and Guo, Hanqi and Wang, Bei},
  title     = {Critical Point Extraction from Multivariate Functional Approximation},
  booktitle = {2024 IEEE Topological Data Analysis and Visualization (TopoInVis)},
  pages     = {12--22},
  year      = {2024},
  doi       = {10.1109/TopoInVis64104.2024.00006}
}

@inproceedings{sun_ldav24,
  author    = {Sun, Jianxin and Lenz, David and Yu, Hongfeng and Peterka, Tom},
  title     = {Adaptive multi-resolution encoding for interactive large-scale volume visualization through functional approximation},
  booktitle = {2024 {IEEE} 14th Symposium on Large Data Analysis and Visualization ({LDAV})},
  pages     = {33--42},
  year      = {2024},
  publisher = {IEEE},
  doi       = {10.1109/LDAV64567.2024.00006}
}

@article{guo_multifidelity_2022,
  author  = {Guo, Mengwu and Manzoni, Andrea and Amendt, Max and Conti, Paolo and Hesthaven, Jan S.},
  title   = {Multi-fidelity regression using artificial neural networks: Efficient approximation of parameter-dependent output quantities},
  journal = {Computer Methods in Applied Mechanics and Engineering},
  volume  = {389},
  pages   = {114378},
  year    = {2022},
  doi     = {10.1016/j.cma.2021.114378}
}

@article{howard_multifidelity_2023,
  author  = {Howard, Amanda A. and Perego, Mauro and Karniadakis, George Em and Stinis, Panagiotis},
  title   = {Multifidelity deep operator networks for data-driven and physics-informed problems},
  journal = {Journal of Computational Physics},
  volume  = {493},
  pages   = {112462},
  year    = {2023},
  doi     = {10.1016/j.jcp.2023.112462}
}

@article{kiener_datadriven_2023,
  author  = {Kiener, Anna and Langer, Stefan and Bekemeyer, Philipp},
  title   = {Data-driven correction of coarse grid {CFD} simulations},
  journal = {Computers {\&} Fluids},
  volume  = {264},
  pages   = {105971},
  year    = {2023},
  doi     = {10.1016/j.compfluid.2023.105971}
}

@article{Kang2023Learning,
  author  = {Kang, Shinhoo and Constantinescu, Emil M.},
  title   = {Learning subgrid-scale models with neural ordinary differential equations},
  journal = {Computers \& Fluids},
  volume  = {261},
  pages   = {105919},
  year    = {2023},
  doi     = {10.1016/j.compfluid.2023.105919}
}

@article{sousa_enhancing_2024,
  author  = {Sousa, Paulo and Rodrigues, Carlos Veiga and Afonso, Alexandre},
  title   = {Enhancing {CFD} solver with machine learning techniques},
  journal = {Computer Methods in Applied Mechanics and Engineering},
  volume  = {429},
  pages   = {117133},
  year    = {2024},
  doi     = {10.1016/j.cma.2024.117133}
}

@misc{kang_differentiable_2025,
  author        = {Kang, Shinhoo and Constantinescu, Emil M.},
  title         = {Enhancing low-order discontinuous {Galerkin} methods with neural ordinary differential equations for compressible {Navier--Stokes} equations},
  year          = {2025},
  eprint        = {2310.18897},
  archivePrefix = {arXiv},
  primaryClass  = {math.NA},
  doi           = {10.48550/arXiv.2310.18897}
}

@misc{Jung2026Learning,
  author        = {Jung, Junoh and Constantinescu, Emil},
  title         = {Learning differentiable weak-form corrections to accelerate finite element simulations},
  year          = {2026},
  eprint        = {2601.20019},
  archivePrefix = {arXiv},
  primaryClass  = {math.NA},
  doi           = {10.48550/arXiv.2601.20019}
}

@article{garg_physics-integrated_2022,
  author  = {Garg, Sarthak and Chakraborty, Souvik and Hazra, Bappaditya},
  title   = {Physics-integrated hybrid framework for model form error identification in nonlinear dynamical systems},
  journal = {Mechanical Systems and Signal Processing},
  volume  = {173},
  pages   = {109039},
  year    = {2022},
  doi     = {10.1016/j.ymssp.2022.109039}
}

@article{raissi2019physics,
  author  = {Raissi, Maziar and Perdikaris, Paris and Karniadakis, George Em},
  title   = {Physics-informed neural networks: A deep learning framework for solving forward and inverse problems involving nonlinear partial differential equations},
  journal = {Journal of Computational Physics},
  volume  = {378},
  pages   = {686--707},
  year    = {2019},
  doi     = {10.1016/j.jcp.2018.10.045}
}

@article{yu2022gradient,
  author  = {Yu, Jeremy and Lu, Lu and Meng, Xuhui and Karniadakis, George Em},
  title   = {Gradient-enhanced physics-informed neural networks for forward and inverse {PDE} problems},
  journal = {Computer Methods in Applied Mechanics and Engineering},
  volume  = {393},
  pages   = {114823},
  year    = {2022},
  doi     = {10.1016/j.cma.2022.114823}
}

@article{shu_physics-informed_2023,
  author  = {Shu, Dule and Li, Zhixuan and Barati Farimani, Amir},
  title   = {A physics-informed diffusion model for high-fidelity flow field reconstruction},
  journal = {Journal of Computational Physics},
  volume  = {478},
  pages   = {111972},
  year    = {2023},
  doi     = {10.1016/j.jcp.2023.111972}
}

@misc{lutjens_flood_2020,
  author        = {L{\"u}tjens, Bj{\"o}rn and Leshchinskiy, Brandon and Requena-Mesa, Christian and Chishtie, Farrukh and D{\'i}az-Rodr{\'i}guez, Natalia and Boulais, Oc{\'e}ane and Pi{\~n}a, Aaron and Newman, Dava and Lavin, Alexander and Gal, Yarin and Ra{\"i}ssi, Chedy},
  title         = {Physics-informed {GANs} for Coastal Flood Visualization},
  year          = {2020},
  eprint        = {2010.08103},
  archivePrefix = {arXiv},
  primaryClass  = {cs.LG},
  doi           = {10.48550/arXiv.2010.08103}
}

@article{chakravarty_hydraulic_2021,
  author  = {Chakravarty, Aditya and Misra, Siddharth and Rai, Chandra S.},
  title   = {Visualization of hydraulic fracture using physics-informed clustering to process ultrasonic shear waves},
  journal = {International Journal of Rock Mechanics and Mining Sciences},
  volume  = {137},
  pages   = {104568},
  year    = {2021},
  doi     = {10.1016/j.ijrmms.2020.104568}
}

@article{banerjee_picv_2024,
  author  = {Banerjee, Chayan and Nguyen, Kien and Fookes, Clinton and Karniadakis, George E.},
  title   = {Physics-Informed Computer Vision: A Review and Perspectives},
  journal = {ACM Computing Surveys},
  volume  = {57},
  number  = {1},
  pages   = {1--38},
  year    = {2024},
  doi     = {10.1145/3689037}
}

@article{ohashi_multifield_2025,
  author  = {Ohashi, Nagahiro and Hwang, Leslie K. and Kwon, Beomjin},
  title   = {Physics-informed neural networks for multi-field visualization with single-color laser induced fluorescence},
  journal = {AI Thermal Fluids},
  volume  = {1},
  pages   = {100005},
  year    = {2025},
  doi     = {10.1016/j.aitf.2025.100005}
}

@article{cox1972bspline,
  author  = {Cox, Maurice G.},
  title   = {The numerical evaluation of {B}-splines},
  journal = {IMA Journal of Applied Mathematics},
  volume  = {10},
  number  = {2},
  pages   = {134--149},
  year    = {1972},
  doi     = {10.1093/imamat/10.2.134}
}

@article{deboor1972bspline,
  author  = {de Boor, Carl},
  title   = {On calculating with {B}-splines},
  journal = {Journal of Approximation Theory},
  volume  = {6},
  number  = {1},
  pages   = {50--62},
  year    = {1972},
  doi     = {10.1016/0021-9045(72)90080-9}
}

@article{Liu_1989,
  author  = {Liu, Dong C. and Nocedal, Jorge},
  title   = {On the limited memory {BFGS} method for large scale optimization},
  journal = {Mathematical Programming},
  volume  = {45},
  number  = {1--3},
  pages   = {503--528},
  year    = {1989},
  doi     = {10.1007/BF01589116}
}

@misc{qiu_lbfgspp,
  author       = {Qiu, Yixuan},
  title        = {{LBFGS++} ({LBFGSpp}): A header-only {C++} library for {L-BFGS} and {L-BFGS-B} algorithms},
  year         = {2025},
  howpublished = {GitHub repository},
  url          = {https://github.com/yixuan/LBFGSpp},
  note         = {Version v0.4.0, commit c524a40}
}

@book{Crank1975Diffusion,
  author    = {Crank, John},
  title     = {The Mathematics of Diffusion},
  edition   = {2},
  publisher = {Clarendon Press},
  address   = {Oxford},
  year      = {1975},
  isbn      = {9780198534112}
}

@book{Haberman2013AppliedPDE,
  author    = {Haberman, Richard},
  title     = {Applied Partial Differential Equations: With Fourier Series and Boundary Value Problems},
  edition   = {5},
  publisher = {Pearson},
  address   = {Boston},
  year      = {2013},
  isbn      = {9780321797056}
}

@article{lenz_fourier-informed_2023,
  author  = {Lenz, David and Marin, Oana and Mahadevan, Vijay S. and Yeh, Raine and Peterka, Tom},
  title   = {Fourier-informed knot placement schemes for {B}-spline approximation},
  journal = {Mathematics and Computers in Simulation},
  volume  = {213},
  pages   = {374--393},
  year    = {2023},
  doi     = {10.1016/j.matcom.2023.05.017}
}

@book{Pareto1971,
  author    = {Pareto, Vilfredo},
  title     = {Manual of Political Economy},
  publisher = {A. M. Kelley},
  address   = {New York},
  year      = {1971}
}

@article{Marler2004Survey,
  author  = {Marler, R. Timothy and Arora, Jasbir S.},
  title   = {Survey of multi-objective optimization methods for engineering},
  journal = {Structural and Multidisciplinary Optimization},
  volume  = {26},
  number  = {6},
  pages   = {369--395},
  year    = {2004},
  doi     = {10.1007/s00158-003-0368-6}
}

@article{Marler2010WeightedSum,
  author  = {Marler, R. Timothy and Arora, Jasbir S.},
  title   = {The weighted sum method for multi-objective optimization: new insights},
  journal = {Structural and Multidisciplinary Optimization},
  volume  = {41},
  number  = {6},
  pages   = {853--862},
  year    = {2010},
  doi     = {10.1007/s00158-009-0460-7}
}

@article{Rohrhofer2021ParetoPINN,
  author  = {Rohrhofer, Franz M. and Posch, Stefan and G{\"o}{\ss}nitzer, Clemens and Geiger, Bernhard C.},
  title   = {Data vs. Physics: The apparent {Pareto} front of physics-informed neural networks},
  journal = {IEEE Access},
  volume  = {11},
  pages   = {86252--86261},
  year    = {2023},
  doi     = {10.1109/ACCESS.2023.3302892},
  eprint  = {2105.00862},
  archivePrefix = {arXiv}
}

@article{Heldmann2023BiObjectivePINN,
  author  = {Heldmann, Fabian and Berkhahn, Sarah and Ehrhardt, Matthias and Klamroth, Kathrin},
  title   = {{PINN} training using biobjective optimization: The trade-off between data loss and residual loss},
  journal = {Journal of Computational Physics},
  volume  = {488},
  pages   = {112211},
  year    = {2023},
  doi     = {10.1016/j.jcp.2023.112211}
}

@article{Bischof2025MultiObjectivePINN,
  author  = {Bischof, Rafael and Kraus, Michael A.},
  title   = {Multi-objective loss balancing for physics-informed deep learning},
  journal = {Computer Methods in Applied Mechanics and Engineering},
  volume  = {439},
  pages   = {117914},
  year    = {2025},
  doi     = {10.1016/j.cma.2025.117914}
}

@article{Cole1951,
  author  = {Cole, Julian D.},
  title   = {On a quasi-linear parabolic equation occurring in aerodynamics},
  journal = {Quarterly of Applied Mathematics},
  volume  = {9},
  number  = {3},
  pages   = {225--236},
  year    = {1951},
  doi     = {10.1090/qam/42889}
}

@book{LeVeque2002FVM,
  author    = {LeVeque, Randall J.},
  title     = {Finite Volume Methods for Hyperbolic Problems},
  publisher = {Cambridge University Press},
  address   = {Cambridge},
  year      = {2002},
  doi       = {10.1017/CBO9780511791253}
}

@article{ClarkDiLeoni2018BurgersGaussians,
  author  = {{Clark di Leoni}, Patricio and Hirata, Alan N. and Mininni, Pablo D.},
  title   = {Dynamics of partially thermalized solutions of the {Burgers} equation},
  journal = {Physical Review Fluids},
  volume  = {3},
  pages   = {014603},
  year    = {2018},
  doi     = {10.1103/PhysRevFluids.3.014603}
}

@book{Evans2010,
  author    = {Evans, Lawrence C.},
  title     = {Partial Differential Equations},
  series    = {Graduate Studies in Mathematics},
  volume    = {19},
  edition   = {2},
  publisher = {American Mathematical Society},
  address   = {Providence, RI},
  year      = {2010},
  isbn      = {9780821849743}
}

@book{Borthwick2017,
  author    = {Borthwick, David},
  title     = {Introduction to Partial Differential Equations},
  publisher = {Springer},
  address   = {Cham},
  year      = {2017},
  doi       = {10.1007/978-3-319-48936-0}
}

@article{KennedyOHagan2001,
  author  = {Kennedy, Marc C. and O'Hagan, Anthony},
  title   = {Bayesian calibration of computer models},
  journal = {Journal of the Royal Statistical Society: Series B (Statistical Methodology)},
  volume  = {63},
  number  = {3},
  pages   = {425--464},
  year    = {2001},
  doi     = {10.1111/1467-9868.00294}
}

@article{Hooker2009,
  author  = {Hooker, Giles},
  title   = {Forcing function diagnostics for nonlinear dynamics},
  journal = {Biometrics},
  volume  = {65},
  number  = {3},
  pages   = {928--936},
  year    = {2009},
  doi     = {10.1111/j.1541-0420.2008.01172.x}
}

@article{MorrisonOliverMoser2018,
  author  = {Morrison, Rebecca E. and Oliver, Todd A. and Moser, Robert D.},
  title   = {Representing model inadequacy: A stochastic operator approach},
  journal = {SIAM/ASA Journal on Uncertainty Quantification},
  volume  = {6},
  number  = {2},
  pages   = {457--496},
  year    = {2018},
  doi     = {10.1137/16M1106419}
}

@article{MasudNasharGoraya2023,
  author  = {Masud, Arif and Nashar, Sohaib and Goraya, Syed A.},
  title   = {Physics-Constrained Data-Driven Variational method for missing physics and model-form error},
  journal = {Computer Methods in Applied Mechanics and Engineering},
  volume  = {417},
  pages   = {116295},
  year    = {2023},
  doi     = {10.1016/j.cma.2023.116295}
}

@manual{FiredrakeUserManual,
  author       = {Ham, David A. and Kelly, Paul H. J. and Mitchell, Lawrence and Cotter, Colin J. and Kirby, Robert C. and Sagiyama, Koki and Bouziani, Nacime and Vorderwuelbecke, Sophia and Gregory, Thomas J. and Betteridge, Jack and Shapero, Daniel R. and Nixon-Hill, Reuben W. and Ward, Connor J. and Farrell, Patrick E. and Brubeck, Pablo D. and Marsden, India and Gibson, Thomas H. and Homolya, Mikl{\'o}s and Sun, Tianjiao and McRae, Andrew T. T. and Luporini, Fabio and Gregory, Alastair and Lange, Michael and Funke, Simon W. and Rathgeber, Florian and Bercea, Gheorghe-Teodor and Markall, Graham R.},
  title        = {Firedrake User Manual},
  organization = {Imperial College London and University of Oxford and Baylor University and University of Washington},
  edition      = {First edition},
  year         = {2023},
  month        = may,
  doi          = {10.25561/104839}
}

@article{GHIA1982387,
  author  = {Ghia, U. and Ghia, K. N. and Shin, C. T.},
  title   = {High-{Re} solutions for incompressible flow using the {Navier--Stokes} equations and a multigrid method},
  journal = {Journal of Computational Physics},
  volume  = {48},
  number  = {3},
  pages   = {387--411},
  year    = {1982},
  doi     = {10.1016/0021-9991(82)90058-4}
}

@article{shen1991hopf,
  author  = {Shen, Jie},
  title   = {Hopf bifurcation of the unsteady regularized driven cavity flow},
  journal = {Journal of Computational Physics},
  volume  = {95},
  number  = {1},
  pages   = {228--245},
  year    = {1991},
  doi     = {10.1016/0021-9991(91)90261-I}
}

@article{Kochkov2021Machine,
  author  = {Kochkov, Dmitrii and Smith, Jamie A. and Alieva, Ayya and Wang, Qing and Brenner, Michael P. and Hoyer, Stephan},
  title   = {Machine learning-accelerated computational fluid dynamics},
  journal = {Proceedings of the National Academy of Sciences},
  volume  = {118},
  number  = {21},
  pages   = {e2101784118},
  year    = {2021},
  doi     = {10.1073/pnas.2101784118}
}

@misc{Pathak2020Using,
  author        = {Pathak, Jaideep and Mustafa, Mustafa and Kashinath, Karthik and Motheau, Emmanuel and Kurth, Thorsten and Day, Marcus},
  title         = {Using machine learning to augment coarse-grid computational fluid dynamics simulations},
  year          = {2020},
  eprint        = {2010.00072},
  archivePrefix = {arXiv},
  primaryClass  = {physics.comp-ph},
  doi           = {10.48550/arXiv.2010.00072}
}

@inproceedings{Li2021Fourier,
  author    = {Li, Zongyi and Kovachki, Nikola and Azizzadenesheli, Kamyar and Liu, Burigede and Bhattacharya, Kaushik and Stuart, Andrew and Anandkumar, Anima},
  title     = {Fourier neural operator for parametric partial differential equations},
  booktitle = {International Conference on Learning Representations},
  year      = {2021},
  url       = {https://openreview.net/forum?id=c8P9NQVtmnO},
  eprint    = {2010.08895},
  archivePrefix = {arXiv},
  doi       = {10.48550/arXiv.2010.08895}
}

@book{hughes1987finite,
  author    = {Hughes, Thomas J. R.},
  title     = {The Finite Element Method: Linear Static and Dynamic Finite Element Analysis},
  publisher = {Prentice-Hall},
  address   = {Englewood Cliffs, NJ},
  year      = {1987},
  isbn      = {9780133170253}
}

@book{ern2004theory,
  author    = {Ern, Alexandre and Guermond, Jean-Luc},
  title     = {Theory and Practice of Finite Elements},
  series    = {Applied Mathematical Sciences},
  volume    = {159},
  publisher = {Springer},
  address   = {New York},
  year      = {2004},
  doi       = {10.1007/978-1-4757-4355-5}
}

@article{raissi2020hidden,
  author  = {Raissi, Maziar and Yazdani, Alireza and Karniadakis, George Em},
  title   = {Hidden fluid mechanics: Learning velocity and pressure fields from flow visualizations},
  journal = {Science},
  volume  = {367},
  number  = {6481},
  pages   = {1026--1030},
  year    = {2020},
  doi     = {10.1126/science.aaw4741}
}

@book{cottrell2009isogeometric,
  author    = {Cottrell, John A. and Hughes, Thomas J. R. and Bazilevs, Yuri},
  title     = {Isogeometric Analysis: Toward Integration of {CAD} and {FEA}},
  publisher = {Wiley},
  address   = {Chichester},
  year      = {2009},
  doi       = {10.1002/9780470749081}
}

 \begin{center}
	\scriptsize \framebox{\parbox{5in}{Government License (will be removed at publication):
			The submitted manuscript has been created by UChicago Argonne, LLC,
			Operator of Argonne National Laboratory (``Argonne").  Argonne, a
			U.S. Department of Energy Office of Science laboratory, is operated
			under Contract No. DE-AC02-06CH11357.  The U.S. Government retains for
			itself, and others acting on its behalf, a paid-up nonexclusive,
			irrevocable worldwide license in said article to reproduce, prepare
			derivative works, distribute copies to the public, and perform
			publicly and display publicly, by or on behalf of the Government. The Department of Energy will provide public access to these results of federally sponsored research in accordance with the DOE Public Access Plan. http://energy.gov/downloads/doe-public-access-plan.
}}
	\normalsize
\end{center}

\end{document}